\documentclass[journal,12pt,onecolumn,draftclsnofoot,]{IEEEtran}
\usepackage{mathtools}
\mathtoolsset{centercolon}
\DeclarePairedDelimiter{\abs}{\lvert}{\rvert}

\DeclarePairedDelimiter{\pra}{\lparen}{\rparen} 

\usepackage{xr}
\usepackage{cite}
\usepackage{amsmath}
\usepackage{amsthm}
\usepackage{graphicx}
\usepackage{textcomp}
\usepackage{physics}
\usepackage{amsfonts}
\usepackage{amssymb}
\usepackage{bbold}
\usepackage{enumitem}
\usepackage{threeparttable}
\usepackage{multirow}
\usepackage{booktabs}
\usepackage[caption=false,font=footnotesize]{subfig}
\usepackage[dvipsnames]{xcolor}
\usepackage{soul}

\theoremstyle{plain}
\newtheorem{thm}{Theorem}

\newtheorem{lem}{Lemma}
\newtheorem{cor}{Corollary}
\newtheorem{claim}{Claim}
\newtheorem{rk}{Remark}

\theoremstyle{definition}
\newtheorem{defn}{Definition}

\newcommand{\mc}[1]{\mathcal{#1}}
\newcommand{\set}[1]{\left\{ #1 \right\}}

\DeclareMathOperator{\supp}{supp}
\DeclareMathOperator*{\argmax}{arg\,max}

\newcommand{\rhoN}{\hat{\rho}^n}
\newcommand{\rhoNotN}{ \hat{\rho}^{\otimes n}_0 }
\newcommand{\rhoNot}{ \hat{\rho}_0 }
\newcommand{\rhoBarN}{\hat{\bar{\rho}}^n}
\newcommand{\rhoAlphaN}{\hat{\rho}^{\otimes n}_{\alpha_n}}
\newcommand{\rhoAlpha}{\hat{\rho}_{\alpha_n}}
\newcommand{\sigmaNotN}{\hat{\sigma}^{\otimes n}_0}

\newcommand{\sigmaNx}{\hat{\sigma}^n (\mathbf{x})}
\newcommand{\sigmaAlphaN}{\hat{\sigma}_{\alpha_n}^{\otimes n}}

\usepackage{soul}

\allowdisplaybreaks
\newcommand{\mtt}[1]{{#1}}

\begin{document}
\title{Fundamental Limits of Covert Communication over Classical-Quantum Channels}

\author{\IEEEauthorblockN{Michael S. Bullock\IEEEauthorrefmark{1},
Azadeh Sheikholeslami\IEEEauthorrefmark{2},
Mehrdad Tahmasbi\IEEEauthorrefmark{3},
Robert C. Macdonald\IEEEauthorrefmark{1},
Saikat Guha\IEEEauthorrefmark{4},
Boulat A. Bash\IEEEauthorrefmark{1}}\\
\IEEEauthorblockA{\IEEEauthorrefmark{1}Dept.~of Electrical and
	Computer Engineering, University of Arizona, Tucson, AZ
	}\\
 \IEEEauthorblockA{\IEEEauthorrefmark{2}Dept.~of Mathematics \& Computer Science, Suffolk University, Boston, MA 
	}\\
\IEEEauthorblockA{\IEEEauthorrefmark{3}Dept.~of Computer Science, University of Illinois at Urbana-Champaign, Urbana, IL
}\\
\IEEEauthorblockA{\IEEEauthorrefmark{4}Dept.~of Electrical and
	Computer Engineering, University of Maryland, College Park, MD
	}

\thanks{This research was funded by the National Science Foundation (NSF) under grants ECCS-1309573, CNS-1564067, and CCF-2006679, and DARPA under contract number HR0011-16-C-0111.
This paper was presented in part at the IEEE International Symposium on Information Theory (ISIT), July 2016 \cite{azadeh16quantumcovert-isit} and at Beyond IID in Information Theory, July 2024 \cite{bullock24biid}.}
}
\maketitle
\begin{abstract}
We investigate covert communication over general memoryless classical-quantum channels with fixed finite-size input alphabets.
We show that the square root law (SRL) governs covert communication in this setting when product a of $n$ input states is used: $L_{\rm SRL}\sqrt{n}+o(\sqrt{n})$ covert bits (but no more) can be reliably transmitted in $n$ uses of classical-quantum channel,  where $L_{\rm SRL}>0$ is a channel-dependent constant that we call \emph{covert capacity}.
We also show that ensuring covertness requires $J_{\rm SRL}\sqrt{n}+o(\sqrt{n})$ bits secret key shared by the communicating parties prior to transmission, where $J_{\rm SRL}\geq0$ is a channel-dependent constant.
We assume a quantum-powerful adversary that can perform an arbitrary joint (entangling) measurement on all $n$ channel uses.
We determine the single-letter expressions for $L_{\rm SRL}$ and $J_{\rm SRL}$, and establish conditions when $J_{\rm SRL}=0$ (i.e., no pre-shared secret key is needed).
Finally, we evaluate the scenarios where covert communication is not governed by the SRL.

\end{abstract}

\section{Introduction}\label{sec:introduction}
Security is critical to communication.  Cryptography \cite{talb2006} and information-theoretic secrecy \cite{liang09itsec,bloch11pls} methods protect against extraction of the information from a message by an unauthorized party, however, they do not prevent the detection of the message transmission. This motivates the exploration of the information-theoretic limits of {\em covert} communications, i.e., communicating with low probability of detection/interception (LPD/LPI). 

\begin{figure}
	\centering
	\includegraphics[width=0.78\textwidth]{setup.png}
	\caption{Covert communication setting. Alice has a noisy channel to legitimate receiver Bob and adversary Willie. Alice encodes message $W$ with blocklength $n$ code and chooses whether to transmit. Willie observes his channel from Alice to determine whether she is quiet (null hypothesis $H_0$) or not (alternate hypothesis $H_1$). Alice and Bob's coding scheme must ensure that any detector Willie uses is close to ineffective (i.e., a random guess between the hypotheses), while allowing Bob to reliably decode the message (if one is transmitted). Alice and Bob may share a resource (e.g., a secret key exchanged prior to transmission.)}
	\label{fig:setup}
\end{figure}

Consider a broadcast channel setting in Figure \ref{fig:setup} typical in the study of the fundamental limits of secure communications, where the intended receiver Bob and adversary Willie receive a sequence of $n$ input symbols from Alice that are corrupted by noise. Let's label one of the input symbols (say, zero) as the ``innocent symbol'' indicating ``no transmission by Alice,'' whereas the other symbols correspond to transmissions, and are, therefore, ``non-innocent.''
Alice must maintain covertness by ensuring that Willie's probability of detection error approaches that resulting from a random guess.
At the same time, Alice's transmission must be reliable in the usual sense of Bob's probability of decoding error vanishing as $n\to\infty$. 
Then, the properties of the channels from Alice to Willie and Bob result in the following numbers of covert and reliably transmissible bits in $n$ channel uses: 
\begin{enumerate}[label=({\Alph*}), ref={\Alph*}]
\item \label{item:srl} \emph{square root law} (SRL): $L_{\rm SRL}\sqrt{n}+o(\sqrt{n})$ covert bits (but no more), where $L_{\rm SRL}>0$,
\item  \label{item:corner}non-SRL covert communication:
\begin{enumerate}[label={\arabic*}., ref={\arabic*}]
\item \label{item:nogo} zero covert bits,
\item \label{item:const} constant-rate covert communication: $L_{\rm lin}n+o(n)$ covert bits, where  $L_{\rm lin}>0$,
\item \label{item:srll} square root-log law: $L_{\log}\sqrt{n}\log n+o(\sqrt{n}\log n)$ covert bits (but no more), with  $L_{\log}>0$. 
\end{enumerate}
\end{enumerate}
The research on the fundamental limits of covert communications has focused on the SRL in (\ref{item:srl}), whereas scenarios in (\ref{item:corner}) are special cases.
Note that, except in case (\ref{item:corner}.\ref{item:const}), the ``standard'' channel capacity is a poor measure of the channel's capability to transmit covert information.
This is because, as $n\to\infty$, the number of covert bits transmitted per channel use decays to zero under both the SRL and the square root-log law.
Therefore, we call $L_{\rm SRL}$ and $L_{\log}$ the \emph{covert capacities} for the respective channels formally discussed in Sections \ref{sec:achievability} and \ref{subsec:sqrtnlogn}.

The authors of \cite{bash12sqrtlawisit, bash13squarerootjsacnonote} examined covert communications when Alice has additive white Gaussian noise (AWGN) channels to both Willie and Bob.
They found that the SRL governs covert communications, and that, to achieve it, Alice and Bob may have to share a resource which is inaccessible by Willie.
When necessary, this \emph{shared secret key} is assumed to be exchanged by Alice and Bob prior to communicating. 
The follow-on work on the SRL for binary symmetric channels (BSCs) \cite{che13sqrtlawbscisit} showed its achievability without the use of a shared resource, provided that Bob has a better channel from Alice than Willie.
The SRL was further generalized to the entire class of discrete memoryless channels (DMCs) \cite{bloch15covert,wang15covert} with \cite{bloch15covert} finding that $J_{\rm SRL}\sqrt{n}+o(\sqrt{n})$ shared secret key bits were sufficient.  
However, the key contribution of \cite{wang15covert, bloch15covert} was the characterization of the covert capacities $L_{\rm SRL}$ and $J_{\rm SRL}$ as functions of the channel parameters (including DMC transition probabilities and AWGN power).
The non-SRL special cases in (\ref{item:corner}) were introduced and characterized in \cite[App.~G]{bloch15covert}.
We note that, while zero is the natural innocent symbol for channels that take continuous-valued input (such as the AWGN channel), in the analysis of the discrete channel setting an arbitrary input is designated as innocent. 
A tutorial overview of this research can be found in \cite{bash15covertcommmag}.

The SRL also governs the fundamental limits of covert communications over a lossy thermal-noise bosonic channel \cite{bash15covertbosoniccomm,bullock20discretemod,gagatsos20codingcovcomm}, which is a quantum description of  optical communications  in many practical scenarios (with vacuum being the innocent input). Notably, the SRL is achievable in this setting even when Willie captures all the photons that do not reach Bob, performs an arbitrary measurement that is only limited by the laws of quantum mechanics, and has access to unlimited quantum storage and computing capabilities.  Furthermore, the SRL cannot be surpassed when Alice and Bob are limited to sharing a classical resource, even if they employ an encoding/measurement/decoding scheme limited only by the laws of quantum mechanics, including the transmission of codewords entangled over many channel uses and making collective measurements \cite{bullock20discretemod,gagatsos20codingcovcomm}. However, a quantum resource such as shared entanglement allows the use of entanglement-assisted (EA) communication methods to improve from SRL scaling to square root-log law and transmission of $L_{\rm EA}\sqrt{n}\log n+o(\sqrt{n}\log n)$ covert bits in $n$ channel uses, where $L_{\rm EA}>0$ \cite{gagatsos20codingcovcomm}. 

The covert capacities with and without entanglement assistance for bosonic channel $L_{\rm EA}$ and $L_{\rm SRL}$ have been characterized in \cite{gagatsos20codingcovcomm}.
The optimal shared secret key sizes for covert bosonic channel with and without entanglement assistance were derived in \cite{wang24resource} and \cite{wang22tracedistance}, respectively.
These facts and the successful demonstration of the SRL for a bosonic channel in \cite{bash15covertbosoniccomm} motivate a generalization to arbitrary quantum channels, which is the focus of this article. 
We study covert communication using product-state inputs over a memoryless classical-quantum channel, which is generalization of the DMC that maps a finite set of discrete classical inputs to quantum states at the output.
We generalize \cite{bloch15covert} by assuming that both Willie and Bob are limited only by the laws of quantum mechanics, and, thus, can perform arbitrary joint measurement over all $n$ channel uses.
We show that the SRL holds when Alice and Bob are restricted to a classical resource and provide expressions for covert capacity $L_{\rm SRL}$ and shared resource requirement $J_{\rm SRL}$.
We also determine these quantities when Bob is restricted to a symbol-by-symbol measurement.
Moreover, we develop explicit conditions that differentiate the special cases of non-SRL covert communication for classical-quantum channels given in (\ref{item:corner}) above and derive the bounds for the corresponding $L_{\log}$.
In contrast to the non-covert private capacity of the classical-quantum channel \cite{cai04, devetak05quantumwiretap}, we obtain single-letter expressions for $L_{\rm SRL}$ and $L_{\log}$.
Although we adapt some of the classical approaches from the proofs in  \cite{bloch15covert, wang15covert}  to classical-quantum channels, the challenges posed by the quantum setting require entirely different set of techniques to obtain our results.

Our characterization of covert communication over a classical-quantum channel with product-state input and classical shared resource motivates important follow-on work on the impact of quantum resources.
In fact, results in this paper have already been used to analyze the impact of EA on covert classical communication over qubit depolarizing channel. 
In \cite{zlotnick23eacovcommdepol}, it was shown that EA yields a scaling gain from SRL to square root-log law and transmission of $L_{\rm EA}\sqrt{n}\log n+o(\sqrt{n}\log n)$ covert bits in $n$ channel uses.
Hence, since the scaling gain from EA for this discrete-values channel is the same as for continuous-valued bosonic channel in \cite{gagatsos20codingcovcomm}, it is the shared entanglement rather than the dimensionality of the input that yields the logarithmic gain.
However, although it is well-known that Holevo capacity is super-additive in general, and that certain classical-quantum channels benefit from inputs that are entangled over all $n$ uses \cite{hastings09superadditivity}, whether covert transmission can derive similar benefit is still an open problem.

The paper is organized as follows: in Section \ref{seq:prereqs}, we present the prerequisite background, our channel model, and covertness metrics as well as provide necessary definitions and lemmas for our results. In Sections \ref{sec:achievability} and \ref{sec:converse}, we state and prove the achievability and converse for the covert capacity of classical-quantum channels. In Section \ref{sec:cornercases}, we examine special cases of covert communication that are not governed by the square root law. We conclude in Section \ref{sec:discussion} with a discussion of our results and avenues for future work. 

\section{Prerequisites}\label{seq:prereqs}

\subsection{Notation}
\subsubsection{Linear operators}
\label{sec:notation_linear_operators}
We employ the standard notation used in quantum information processing, found in, e.g., \cite[Ch.~2.2.1]{tomamichel15finiteresourcesQIP}, \cite{hayashi2006quantum}. For a finite-dimensional Hilbert space $\mathcal{H}$, we denote its dimension by $\dim \mathcal{H}$. The space of linear operators (resp. density operators) on $\mathcal{H}$ is denoted by $\mathcal{L}(\mathcal{H})$ (resp. $\mathcal{D}(\mathcal{H})$). We use hats for operators, e.g., $\hat{A} \in \mathcal{L}(\mathcal{H})$. Trace of $\hat{A}$ is $\Tr[\hat{A}]$.  The kernel $\ker(\hat{A})$ of $\hat{A}$ is the subspace of $\mathcal{H}$ spanned by vectors $\ket{v}\in\mathcal{H}$ satisfying $\hat{A}\ket{v}=0$. The support $\supp(\hat{A})$ of $\hat{A}$ is the orthogonal complement of $\ker(\hat{A})$ in $\mathcal{H}$. For $\hat{A}, \hat{B} \in \mathcal{L}(\mathcal{H})$, we use $\hat{A} \succ \hat{B}$  (resp. $\hat{A} \succeq \hat{B}$) to specify that the operator $\hat{A}-\hat{B}$ is positive definite (resp. positive semi-definite).
The $i^{\text{th}}$ eigenvalue and singular value of $\hat{A}$ are denoted by $\lambda_i(\hat{A})$ and $\sigma_i(\hat{A})$, respectively.  The eigenvector corresponding to $\lambda_i(\hat{A})$ is denoted by $\ket{\lambda_i(\hat{A})}$.  When the context is clear, we drop explicit specification of $\hat{A}$. We call an operator $\hat{A}$ Hermitian if $\hat{A} = \hat{A}^\dagger$ where $\hat{A}^\dagger$ denotes the adjoint of $\hat{A}$. 
For a Hermitian operator $\hat{A}$ with spectral decomposition $\hat{A} = \sum_{i} \lambda_i(\hat{A}) \ket{\lambda_i(\hat{A})}\bra{\lambda_i(\hat{A})}$, 
$\lambda_{\min}(\hat{A})$ and $\lambda_{\max}(\hat{A})$ denote the minimum and maximum non-zero eigenvalues of $\hat{A}$.
Furthermore, we define the projection to the eigenspace corresponding to non-negative eigenvalues of $\hat{A}$ as
\begin{align}
\{ \hat{A} \succeq 0 \} = \sum_{i:\lambda_i(\hat{A}) \geq 0} \ket{\lambda_i(\hat{A})}\bra{\lambda_i(\hat{A})}, \label{eq:projDefn}
\end{align}
with projections $\{ \hat{A} \succ 0\}$, $\{ \hat{A} \preceq 0 \}$, and $\{ \hat{A} \prec 0 \}$ defined similarly as in \cite{hayashi2003general}. The trace norm  (or Schatten 1-norm) of $\hat{A}$ is $\|\hat{A}\|_1=\Tr\left[\sqrt{\hat{A}^\dagger\hat{A}}\right]=\sum_i\sqrt{\lambda_i(\hat{A}^\dagger\hat{A})}$ \cite[Def. 9.1.1]{wilde16quantumit2ed}, whereas the supremum norm $\|\hat{A}\|_\infty=\sqrt{\lambda_{\max}(\hat{A}^\dagger\hat{A})}$ is the largest singular value of $\hat{A}$.

\subsubsection{Random variables}
We use capital letters for scalar random variables and corresponding small letters for their realizations (e.g., $X$ and $x$), and we use bold capital letters for random vectors and corresponding bold small letters for their realizations (e.g., $\mathbf{X}$ and $\mathbf{x}$). 

\subsubsection{Asymptotics}
We employ the standard asymptotic notation \cite[Ch. 3.1]{clrs2e}, where 
\begin{align}
\mathcal{O}(g(n)) &\triangleq \{f(n) : \exists m,n_0 >0 \text{ s.t. } 0\leq f(n) \leq m g(n) \text{ } \forall n \geq n_0\}\\
&=\left\{f(n) : \limsup_{n\to\infty}\left|\frac{f(n)}{g(n)}\right| <\infty\right\}\\
o(g(n))&\triangleq \{f(n) : \forall m>0 \text{,	 } \exists n_0 > 0 \text{ s.t. } 0\leq f(n) < m  g(n)\text{ } \forall n \geq n_0 \}\\
&=\left\{f(n) : \lim_{n\to\infty}\frac{f(n)}{g(n)} =0\right\}\\
\Omega(g(n))&\triangleq\{f(n) : \exists m, n_0 >0 \text{ s.t. } 0\leq m g(n) \leq f(n)\text{ }\forall	n \geq n_0\}\\
&=\left\{f(n) : \liminf_{n\to\infty}\frac{f(n)}{g(n)} >0\right\}\\	
\omega(g(n))&\triangleq \{f(n) : \forall m>0, \exists n_0 >0 \text{ s.t. }0\leq m g(n) < f(n)\text{ }\forall	n \geq n_0\}\\
&=\left\{f(n) : \lim_{n\to\infty}\left|\frac{f(n)}{g(n)}\right| = \infty\right\}.
\end{align}
Thus, $\mathcal{O}(g(n))$ and $\Omega(g(n))$ denote asymptotically tight upper and lower bounds on $g(n)$, whereas $o(g(n))$ and $\omega(g(n))$ denote upper and lower bounds on $g(n)$ that are not asymptotically tight.
Finally, $\Theta(g(n))\triangleq\mathcal{O}(g(n))\cap\Omega(g(n))$.
	
\subsection{Channel Model}\label{sec:channel_model}
\begin{figure}
	\centering
	\includegraphics[width=0.95\textwidth]{cqchannel-model.png}
	\caption{Covert classical-quantum channel setting. Alice encodes message $m$ drawn from random variable $W$ by using the pre-shared secret key $k$ drawn from random variable $S$ into $\mathbf{x}(m,k) \in \mathcal{X}^n$. She then transmits $\mathbf{x}(m,k)$ in $n$ uses of the classical-quantum channel. Bob uses pre-shared secret key $k$ to select POVM $\left\{\hat{\Lambda}^n_{m,k}\right\}_{m\in \{1,\ldots, M\}}$, and obtain an estimate of the message $\check{W}$ from his received quantum state $\hat{\sigma}_B^n(\mathbf{x}(m,k))$. Willie performs a measurement to determine whether his quantum state $\hat{\rho}_W^n(\mathbf{x}(m,k))$ corresponds to innocent input $\mathbf{0}$ (null hypothesis $H_0$) or not (alternate hypothesis $H_1$).}
	\label{fig:channelmodel}
\end{figure}
 We focus on the covert memoryless classical-quantum channel described in Figure \ref{fig:channelmodel}. Consider a discrete input alphabet $\mathcal{X}=\{0,1,2,\ldots,N\}$.  For a single use of the channel, Alice maps her classical input $x\in\mathcal{X}$ to a quantum state $\hat{\phi}_A(x) \in \mathcal{D}(\mathcal{H}_A)$ and transmits via a quantum channel represented by a completely positive trace-preserving (CPTP) map $\mathcal{N}_{A\to BW}$, resulting in the output state $\hat{\tau}_{BW}(x) = \mathcal{N}_{A\to BW}(\hat{\phi}_A(x)) \in \mathcal{D}(\mathcal{H}_B\otimes\mathcal{H}_W)$. Thus, a single use of the channel takes a classical input $x \in \mathcal{X}$ to quantum outputs $\Tr_W\left[\hat{\tau}_{BW}(x)\right]=\hat{\sigma}_B(x) = \hat{\sigma}_x \in \mathcal{D}(\mathcal{H}_B)$ and $\Tr_B\left[\hat{\tau}_{BW}(x)\right]=\hat{\rho}_W(x) = \hat{\rho}_x\in \mathcal{D}(\mathcal{H}_W)$ at Bob and Willie, respectively. We drop system labels and relegate the classical input label to the subscript for brevity when the context is clear.

\subsection{Covert Communication Scheme and Decoding Reliability}
A covert communication    scheme is a set $\mathcal{K}_n=\left\{\mathcal{K}_{k,n}\right\}_{k\in\{1,\ldots, K\}}$ of $K$ codes, where $\mathcal{K}_{k,n}=\left\{\mathbf{x}(m,k),\hat{\Lambda}^n_{m,k}\right\}_{m \in \{1,\ldots, M\}}$ is an $(M,n)$ classical-quantum code containing $n$-symbol encoding vectors $\mathbf{x}(m,k)$ and corresponding elements of positive operator-valued measure (POVM) $\hat{\Lambda}^n_{m,k}$.  
Alice and Bob use their pre-shared secret key $k\in\{1,\ldots, K\}$ to select a code from $\mathcal{K}_n$.
Suppose Alice desires to transmit message $m \in \{1,\ldots, M\}$ to Bob. Given $k$, her encoder maps $m$ to vector $\mathbf{x}(m,k)$, and transmits it on the classical-quantum channel described above. Bob receives quantum state
\begin{align}
\hat{\sigma}_B^n(\mathbf{x}(m,k)) = \hat{\sigma}^n(m,k)=\bigotimes_{i=1}^n \hat{\sigma}_{x_i(m,k)},
\end{align}
where $x_i(m,k)$ denotes the $i^{\text{th}}$ element of $\mathbf{x}(m,k)$, and we drop system labels and relegate the classical input label to the subscript for brevity.
Bob uses $k$ to select a POVM $\left\{\hat{\Lambda}^n_{m,k}\right\}_{m\in \{1,\ldots, M\}}$ and estimate $m$ from his received state. If the message and secret key are selected uniformly at random, Bob's average probability of decoding error is 
\begin{align}
P_{\rm e}^{(b)} = \frac{1}{KM}\sum_{m=1}^M\sum_{k=1}^K\left(1-\Tr\left[\hat{\sigma}^n(m,k)\hat{\Lambda}^n_{m,k}\right]\right).\label{eq:BobDecodingErrorP}
\end{align}
We call a communication system reliable if, for a sequence of schemes $\left(\mathcal{K}_n\right)$ with increasing blocklength $n$, $\lim_{n\to\infty} P_{\rm e}^{(b)} = 0$.

\subsection{Hypothesis Testing and Covertness Criteria} \label{sec:hypothesis_testing}
Suppose $x=0$ corresponds to the innocent input, indicating that Alice not transmitting. 
Willie has access to the sequence of covert communication schemes $\left(\mathcal{K}_n\right)$, but no knowledge of $k$ and $m$. Therefore, he must distinguish between the state that he receives when no communication occurs (null hypothesis $H_0$):
\begin{align}
	\rhoNotN = \hat{\rho}_{0}\otimes\cdots\otimes \hat{\rho}_{0},
	\label{eq:W_notransmissionstate}
\end{align}
and the average state that he receives when Alice transmits (alternate hypothesis $H_1$):
\begin{equation}
\rhoBarN=\frac{1}{KM}\sum_{k=1}^K \sum_{m=1}^M \rhoN (m,k).\label{eq:W_transmissionstate}
\end{equation}
The quantum state that Willie receives when Alice uses shared secret key $k$ to transmit message $m$ over $n$ channel uses is
\begin{align}
\hat{\rho}^n(m,k)=\hat{\rho}_W^n(\mathbf{x}(m,k)) = \bigotimes_{i=1}^n \hat{\rho}_{x_i(m,k)}.
\end{align}
Willie fails by either accusing Alice of transmitting when she is not (false alarm), or missing Alice's transmission (missed detection).
Denoting the probabilities of these respective errors by $P_{\rm FA}=P(\text{choose~}H_1|H_0\text{~is true})$ and $P_{\rm MD}=P(\text{choose~}H_0|H_1\text{~is true})$, and assuming equally likely hypotheses $P(H_0) = P(H_1) = \frac{1}{2}$, Willie's probability of error is:
\begin{align} 
P_{\rm e}^{(w)} &=\frac{P_{\rm FA}+P_{\rm MD}}{2}.
\end{align}
Randomly choosing whether to accuse Alice yields an ineffective detector with $P_{\rm e}^{(w)}=\frac{1}{2}$. The goal of covert communication is to design a sequence of codes such that Willie's detector is forced to be arbitrarily close to ineffective. That is, 
\begin{align}
\lim_{n\to\infty}P_{\rm e}^{(w)}&=\frac{1}{2}. \label{eq:PWillie_condition}
\end{align}
The minimum $P_{\rm e}^{(w)}$ is related to the trace distance $\|\rhoBarN -\rhoNotN \|_1$ between the states $\rhoBarN$ and $
\rhoNotN$ as follows  \cite[Sec. 9.1.4]{wilde16quantumit2ed}:
\begin{align}
\min P_{\rm e}^{(w)}=\frac{1}{2}\left( 1 - \frac{1}{2}\|\rhoBarN -\rhoNotN \|_1\right), \label{eq:error}
\end{align}
where $\|\hat{A}\|_1=\Tr\left[\sqrt{\hat{A}^\dagger\hat{A}}\right]$ is the trace norm of $\hat{A}$ \cite[Def. 9.1.1]{wilde16quantumit2ed} discussed in Section \ref{sec:notation_linear_operators}. Note that \eqref{eq:error} can be reformulated in terms of trace norm with non-uniform priors $P(H_1)$, $P(H_0)$ \cite[Sec.~II.B]{sobers17jammer}.
The quantum relative entropy (QRE) $D(\hat{\rho} \| \hat{\sigma}) = \Tr \left[\hat{\rho}\ln \hat{\rho} - \hat{\rho} \ln \hat{\sigma} \right]$ is a convenient covertness measure because it upper-bounds the trace distance in \eqref{eq:error} and is additive over product states. By the quantum Pinsker's inequality \cite[Th. 11.9.2]{wilde16quantumit2ed}, 
\begin{align}
 \frac{1}{2}\left(\|\rhoBarN - \rhoNotN \|_1\right)^2\leq 	D\left(\rhoBarN \| \rhoNotN \right).
 \label{eq:qPinsker}
\end{align}
Therefore, we call a sequence of schemes covert if
\begin{align} \label{eq:covertnessCriteria}
\lim_{n\to\infty}D\left(\rhoBarN \|\rhoNotN\right) =0.
\end{align}
We choose \eqref{eq:covertnessCriteria} as our covertness criterion for its mathematical tractability, as was done in \cite{wang15covert, bloch15covert, bullock20discretemod, gagatsos20codingcovcomm}. Combining \eqref{eq:error} and \eqref{eq:qPinsker}, we note that satisfying \eqref{eq:covertnessCriteria} also necessarily satisfies \eqref{eq:PWillie_condition}. 

\subsection{Quantum-secure Covert State} \label{sec:QSCS}
Suppose that the support of the Willie's output state $\hat{\rho}_x$ corresponding to non-innocent input $x\in\mathcal{X}\setminus\{0\}$ is contained in the support of the innocent state $\hat{\rho}_0$, i.e., $\supp(\hat{\rho}_x) \subseteq \supp(\hat{\rho}_0)$.
We show in Section \ref{subsec:nocovcomms} that, if for all $x\in\mathcal{X}\setminus\{0\}$, $\supp\left(\hat{\rho}_x\right) \nsubseteq\supp\left(\hat{\rho}_0\right)$, then covert communication is impossible. 
We also assume that innocent output state $\hat{\rho}_0$ is not a mixture of non-innocent ones $\{\hat{\rho}_x\}_{x\in \mathcal{X} \backslash \{0\}}$, since constant-rate covert communication is achieved trivially otherwise, per Section \ref{subsec:constantrate}.

Suppose that Alice transmits $x\in\mathcal{X}$ randomly with the following distribution:
\begin{align} \label{eq:probDist}
p_X(x) &=
\begin{cases}
	1- \alpha_n, & x = 0 \\
	\alpha_n \pi_x, & x = 1,2,\ldots,N,\\
\end{cases}
\end{align}
where $\{\pi_x\}$ is an arbitrary non-innocent state distribution such that $\pi_x\in[0,1]$ and $\sum_{x\in\mathcal{X}\setminus\{0\}}\pi_x = 1$.  Then Willie observes a mixed state:
\begin{align}
\label{eq:rhoalphan}\hat{\rho}_{\alpha_n}&= \sum_{x \in \mathcal{X}} p_X(x)\hat{\rho}_x=(1-\alpha_n)\hat{\rho}_0+\alpha_n\sum_{x\in\mathcal{X}\setminus\{0\}}\pi_x\hat{\rho}_x.
\end{align}
We use the following lemma to characterize the asymptotic behavior of QRE $D\left(\hat{\rho}_{\alpha_n}\|\hat{\rho}_0\right)$:
 \begin{lem} \label{lem:etaDefn}
\cite[Lem. 5]{tahmasbi2021covertQSensing}
Let $\hat{\rho}_0$ and $\hat{\rho}_1$ be density operators such that $\hat{\rho}_0$ is invertible. For spectral decomposition of $\hat{\rho}_0 = \sum_i \lambda_i \hat{P}_i$, define
\begin{align}
\label{eq:etaDefn}\eta ( \hat{\rho}_1 \| \hat{\rho}_0) &= \sum_{i \neq j} \frac{\log \lambda_i - \log \lambda_j}{\lambda_i - \lambda_j} \Tr \left[(\hat{\rho}_1 - \hat{\rho}_0 ) \hat{P}_i (\hat{\rho}_1 - \hat{\rho}_0) \hat{P}_j \right] \nonumber\\
&\phantom{=}+\sum_i \frac{1}{\lambda_i} \Tr \left[ (\hat{\rho}_1 - \hat{\rho}_0 ) \hat{P}_i (\hat{\rho}_1 - \hat{\rho}_0) \hat{P}_i \right].
\end{align}
For small $\alpha > 0$,
\begin{align}
D ((1-\alpha) \hat{\rho}_0 + \alpha \hat{\rho}_1 \| \hat{\rho}_0 ) = \frac{1}{2} \alpha^2 \eta ( \hat{\rho}_1 \| \hat{\rho}_0) + R(\alpha), \label{eq:etaremainder}
\end{align}
where $R(\alpha) \in \mathcal{O} (\alpha^3)$.
\end{lem}
Note that $\eta(\hat{\rho}\|\hat{\rho})=0$.  The proof of Lemma \ref{lem:etaDefn} is in \cite[App. A]{tahmasbi2021covertQSensing}.
As highlighted in \cite[Rem. 5]{tahmasbi2021covertQSensing}, Lemma \ref{lem:etaDefn} is similar to \cite[Lem. 1]{wang16cq-srlconverse} but the definition therein of $\eta(\cdot || \cdot)$ involves integration. As in \cite{tahmasbi2021covertQSensing}, it is key that the remainder term in \eqref{eq:etaremainder} is independent of $\hat{\rho}_1$, which is not shown for \cite[Lem. 1]{wang16cq-srlconverse}. 

 Now, define a product state $\rhoAlphaN = \hat{\rho}_{\alpha_n} \otimes \ldots \otimes \hat{\rho}_{\alpha_n}$. 
By the additivity of relative entropy \cite[Ex. 11.8.7]{wilde16quantumit2ed},
\begin{align}
D(\rhoAlphaN \| \rhoNotN) = nD\left(\rhoAlpha \middle\| \hat{\rho}_0\right). \label{eq:QRE_bound_memoryless}
\end{align}
Combining Lemma \ref{lem:etaDefn} with \eqref{eq:QRE_bound_memoryless}, and choosing $\alpha_n=\frac{\gamma_n}{\sqrt{n}}$, where
\begin{align}
\label{eq:gamman}\gamma_n \in o(1) \cap \omega\left(\frac{\pra{\log n}^{\frac{4}{3}}} {n^{\frac{1}{6}}}\right).
\end{align}
yields
	\begin{align}
	D\left(\rhoAlphaN \| \rhoNotN \right)&=\frac{1}{2}\gamma_n^2\eta\left(\sum_{x\in\mathcal{X}\setminus\{0\}}\pi_x\hat{\rho}_{x}\|\hat{\rho}_0\right)+R\left(\frac{\gamma_n}{\sqrt{n}}\right)\\&\in \mathcal{O}\left(\gamma_n^2\right).
	\end{align}
	\color{black}	
We note that our constraint $\gamma_n \in {\omega\left(\frac{\pra{\log n}^{\frac{4}{3}}} {n^{\frac{1}{6}}}\right)}$ in \eqref{eq:gamman} is more restrictive than that in classical setting, which is $\gamma_n \in \omega\pra*{\frac{\log n}{\sqrt{n}}}$ \cite{bloch15covert,bloch20errata}. This constraint is a technical artifact of using quantum channel resolvability results in the proof of Lemma~\ref{lem:trace-expected}.
Nevertheless, the QRE between $\rhoAlphaN$ and $\rhoNotN$ tends to 0 as $n\to\infty$, and $\rhoAlphaN$ becomes indistinguishable from $\rhoNotN$. We, therefore, call $\rhoAlphaN$ a ``quantum-secure covert state,'' analogous to a covert random process introduced in \cite[Sec. III.A]{bloch15covert}. Our quantum-secure covert communication protocols ensure that Willie's output state is arbitrarily close to $\rhoAlphaN$. 

\subsection{Lemmas and definitions}
\label{sec:lemmas}
Here we include lemmas and definitions needed for our proofs that follow.

\begin{defn}[Pinching Maps {\cite[Ch. IV.2]{bhatia2013matrix}}] \label{defn:pinchingmaps}

Let $\hat{A}$ be a Hermitian operator with spectral decomposition \mtt{ $\hat{A}=\sum_{i}\lambda_i \hat P_i$, where $\{\lambda_i\}$ are \emph{distinct} eigenvalues of $\hat{A}$, and  $\hat P_i$ is orthogonal projection onto the eigenspace corresponding to $\lambda_i$.
For an arbitrary operator $\hat{B}$, the following map is called a pinching of $\hat{B}$ with respect to $\hat{A}$:
\begin{align}
\mc{E}_{\hat{A}}:\hat{B}\to\mc{E}_{\hat{A}}(\hat{B})=\sum_i \hat P_i\hat{B}\hat P_i \nonumber
\end{align}}
Lemmas \ref{lem:pinchingMapCommute}-\ref{lem:pinchingMapHayashi} describe properties of pinching maps.
\end{defn}
\begin{lem} \label{lem:pinchingMapCommute}
For a Hermitian operator $\hat{A}$ and an arbitrary operator $\hat{B}$, $\mc{E}_{\hat{A}}(\hat{B})$ commutes with $\hat{A}$.
\end{lem}
\begin{lem} \label{lem:pinchingMapTr}
For an arbitrary operator $\hat{C}$ commuting with a Hermitian operator $\hat{A}$, $\Tr\left[\hat{B}\hat{C}\right]=\Tr\left[ \mc{E}_{\hat{A}}(\hat{B})\hat{C}\right]$.
\end{lem}
Lemmas \ref{lem:pinchingMapCommute} and \ref{lem:pinchingMapTr} follow immediately from the  definition of $\mc{E}_{\hat{A}}$.

\begin{lem}\label{lem:pinchingMapConvex}
Let $\hat A$ be a Hermitian operator with spectral decomposition $\hat{A}=\sum_{i}\lambda_i \hat P_i$, where $\{\lambda_i\}$ are distinct eigenvalues of $\hat{A}$, and  $\hat P_i$ is orthogonal projection onto the eigenspace corresponding to $\lambda_i$. Let $\hat B$ be any positive semi-definite operator. For positive semi-definite operator $\hat C = \sum_i\gamma_i\hat P_i$ and natural number $n$, 
\begin{align}
    \Tr\left[\left(\mathcal{E}_{\hat A}(\hat B) \right)^n\hat C\right]\leq \Tr\left[\mathcal{E}_{\hat A}(\hat{B}^n)\hat C\right]\label{eq:pinchingMapConvex},
\end{align}
 where $\mc{E}_{\hat{A}}(\hat{B})$ is a pinching of $\hat{B}$ with respect to $\hat{A}$.
\end{lem}
The proof of Lemma \ref{lem:pinchingMapConvex} is in Appendix \ref{ap:pinchingMapConvex}.

\begin{lem} \label{lem:pinchingMapHayashi}
(Hayashi's pinching inequality): For any Hermitian operator $\hat{A}$  with $N_{\hat{A}}$ distinct eigenvalues, $\hat{B}\preceq N_{\hat{A}}\mc{E}_{\hat{A}}(\hat{B}) $, where $\mc{E}_{\hat{A}}(\hat{B})$ is a pinching of arbitrary operator $\hat{B}$ with respect to $\hat{A}$.
\end{lem}
The proof of Lemma \ref{lem:pinchingMapHayashi} is in \cite[Lem. 9]{hayashi2002optimal}.

\begin{lem} \label{lem:hayashinagaoka}
For arbitrary operators $0 \preceq \hat{A} \preceq \hat{I}$ and $\hat{B} \succeq 0$,
\begin{align}
\hat{I} - (\hat{A} + \hat{B})^{-1/2}\hat{A}(\hat{A}+\hat{B})^{-1/2} \preceq (1+c)(\hat{I}-\hat{A})+(2+c+c^{-1})\hat{B},
\end{align}
where $c > 0$ is a real number and $\hat{I}$ is the identity operator.
\end{lem}
The proof of Lemma \ref{lem:hayashinagaoka} is in \cite[Lem. 2]{hayashi2003general}.

\begin{lem} \label{lem:ogawaHayashi}
For arbitrary product states $\hat{\phi}^n$ and $\hat{\tau}^n$ and arbitrary $t > 0$ and $0 \leq r \leq 1$,
\begin{align}
\Tr \left[\hat{\phi}^n \{ \mathcal{E}_{\hat{\tau}^n} (\hat{\phi}^n) - t \hat{\tau}^n \preceq 0 \} \right] \leq (n+1)^d t^r \Tr \left[ \hat{\phi}^n (\hat{\tau}^n)^{r/2} (\hat{\phi}^n)^{-r} (\hat{\tau}^n)^{r/2} \right],\label{eq:ogawaHayashi}
\end{align} 
where $d$ is the dimension of the Hilbert space that $\hat{\tau}$ acts on.
\end{lem}
The proof of Lemma \ref{lem:ogawaHayashi} is in \cite[Th. 2]{ogawa2004error}. 

\begin{lem} \label{lem:phi_bound}
Consider $\phi(s,\alpha_n)$ defined as in \cite[(9.53)]{hayashi2006quantum} such that 
\begin{align}
\phi(s, \alpha_n)=\log \left( \sum_{x\in\mathcal{X}}p_{X}(x)\left(\Tr\left[\hat{\rho}_x^{1-s}\hat{\rho}_{\alpha_n}^s\right]\right)\right), \label{eq:phiSAlphaDef}
\end{align}
where $s\in [s_0, 0]$, $s_0<0$ is an arbitrary constant, $p_X(x)$ is defined in \eqref{eq:probDist} and $\hat{\rho}_{\alpha_n}$ is defined in Section \ref{sec:QSCS}.
Then there exist  constants $\vartheta_1, \vartheta_2>0$ 
independent of $s$ and $\alpha_n$ such that
\begin{align}
\phi(s, \alpha_n) \leq -\alpha_n s \sum_{x\in\mathcal{X}\setminus\{0\}} \pi_x D(\hat{\rho}_x || \hat{\rho}_0) +  \vartheta_1\alpha_n s^2 -\vartheta_2 s^3, \label{eq:phi_bound}
\end{align}
where $D(\hat{\rho} \| \hat{\sigma}) = \Tr \left[\hat{\rho}\ln \hat{\rho} - \hat{\rho} \ln \hat{\sigma} \right]$ is quantum relative entropy.
\end{lem}
The proof of Lemma \ref{lem:phi_bound} is in Appendix \ref{ap:phi_bound}.

    \begin{lem}
    \label{lem:lambda-min-alpha}
           Let $\{\hat{\rho}_x\}$ be a set of density operators with supports such that $\supp(\hat{\rho}_x) \subseteq \supp\pra{\rhoNot}$ for all $x\in \mathcal{X}\setminus \set{0}$.  Then, 
        \begin{align}
            \lambda_{\min}(\rhoAlpha) \geq (1-\alpha_n) \lambda_{\min}(\rhoNot).
        \end{align}
        In particular, for large enough $n$, 
        \begin{align}
            \lambda_{\min}(\rhoAlpha) \geq \frac{1}{2} \lambda_{\min}(\rhoNot).
        \end{align}
    \end{lem}

The proof of Lemma \ref{lem:lambda-min-alpha} is in Appendix \ref{ap:lambda-min-alpha}.

We now define von Neumann entropy and Holevo information.

\begin{defn}[von Neumann entropy] For a quantum state $\hat{\rho}$, the von Neumann entropy is $H\left(\hat{\rho}\right)=-\Tr\left[\hat{\rho}\log\hat{\rho}\right]$.
\end{defn}

\begin{defn}[Holevo information]For an ensemble of quantum states $\left\{p_x,\hat{\rho}_x\right\}$, $x\in\mathcal{X}=\{0,1,\ldots,N\}$, $\sum_x p_x=1$, the Holevo information is $\chi\left(\left\{p_x,\hat{\rho}_x\right\}\right)=H\left(\sum_x p_x\hat{\rho}_x\right)-\sum_xp_xH\left(\rho_x\right)$.
\end{defn}

Finally we need \cite[Lemma 18]{tahmasbi20covertqkd} re-stated as follows.
\begin{lem}
    \label{lem:qre-vs-trace}
    Let $\hat{\rho}$ and $\hat{\sigma}$ be two quantum states over Hilbert space $\mathcal{H}$. Let $\supp(\hat{\rho}) \subseteq \supp(\hat{\sigma})$ and $\norm{\hat{\rho}- \hat{\sigma}}_1 \leq e^{-1}$. Then,
    \begin{align}
        D(\hat{\rho}\|\hat{\sigma}) \leq \norm{\hat{\rho}- \hat{\sigma}}_1 \log \left(\frac{\dim \mathcal{H}}{\lambda_{\min}(\hat{\sigma})\norm{\hat{\rho}- \hat{\sigma}}_1}\right).
    \end{align}
\end{lem}

\section{Square Root Law}

\subsection{Achievability} \label{sec:achievability}
After stating the achievability in Theorem \ref{thm:achievability}, we define the covert capacity and the pre-shared secret key requirement.
This allows us to show their lower and upper bounds, before proceeding with the proof.
In Section \ref{sec:converse} we prove Theorem \ref{thm:converse}, providing the matching upper and lower bounds on these quantities.

\begin{thm}[Achievability] \label{thm:achievability}
Consider a covert memoryless classical-quantum channel such that, for inputs $x\in\mathcal{X}=\left\{0,1,2,\ldots,N\right\}$, the output state $\hat{\rho}_0$ corresponding to innocent input $x=0$ is not a mixture of non-innocent ones $\left\{\hat{\rho}_x\right\}_{x\in\mathcal{X}\setminus\{0\}}$, and, $\forall x\in\mathcal{X}\setminus\{0\}$, $\supp (\hat{\sigma}_{x}) \subseteq \supp (\hat{\sigma}_0)$ and $\supp (\hat{\rho}_{x}) \subseteq \supp (\hat{\rho}_0)$. Let non-innocent input distribution $\left\{\pi_x\right\}_{x\in\mathcal{X}\setminus\{0\}}$ be arbitary such that $\pi_x\in[0,1]$ and $\sum_{x\in\mathcal{X}\setminus\{0\}}\pi_x=1$.
Let $\alpha_n=\frac{\gamma_n}{\sqrt{n}}$ with $\gamma_n$ defined in \eqref{eq:gamman}.  Then, for any $\varsigma_n\in o(1)\cap\omega\left(\frac{1}{\pra{\log n}^{\frac{2}{3}}}\right)$, there exist $\varsigma_{n}^{(1)}\in \omega\left(\frac{1}{\log(n)^{\frac{4}{3}}n^{\frac{1}{3}}}\right)$, $\varsigma_n^{(2)}\in\omega\left(\frac{1}{\pra{\log n}^2}\right)$,  and a sequence of covert communication schemes $\left(\mathcal{K}_n\right)$ such that, for $n$ sufficiently large,
\begin{align}
\label{eq:messageSize}\log M &= (1-\varsigma_n)\gamma_n\sqrt{n} \sum_{x\in\mathcal{X}\setminus\{0\}}\pi_x D(\hat{\sigma}_{x} \| \hat{\sigma}_0), \\
\label{eq:secretsize}{\log K} &= \gamma_n \sqrt{n} \left[ \sum_{x\in\mathcal{X}\setminus\{0\}}\pi_x \left(\left(1+\varsigma_n\right)D(\hat{\rho}_{x} \| \hat{\rho}_0 ) - (1-\varsigma_n) D(\hat{\sigma}_{x} \| \hat{\sigma}_0) \right)\right]^{+}, 
\end{align}
and,
\begin{align}
P_{\rm e}^{(b)} &\leq e^{-\varsigma_n^{(1)} \gamma_n \sqrt{n}}, \label{eq:reliabilityCond} \\
\abs{D(\hat{\bar{\rho}}^n\| \rhoNotN) - D(\rhoAlphaN \| \rhoNotN)} &\leq e^{-\varsigma_n^{(2)} \gamma_n^{\frac{3}{2}} n^{\frac{1}{4}}} . \label{eq:covertnessCond}
\end{align}
where $[a]^+=\max(0,a)$ and Willie's average state $\hat{\bar{\rho}}^n$ is defined in \eqref{eq:W_transmissionstate}.
\end{thm}

Theorem \ref{thm:achievability} states that, with $\log K \in \mathcal{O}(\sqrt{n})$ bits of pre-shared secret key, reliable transmission of $\log M \in \mathcal{O}(\sqrt{n})$ covert bits is achievable in $n$ uses of the classical-quantum channel. Recall that channel capacity is the supremum of \emph{achievable rates}, where the rate of a code is given by $R_n=\frac{1}{n}\log M$, and a rate $R$ is said to be achievable for a channel if there exists a sequence of codes that satisfy $\lim_{n\to\infty} P_{\rm e}^{(b)} = 0$ and  $R=\lim_{n\to\infty}R_n$ \cite{cover02IT,Csiszar_Korner_2011}. Thus, capacity is measured in bits per channel use. Note that since $\log M \in \mathcal{O}(\sqrt{n})$ due to the square root law, any achievable rate that maintains covertness converges to zero, and thus, the capacity of our covert channel is zero. Therefore, as in \cite{bloch15covert,wang15covert}, the covert rate is defined as the number of reliably transmissible covert bits regularized by $\sqrt{n}$ instead of $n$. In keeping with \cite{bloch15covert}, we also regularize by the covertness metric $D(\rhoBarN \| \rhoNotN)$ to account for its dependence on $n$ in the following definitions of covert capacity and pre-shared secret key requirement.  

\begin{defn}[SRL covert rate]\label{def:RSRL}
The rate $R_{n, \rm{SRL}}$ of a covert communication scheme $\mathcal{K}_n$ over the memoryless SRL classical-quantum channel is 
\begin{align}
    R_{n, \rm{SRL}} = \frac{\log M}{\sqrt{n D(\rhoBarN \| \rhoNotN)}},
\end{align}
where $M$ is the size of the message set and $n$ is blocklength. 
\end{defn}

\begin{defn}[SRL pre-shared secret key requirement rate]
    \label{def:SRLkeyreqrate}The SRL pre-shared secret key requirement for a given SRL covert rate $R_{n,\rm{SRL}}$ over the memoryless SRL classical-quantum channel is 
    \begin{align}
        J_{n, \rm SRL}(R_{n,\rm{SRL}})= \inf_{\mathcal{K}_n}\frac{\log K}{\sqrt{n D(\rhoBarN \| \rhoNotN)}},\label{eq:Jnrate}
    \end{align}
    where the infimum is taken over the set of covert schemes with covert rate $R_{n,\rm{SRL}}$.
\end{defn}

\begin{defn}[SRL achievable covert rate]
    A covert rate $R_{\rm SRL}$ is achievable if there exists a sequence of covert communication schemes $\left(\mathcal{K}_n\right)$ such that 
    \begin{align}
        \label{eq:RSRL}
        R_{\rm SRL}&=\lim_{n\to\infty} R_{n, \rm{SRL}} \text{, with }\\
\label{eq:RSRLconds}\lim_{n \to \infty} D(\rhoBarN \| \rhoNotN) &= 0, \lim_{n \to \infty} P_{\rm e}^{(b)} = 0.
    \end{align}
\end{defn}

\begin{defn}[SRL pre-shared secret key requirement]\label{def:JSRL}
    For each SRL achievable covert rate $R_{\rm SRL}$ and corresponding sequence of covert communication schemes $\left(\mathcal{K}_n\right)$ each satisfying \eqref{eq:Jnrate}, the corresponding SRL pre-shared secret key requirement is 
    \begin{align}
        J_{\rm SRL}(R_{\rm SRL}) = \lim_{n\to\infty}J_{n, \rm SRL}(R_{n,\rm{SRL}}).
    \end{align}
\end{defn}

\begin{defn}[Square root law (SRL) covert capacity]
    \label{def:LSRL}The capacity $L_{\rm SRL}$ of covert communication over the memoryless SRL classical-quantum channel is the supremum of achievable covert rates defined in \eqref{eq:RSRL}.
\end{defn}

We now derive the lower bound for $L_{\rm SRL}$ and the upper bound for $J_{\rm SRL}(L_{\rm SRL})\triangleq J_{\rm SRL}$.
\begin{cor} \label{cor:scalingConstants}
 Consider a covert memoryless classical-quantum channel.
Then, there exists a covert communication scheme meeting the conditions in \eqref{eq:RSRLconds}
with covert capacity and pre-shared secret key requirement:
\begin{align}
\label{eq:achL} L_{\rm SRL}  &\geq \frac{ \sum_{x\in\mathcal{X}\setminus\{0\}}  \pi^\ast_x D(\hat{\sigma}_x \| \hat{\sigma}_0) }{\sqrt{\frac{1}{2} \eta( \hat{\rho}^\ast_{\neg 0}\| \hat{\rho}_0) }} \\
\label{eq:achJ} J_{\rm SRL} &\leq \frac{\left[ \sum_{x\in\mathcal{X}\setminus\{0\}}  \pi^\ast_x \left(D(\hat{\rho}_x \| \hat{\rho}_0) - D(\hat{\sigma}_x \| \hat{\sigma}_0) \right) \right]^{+}}{\sqrt{\frac{1}{2} \eta( \hat{\rho}^\ast_{\neg 0} \| \hat{\rho}_0) }},
\end{align}
where $\{{\pi}^\ast_x\}=\argmax_{\{{\pi}_x\}} \frac{\sum_{x\in\mathcal{X}\setminus\{0\}}\pi_x D(\hat{\sigma}_x \| \hat{\sigma}_0)}{\sqrt{\frac{1}{2}\eta(\hat{\rho}_{\neg 0} \| \hat{\rho}_0)}}$,  $\hat{{\rho}}^\ast_{\neg 0}\triangleq\sum_{x\in\mathcal{X}\setminus\{0\}}{\pi}^\ast_x\hat{\rho}_x$ is the corresponding Willie's average non-innocent state, and $\hat{\bar{\rho}}^n$ is Willie's average state defined in \eqref{eq:W_transmissionstate}.
\end{cor}

\begin{IEEEproof}[Proof (Corollary \ref{cor:scalingConstants})]
Theorem \ref{thm:achievability} proves existence of a covert communication scheme such that \eqref{eq:covertnessCond} holds. Hence,
\begin{align}
\label{eq:QREUpperBound}D(\rhoBarN \| \rhoNotN) &\leq nD(\rhoAlpha \| \hat{\rho}_0) +e^{-\varsigma_{n}^{(2)}\gamma_n^{\frac{3}{2}}n^{\frac{1}{4}}}, \\
\label{eq:QRELowerBound}D(\rhoBarN \| \rhoNotN) &\geq nD(\rhoAlpha \| \hat{\rho}_0) - e^{-\varsigma_{n}^{(2)} \gamma_n^{\frac{3}{2}}n^{\frac{1}{4}}}.
\end{align}
By Lemma \ref{lem:etaDefn},
\begin{align}
nD\left(\hat{\rho}_{\alpha_n}\middle\|\hat{\rho_0}\right)
 &= n \frac{\alpha_n^2}{2} \eta(\hat{\rho}_{\neg 0} \| \hat{\rho}_0) + nR(\alpha_n) \\
&= \frac{\gamma_n^2}{2} \eta(\hat{\rho}_{\neg 0} \| \hat{\rho}_0) + nR(\alpha_n), \label{eq:QRErhoBarNrhoNotN}
\end{align}
where $R(\alpha_n) \in \mathcal{O}(\alpha_n^3)$. 
Using Definition \ref{def:LSRL} with \eqref{eq:messageSize}, \eqref{eq:QREUpperBound}, and \eqref{eq:QRErhoBarNrhoNotN}, we have, 
\begin{align}
L_{\rm SRL} &\geq \lim_{n \to \infty} \frac{(1-\varsigma_n) \gamma_n\sqrt{n} \sum_{x\in\mathcal{X}\setminus\{0\}}\pi_x D(\hat{\sigma}_x \| \hat{\sigma}_0)  }{\sqrt{ n\frac{\gamma_n^2}{2} \eta(\hat{\rho}_{\neg 0} \| \hat{\rho}_0)  + n^2R(\alpha_n)+ne^{-\varsigma_2 \gamma_n^{\frac{3}{2}}n^{\frac{1}{4}}}
}} \\
&= \frac{\sum_{x\in\mathcal{X}\setminus\{0\}} \pi_x D(\hat{\sigma}_x \| \hat{\sigma}_0)}{\sqrt{\frac{1}{2} \eta(\hat{\rho}_{\neg{0}} \| \hat{\rho}_0 )}}.\label{eq:LSRLgeneraldist}
\end{align}
Using Definition \ref{def:JSRL} with \eqref{eq:secretsize}, \eqref{eq:QRELowerBound}, and \eqref{eq:QRErhoBarNrhoNotN}, we also have,
\begin{align}
J_{\rm SRL}&\leq \lim_{n \to \infty} \frac{\left[ \gamma_n \sqrt{n} \sum_{x\in\mathcal{X}\setminus\{0\}}\pi_x \left( (1+\varsigma_n) D(\hat{\rho}_x \| \hat{\rho}_0) - (1-\varsigma_n) D(\hat{\sigma}_x \| \hat{\sigma}_0) \right) \right]^+ }{\sqrt{n\frac{\gamma_n^2}{2} \eta (\hat{\rho}_{\neg 0} \| \hat{\rho}_0) + n^2R(\alpha_n)-ne^{-\varsigma_2 \gamma_n^{\frac{3}{2}}n^{\frac{1}{4}}}}  } \\
&= \frac{\left[ \sum_{x\in\mathcal{X}\setminus\{0\}}\pi_x  \left( D(\hat{\rho}_x \| \hat{\rho}_0) -  D(\hat{\sigma}_x \| \hat{\sigma}_0) \right) \right]^+ }{\sqrt{\frac{1}{2} \eta (\hat{\rho}_{\neg{0}} \| \hat{\rho}_0)}} \label{eq:JSRLgeneraldist}.
\end{align}
Maximizing $\frac{\sum_{x\in\mathcal{X}\setminus\{0\}}\pi_x D(\hat{\sigma}_x \| \hat{\sigma}_0)}{\sqrt{\frac{1}{2} \eta( \hat{\rho}_{\neg 0} \| \hat{\rho}_0) }}$ over $\pi_x$ completes the proof.
\end{IEEEproof}

\begin{rk}
Although we choose the non-innocent distribution $\{\pi_x^\ast\}$ that maximizes the covert rate, this does not guarantee that the pre-shared secret key requirement is minimized. By construction, \eqref{eq:LSRLgeneraldist} and \eqref{eq:JSRLgeneraldist} hold for any non-innocent distribution. Thus, Theorem \ref{thm:achievability} yields an achievability region characterized by choice of non-innocent distribution. 
\end{rk}

Now we proceed with the proof of Theorem \ref{thm:achievability}.

\begin{IEEEproof}[Proof (Theorem \ref{thm:achievability})] {\bf Construction:}
For each $(m,k)$, where $m\in\left\{1,\ldots, M\right\}$ is a message and $k\in\left\{1,\ldots,K\right\}$ is a pre-shared secret key, Alice generates an i.i.d. random sequence $\mathbf{x} (m,k)\in \mathcal{X}^n$ from the distribution in \eqref{eq:probDist}, where $\alpha_n$ satisfies the requirements for a covert quantum-secure state given in Section \ref{sec:QSCS}. 
Alice chooses a codeword $\mathbf{x}(m,k)$ based on the message $m$ she wants to send and secret key $k$ that is pre-shared with Bob. The codebook is used only once to transmit a single message, and Willie does not know the secret key. Willie's output state corresponding to a single use of the classical-quantum channel by Alice is given in \eqref{eq:rhoalphan}.
Similarly, Bob's output state is:
\begin{align}
 \hat{\sigma}_{\alpha_n} = \sum_{x \in \mathcal{X}} p_X(x)\hat{\sigma}_x = (1-\alpha_n)\hat{\sigma}_0 + \alpha_n\sum_{x\in\mathcal{X}\setminus\{0\}}\pi_x \hat{\sigma}_{x}. \label{eq:bobState}
\end{align}
Elements of the codebook used to generate the transmission are instances of the quantum-secure covert random state defined in Section \ref{sec:QSCS}. Therefore, the codebook is an instance of a set of random vectors
\begin{align} \label{eq:codebook}
\mathcal{C} = \bigcup_{m}^{M} \bigcup_{k}^{K} \left\{ \mathbf{X}(m,k) \right\},
\end{align}
where $\mathbf{X}(m,k)$ describes the codeword corresponding to the message $m$ and secret key $k$. $\mathbf{X}(m,k)$ is distributed according to $p_{\mathbf{X}(m,k)}(\mathbf{x}(m,k))=\prod_{l=1}^n p_{X}(x_l(m,k))$, where $p_X(x)$ is defined in \eqref{eq:probDist}. We now show that this construction admits a decoding scheme that satisfies reliability condition \eqref{eq:reliabilityCond} for $M$ given in \eqref{eq:messageSize}.

\noindent{\bf Reliability analysis:} 
For a given codebook instance, we employ a square root measurement POVM for $n$ channel uses defined as 
\begin{align}
\hat{\Lambda}_{m,k}^{n}=\left(\sum_{m=1}^M\hat{\Pi}_{m,k}^n\right)^{-1/2}\hat{\Pi}_{m,k}^n\left(\sum_{m=1}^M\hat{\Pi}_{m,k}^n\right)^{-1/2},
\end{align}
with $\hat{\Pi}_{m,k}^n =\hat{\Pi}_{\mathbf{x}(m,k)}^n$ for the following projector corresponding to input vector $\mathbf x \in \mathcal{X}^n$:
\begin{align}
    \hat{\Pi}_{\mathbf{x}}^n=\left\{\mathcal{E}_{\hat{\sigma}_0^{\otimes n}}\left(\hat{\sigma}^n\left(\mathbf{x}\right)\right)-e^a\hat{\sigma}_0^{\otimes n} \succ 0\right\}, \label{eq:projX}
\end{align} where $\mathcal{E}_{\hat{\sigma}_0^{\otimes n}}\left(\hat{\sigma}^n\left(\mathbf{x}\right)\right)$ is a pinching of  $\hat{\sigma}^n\left(\mathbf{x}\right)$ with respect to $\hat{\sigma}_0^{\otimes n}$, $\{\hat{A}\succ 0\}$ is a projection onto the positive eigenspace and $a>0$ is a real number to be determined later. The parameter $a$ in the pinching ultimately defines $\log M$ that satisfies \eqref{eq:reliabilityCond}.  By adapting standard methods \cite{hayashi2003general}, we bound the expectation over codebook $\mathcal{C}$ of Bob's probability of error defined in \eqref{eq:BobDecodingErrorP} as follows, providing a detailed derivation in Appendix \ref{ap:ECPebBoundDerivation} for completeness:
\begin{align}
    E_\mathcal{C}\left[P_{\rm{e}}^{(b)}\right]&\leq\sum_{\mathbf{x}\in\mathcal{X}^n} p_{\mathbf{X}} (\mathbf{x}) \left[ 2\Tr \left[ \sigmaNx (\hat{I} - \hat{\Pi}_{\mathbf{x}}^n)\right] + 4(M-1) \Tr \left[  \hat{\sigma}^n (\mathbf{x}) \sum_{\mathbf{x'}\in\mathcal{X}^n}p_{\mathbf{X'}} (\mathbf{x'})\hat{\Pi}_{\mathbf{x'}}^n \right]\right]\label{eq:expectationIID}.
\end{align}
Since $\sum_{\mathbf{x}\in\mathcal {X}^n} p_{\mathbf{X}} (\mathbf{x})\hat{\sigma}^n (\mathbf{x}) = \hat{\sigma}_{\alpha_n}^{\otimes n}$ due to bilinearity and associativity of the tensor product as well as the definition of $\hat{\sigma}_{\alpha_n}$ in \eqref{eq:bobState}, we have
\begin{align}
E_{\mathcal{C}} \left[ P_{\rm e}^{(b)} \right]&\leq 
{2\sum_{\mathbf{x}} p_{\mathbf{X}} (\mathbf{x}) \Tr \left[ \sigmaNx (\hat{I} - \hat{\Pi}_{\mathbf{x}}^n)\right]} + {4(M-1) \sum_{\mathbf{x'}} p_{\mathbf{X'}} (\mathbf{x'})\Tr \left[\hat{\sigma}_{\alpha_n}^{\otimes n} \hat{\Pi}_{\mathbf{x'}}^n \right]}, \label{eq:sigmaAlphaNDef}
\end{align}
We bound the first term on the right hand side of \eqref{eq:sigmaAlphaNDef} by adapting the typical set construction from \cite{bloch15covert} in the proof of Lemma \ref{lem:c1} in Appendix \ref{ap:lemc1}. We subsequently bound the second term by using the properties of pinching and projectors in the proof of Lemma \ref{lem:c2} in Appendix \ref{ap:lemc2}.
\begin{lem}\label{lem:c1}
For the random codebook construction described in Section \ref{sec:achievability} and sequences $\nu_n^{(1)}\in o(1)\cap\omega\left(\frac{1}{\log(n)^{\frac{1}{3}}n^{1/12}}\right),$ $r_n\in o(1)\cap\omega\left(\frac{1}{n^{1/8}}\right)$, $\delta_n\in o\left(\frac{1}{(\log n)^{2/3}n^{1/8}}\right) \cap \omega\left(\frac{1}{(\log n)^{2/3}n^{1/6}}\right)$ there exists $\nu_n^{(2)}\in o(1)$ such that 
    \begin{align}
        \sum_{\mathbf{x}}& p_{\mathbf{X}} (\mathbf{x}) \Tr \left[ \sigmaNx (\hat{I} - \hat{\Pi}_{\mathbf{x}}^n)\right]\nonumber\\ \nonumber \leq &  (n+1)^{d_b} \exp\left(ar_n - \left(1-\sqrt{\nu_n^{(1)}}\right) \gamma_n\sqrt{n}\left((1-\nu_n^{(2)})r_n\sum_{x\in\mathcal{X}\setminus\{0\}}\pi_x D\left(\hat{\sigma}_x\middle\|\hat{\sigma}_0\right)-\delta_n\right) \right)    \\
&+ e^{-\nu_n^{(1)}\gamma_n \sqrt{n}/2},\label{eq:C1simple}
    \end{align}
    where $\gamma_n$ is defined in Section \ref{sec:QSCS}.
\end{lem}
\begin{lem}\label{lem:c2}
For the random codebook construction described in Section \ref{sec:achievability},
    \begin{align}
\sum_{\mathbf{x}} p_{\mathbf{X}} (\mathbf{x})\Tr \left[\hat{\sigma}_{\alpha_n}^{\otimes n} \hat{\Pi}_{\mathbf{x}}^n \right] \leq e^{-a+\frac{\gamma_n^2}{2}\beta_1 +\frac{\gamma_n^3}{2\sqrt{n}}\beta_2}, \label{eq:defnOfAlphaN2}
    \end{align}
    where $\beta_1,\beta_2$ are positive constants and $\gamma_n$ is defined in Section \ref{sec:QSCS}. 
\end{lem}

Substitution of \eqref{eq:C1simple} and \eqref{eq:defnOfAlphaN2} into \eqref{eq:sigmaAlphaNDef} gives us 
\begin{align}
& \begin{aligned}  E_{\mathcal{C}} \left[ P_{\rm e}^{(b)} \right]  & \leq 2(n+1)^{d_b} \exp\left(ar_n - \left(1-\sqrt{\nu_n^{(1)}}\right) \gamma_n\sqrt{n}\left((1-\nu_n^{(2)})r_n\sum_{x\in\mathcal{X}\setminus\{0\}}\pi_x D\left(\hat{\sigma}_x\middle\|\hat{\sigma}_0\right)-\delta_n\right) \right)    \\
&\phantom{\leq}+ 2e^{-\nu_n^{(1)}\gamma_n \sqrt{n}/2} + 4(M-1)e^{-a+\frac{\gamma_n^2}{2}\beta_1 +\frac{\gamma_n^3}{2\sqrt{n}}\beta_2}.
\end{aligned}
\end{align}
Selecting $a = (1-\nu_n^{(3)})(1-\nu_n^{(2)})\left(1-\sqrt{\nu_n^{(1)}}\right)\gamma_n \sqrt{n} \sum_{x\in\mathcal{X}\setminus\{0\}}\pi_x D\left(\hat{\sigma}_x\middle\|\hat{\sigma}_0\right)$ for $\nu_n^{(3)}\in o(1)\cap\omega\left(\frac{1}{\log(n)^{\frac{2}{3}}}\right)$, we obtain: 
\begin{align}
E_{\mathcal{C}} \left[ P_{\rm e}^{(b)} \right] 
&\leq  2(n+1)^{d_b} e^{-\nu_n^{(3)}(1-\nu_n^{(2)}\left(1-\sqrt{\nu_n^{(1)}}\right)r_n\gamma_n\sqrt{n}\sum_{x\in\mathcal{X}\setminus\{0\}}\pi_x D\left(\hat{\sigma}_x\middle\|\hat{\sigma}_0\right)+\left(1-\sqrt{\nu_n^{(1)}}\right)\gamma_n\sqrt{n}\delta_n} +2 e^{-\nu_n^{(1)}\gamma_n \sqrt{n}/2}  \nonumber \\
&\phantom{\leq}+ 4Me^{-(1-\nu_n^{(3)})(1-\nu_n^{(2)})\left(1-\sqrt{\nu_n^{(1)}}\right)}\gamma_n\sqrt{n} \sum_{x\in\mathcal{X}\setminus\{0\}}\pi_x D\left(\hat{\sigma}_x\middle\|\hat{\sigma}_0\right)e^{\frac{\gamma_n^2}{2}\beta_1 +\frac{\gamma_n^3}{2\sqrt{n}}\beta_2}. \label{eq:beforeSubM}
\end{align}
Note that we can choose $\nu_n^{(1)}$, $\nu_n^{(2)}$, $\nu_n^{(3)}$, and $\nu_n^{(4)}\in o(1)\cap\omega\left(\frac{1}{\log(n)^{\frac{4}{3}}n^{\frac{1}{3}
}}\right)$ such that $(1-\varsigma_n) = \left(1-\sqrt{\nu_n^{(1)}}\right)(1-\nu_n^{(1)})(1-\nu_n^{(2)})(1-\nu_n^{(3)})$ for an arbitrary $\varsigma_n\in o(1)\cap\omega\left(\frac{1}{\pra{\log n}^{\frac{2}{3}}}\right)$, as in the statement of the theorem.
Substitution of \eqref{eq:messageSize} in \eqref{eq:beforeSubM} then yields:
\begin{align}
E_{\mathcal{C}} \left[ P_{\rm e}^{(b)} \right]  
& \leq 2(n+1)^{d_b} e^{-\nu_n^{(3)}(1-\nu_n^{(1)})\left(1-\sqrt{\nu_n^{(1)}}\right)r_n\gamma_n\sqrt{n}\sum_{x\in\mathcal{X}\setminus\{0\}}\pi_x D\left(\hat{\sigma}_x\middle\|\hat{\sigma}_0\right)+\left(1-\sqrt{\nu_n^{(1)}}\right)\gamma_n\sqrt{n}\delta_n} + 2e^{-\nu_n^{(1)}\gamma_n \sqrt{n}/2} \nonumber \\ &\phantom{\leq}+ 4e^{-\nu_n^{(4)}(1-\nu_n^{(3)})(1-\nu_n^{(1)})\left(1-\sqrt{\nu_n^{(1)}}\right)\gamma_n\sqrt{n} \sum_{x\in\mathcal{X}\setminus\{0\}}\pi_x D\left(\hat{\sigma}_x\middle\|\hat{\sigma}_0\right)}e^{\frac{\gamma_n^2}{2}\beta_1 +\frac{\gamma_n^3}{2\sqrt{n}}\beta_2}. \label{eq:reliabilityPenultimate} 
\end{align}
Thus, given the earlier definitions of $r_n$, $\nu_n^{(1)}$, $\nu_n^{(2)}$, $\nu_n^{(3)}$, and $\nu_n^{(4)}$, for large enough $n$ there exists a sequence $\zeta_{n}^{(1)}\in \omega\left(\frac{1}{\log(n)^{\frac{4}{3}}n^{\frac{1}{3}}}\right)$ such that
\begin{align}
E_{\mathcal{C}} \left[ P_{\rm e}^{(b)} \right] \leq e^{-\zeta_{n}^{(1)} \gamma_n \sqrt{n}}.\label{eq:achievableEPeb}
\end{align}
\noindent{\bf Covertness analysis:} 
We need the following two lemmas for our covertness analysis: Lemma \ref{lem:qre-to-trace} relates the difference between $D(\rhoBarN\| \rhoNotN)$ and $D(\rhoAlphaN\| \rhoNotN)$ for any code to the trace distance between $\rhoBarN$ and $\rhoAlphaN$. Lemma \ref{lem:trace-expected} proves an upper bound on the expected value of the trace distance between $\rhoBarN$ and $\rhoAlphaN$  for random codes when $\log M + \log K$ is sufficiently large.
\begin{lem}
\label{lem:qre-to-trace}
For any code and large enough $n$, we have 
\begin{align}
     \abs{D(\rhoBarN\| \rhoNotN) - D(\rhoAlphaN\| \rhoNotN)}  \leq \norm{\rhoBarN- \rhoAlphaN}_1\left( n\log \left(\frac{4\dim \mathcal{H}_W}{\lambda_{\min}(\rhoNot)^3}\right) + \log \frac{1}{\norm{\rhoBarN- \rhoAlphaN}_1}\right).
\end{align}
\end{lem}
\begin{lem}
\label{lem:trace-expected}
Consider a random coding scheme with $\varsigma_n\in o(1)\cap\omega\left(\frac{1}{(\log n)^{\frac{2}{3}}}\right)$
\begin{align}
\log M + \log K = (1+\varsigma_n) \alpha_n n\sum_{x\in\mathcal{X}\setminus\{0\}}\pi_x D(\hat{\rho}_{x} || \hat{\rho}_0).
\label{eq:covertCoding}
\end{align}
We have, for some $\zeta_n\in \omega\left(\frac{1} {\log n}\right)$,
\begin{align}
    E_{\mathcal{C}} \bigg[ \norm{\rhoBarN- \rhoAlphaN}_1  \bigg]  \leq e^{-\zeta_n \alpha_n^{\frac{3}{2}}n}.
\end{align}
\end{lem}
Before proving these lemmas, we show how they are used in covertness analysis. Note that
\begin{IEEEeqnarray}{rCl}
\IEEEeqnarraymulticol{3}{l}{E_{\mathcal{C}} \left[\abs{D(\rhoBarN\| \rhoNotN) - D(\rhoAlphaN\| \rhoNotN)}\right]}\IEEEnonumber\\
    & \leq & E_{\mathcal{C}}\left[ \norm{\rhoBarN- \rhoAlphaN}_1\left( n\log \left(\frac{4\dim \mathcal{H}_W}{\lambda_{\min}(\rhoNot)^3}\right) + \log \frac{1}{\norm{\rhoBarN- \rhoAlphaN}_1}\right)\right]\label{eq:110} \\
    &\label{eq:111} \leq & E_{\mathcal{C}}\left[ \norm{\rhoBarN- \rhoAlphaN}_1\right]\left( n\log \left(\frac{4\dim \mathcal{H}_W}{\lambda_{\min}(\rhoNot)^3}\right) + \log \frac{1}{E_{\mathcal{C}}\left[\norm{\rhoBarN- \rhoAlphaN}_1\right]}\right)\\
    &\label{eq:112} \leq & e^{-\zeta_n \alpha_n^{\frac{3}{2}}n} \pra*{n\log \left(\frac{4\dim \mathcal{H}_W}{\lambda_{\min}(\rhoNot)^3}\right) + \zeta_n \alpha_n^{\frac{3}{2}}n },
\end{IEEEeqnarray}
where \eqref{eq:110} follows from Lemma~\ref{lem:qre-to-trace}, \eqref{eq:111} follows from Jensen's inequality, and \eqref{eq:112} follows from Lemma~\ref{lem:trace-expected}. Therefore, since $\alpha_n \in \omega\pra*{\frac{\pra{\log n}^{\frac{4}{3}}}{n^{\frac{2}{3}}}}$, for large enough $n$, there exists $\zeta_{n}^{(2)}\in \omega\pra*{\frac{1}{\pra{\log n}^{2}}}$ such that
\begin{align}
    E_{\mathcal{C}} \left[\abs{D(\rhoBarN\| \rhoNotN) - D(\rhoAlphaN\| \rhoNotN)}\right] \leq e^{-\zeta_{n}^{(2)}\alpha_n^{\frac{3}{2}} n}\label{eq:achievableEQRE}
\end{align}

\begin{IEEEproof}[Proof of Lemma~\ref{lem:qre-to-trace}]
    We first note that by the definition of the QRE, 
\begin{align}
    \abs{D(\rhoBarN\| \rhoNotN) - D(\rhoAlphaN\| \rhoNotN)}  
    &= \abs{D(\hat{\bar{\rho}}^n\| \rhoAlphaN) + \Tr\left[(\rhoBarN - \rhoAlphaN)(\log(\rhoAlphaN) - \log(\rhoNotN)) \right]}\\
    &\leq D(\hat{\bar{\rho}}^n\| \rhoAlphaN) + \abs{\Tr\left[(\rhoBarN - \rhoAlphaN)(\log(\rhoAlphaN) - \log(\rhoNotN)) \right]} \label{eq:qre-diff}
\end{align}
We upper-bound the first term in the right hand side (RHS) of \eqref{eq:qre-diff} as
\begin{align}
    D(\rhoBarN\| \rhoAlphaN) 
    &\leq \norm{\rhoBarN- \rhoAlphaN}_1 \log \left(\frac{\dim \mathcal{H}_W^{\otimes n}}{\lambda_{\min}(\rhoAlphaN)\norm{\rhoBarN- \rhoAlphaN}_1}\right)\label{eq:qre-trace}\\
    &= \norm{\rhoBarN- \rhoAlphaN}_1 \log \left(\frac{(\dim \mathcal{H}_W)^n}{\lambda_{\min}(\rhoAlpha)^n\norm{\rhoBarN- \rhoAlphaN}_1}\right)\\
    &\leq \norm{\rhoBarN- \rhoAlphaN}_1 \log \left(\frac{(2\dim \mathcal{H}_W)^n}{\lambda_{\min}(\rhoNot)^n\norm{\rhoBarN- \rhoAlphaN}_1}\right),\label{eq:eigen-alpha2}
\end{align}
where \eqref{eq:qre-trace} follows from Lemma~\ref{lem:qre-vs-trace} and \eqref{eq:eigen-alpha2} follows from Lemma~\ref{lem:lambda-min-alpha}.
Furthermore, we have the following chain of inequalities for the second term in the RHS of \eqref{eq:qre-diff}:
\begin{IEEEeqnarray}{rCl}
\IEEEeqnarraymulticol{3}{l}{\abs{\Tr\left[(\rhoBarN - \rhoAlphaN)(\log(\rhoAlphaN) - \log(\rhoNotN)) \right]}}\IEEEnonumber\\
& \leq & \norm{\rhoBarN - \rhoAlphaN}_1 \norm{\log(\rhoAlphaN) - \log(\rhoNotN)}_\infty\\
 & \leq & \norm{\rhoBarN - \rhoAlphaN}_1 (\norm{\log(\rhoAlphaN)}_\infty + \norm{\log(\rhoNotN)}_\infty)\\
 & = & \norm{\rhoBarN - \rhoAlphaN}_1 \left(n\log\left(\frac{1}{\lambda_{\min}(\rhoAlpha)}\right) + n \log \left(\frac{1}{\lambda_{\min}(\rhoNot)}\right)\right)\\
&\label{eq:eigen-alpha} \leq &\norm{\rhoBarN - \rhoAlphaN}_1 n\log\left(\frac{2}{\lambda_{\min}(\rhoNot)^2}\right),
\end{IEEEeqnarray}
where \eqref{eq:eigen-alpha} follows from Lemma~\ref{lem:lambda-min-alpha}.
Combining the above inequalities concludes the proof.
\end{IEEEproof}

\begin{IEEEproof}[Proof of Lemma~\ref{lem:trace-expected}]
By quantum channel resolvability \cite[Lemma 9.2]{hayashi2006quantum}, we have for any $s_n\leq 0$ and $\beta_n \in \mathbb{R}$
\begin{align}
    E_{\mathcal{C}} \bigg[ \norm{\rhoBarN- \rhoAlphaN}_1  \bigg] \leq 2\sqrt{e^{\beta_n s_n + n \phi(s_n, \alpha_n)}} + \sqrt{\frac{e^{\beta_n }\nu_n}{MK}},\label{eq:hayashires}
\end{align}
where $\nu_n$ is the number of distinct eigenvalues of $\rhoAlphaN$ and $\phi(s_n, \alpha_n)$ is defined in \eqref{eq:phiSAlphaDef}. Choosing
\begin{align}
    \beta_n = \pra*{1+\frac{\varsigma_n}{2}}\alpha_n  
 n\sum_{x\in\mathcal{X}\setminus\{0\}} \pi_x D(\hat{\rho}_x || \hat{\rho}_0),
\end{align}
we upper-bound the exponent in the first term in the RHS of \eqref{eq:hayashires} as
\begin{align}
    {{\beta_n s_n + n \phi(s_n, \alpha_n)}}
    &\leq \beta_n s_n + n \pra*{-\alpha_n s_n \sum_{x\in\mathcal{X}\setminus\{0\}} \pi_x D(\hat{\rho}_x \| \hat{\rho}_0) + \vartheta_1\alpha_n s_n^2 -\vartheta_2 s_n^3}\label{eq:phi-bound-app}\\
    &= s_n\alpha_n n\left( {\frac{\varsigma_n}{2}}\sum_{x\in\mathcal{X}\setminus\{0\}} \pi_x D(\hat{\rho}_x || \hat{\rho}_0)  +\vartheta_1 s_n - \vartheta_2s_n^2 \alpha_n^{-1} \right)\label{eq:sub-gamma-algebra},
\end{align}
where \eqref{eq:phi-bound-app} follows from Lemma~\ref{lem:phi_bound} for $s_n\in[s_0, 0]$ with arbitrary constant $s_0<0$ and constants $\vartheta_1,\vartheta_2>0$ defined in Lemma~\ref{lem:phi_bound}, and \eqref{eq:sub-gamma-algebra} follows by substituting the value of $\beta_n$ and rearranging terms. We next set
\begin{align}
    s_n &= -\sqrt{\frac{\alpha_n \varsigma_n\sum_{x\in\mathcal{X}\setminus\{0\}} \pi_x D(\hat{\rho}_x || \hat{\rho}_0)}{4\vartheta_2}}.\label{eq:s}
\end{align}
Since $\alpha_n\varsigma_n\in o\left(\frac{1}{\sqrt{n}}\right)$ per the definition of $\alpha_n$ in Section \ref{sec:QSCS} and our choice of $\varsigma_n$, \eqref{eq:s} guarantees that, for large enough $n$, $s_n\in[s_0,0]$ for any constant $s_0<0$, and ensuring that Lemma \ref{lem:phi_bound} holds. Furthermore, \eqref{eq:s} implies that: 
\begin{align}
    -s_n\alpha_n n&\left( {\frac{\varsigma_n}{2}}\sum_{x\in\mathcal{X}\setminus\{0\}} \pi_x D(\hat{\rho}_x || \hat{\rho}_0)  + {   \vartheta_1 s_n - \vartheta_2 s_n^2 \alpha_n^{-1}} \right)   
    \\ \nonumber&=
    -{{s_n\alpha_n n\left( {\frac{\varsigma_n}{4}}\sum_{x\in\mathcal{X}\setminus\{0\}} \pi_x D(\hat{\rho}_x || \hat{\rho}_0)   + \vartheta_1 s_n \right)  }} \\
    &\in\Theta\left(\varsigma_n^{\frac{3}{2}}\alpha_n^{\frac{3}{2}}n\right)
    \label{eq:first-res},
\end{align}
where \eqref{eq:first-res} follows since $\vartheta_1 s\in o(1)$.
Moreover, the expression under the square root  in the second term in the RHS of \eqref{eq:hayashires} is
\begin{align}
    \frac{e^{\beta_n }\nu_n}{MK} 
    &= \frac{e^{\pra*{1+\frac{\varsigma_n}{2}}\alpha_n n \sum_{x\in\mathcal{X}\setminus\{0\}} \pi_x D(\hat{\rho}_x || \hat{\rho}_0)} \nu_n}{MK}\\
    &\leq\frac{e^{\pra*{1+\frac{\varsigma_n}{2}}\alpha_n  n \sum_{x\in\mathcal{X}\setminus\{0\}} \pi_x D(\hat{\rho}_x || \hat{\rho}_0)} (n+1)^{\textnormal{dim} \mathcal{H}_W}}{MK}\label{eq:dis-eigen}\\
    &= \frac{e^{\pra*{1+\frac{\varsigma_n}{2}}\alpha_n n \sum_{x\in\mathcal{X}\setminus\{0\}} \pi_x D(\hat{\rho}_x || \hat{\rho}_0)} (n+1)^{\textnormal{dim} \mathcal{H}_W}}{e^{\pra*{1+{\varsigma_n}}\alpha_n n \sum_{x\in\mathcal{X}\setminus\{0\}} \pi_x D(\hat{\rho}_x || \hat{\rho}_0)}}\label{eq:mk}\\
    &= e^{-\frac{\varsigma_n}{2}\alpha_n  n\sum_{x\in\mathcal{X}\setminus\{0\}} \pi_x D(\hat{\rho}_x || \hat{\rho}_0) + \log(n+1) \dim \mathcal{H}_W}\\
    &\in \Theta\left(e^{-\varsigma_n\alpha_n n}\right)\label{eq:second-res}
\end{align}
where \eqref{eq:dis-eigen} follows from \cite[Lemma 3.7]{hayashi2006quantum}, and \eqref{eq:mk} follows from our choice of $MK$. Substituting \eqref{eq:first-res} and \eqref{eq:second-res} in \eqref{eq:hayashires}, we conclude that there exists a sequence $\zeta_n\in\omega\pra{\frac{1}{\log n}}$, for $n$ large enough, such that
\begin{align}
    E_{\mathcal{C}} \bigg[ \norm{\rhoBarN- \rhoAlphaN}_1  \bigg]  \leq e^{-\zeta_n \alpha_n^{\frac{3}{2}}n}.
\end{align}
\end{IEEEproof}

\noindent{\bf Identification of a specific code:}
Let's choose $\varsigma_n$, $M$, and $K$ satisfying \eqref{eq:messageSize} and \eqref{eq:secretsize}.  This implies that \eqref{eq:achievableEPeb} and \eqref{eq:achievableEQRE} hold for sequences $\zeta_{n}^{(1)}\in \omega\left(\frac{1}{\log(n)^{\frac{4}{3}}n^{\frac{1}{3}}}\right)$ and $\zeta_n^{(2)}\in\omega\left(\frac{1}{\pra{\log n}^2}\right)$ for sufficiently large $n$. Now,
\begin{IEEEeqnarray}{rCl}
\IEEEeqnarraymulticol{3}{l}{P\left(P_{\rm e}^{(b)} \leq e^{-\varsigma_n^{(1)} \gamma_n \sqrt{n}} \cap \abs{D(\rhoBarN\| \rhoNotN) - D(\rhoAlphaN\| \rhoNotN)} \leq e^{-\varsigma_n^{(2)}\gamma_n^{\frac{3}{2}}n^{\frac{1}{4}}}\right)}\IEEEnonumber\\
&\geq & 1-P\left(P_{\rm e}^{(b)} \geq e^{-\varsigma_n^{(1)} \gamma_n \sqrt{n}}\right)-P\left(\abs{D(\rhoBarN\| \rhoNotN) - D(\rhoAlphaN\| \rhoNotN)} \geq e^{-\varsigma_n^{(2)}\gamma_n^{\frac{3}{2}}n^{\frac{1}{4}}}\right)\\
&\geq&1-e^{-(\zeta_n^{(1)}-\varsigma_n^{(1)})\gamma_n\sqrt{n}}- e^{-(\zeta_n^{(2)}-\varsigma_n^{(2)})\gamma_n^{\frac{3}{2}}n^{\frac{1}{4}}},\label{eq:errorBounds}
\end{IEEEeqnarray}
where \eqref{eq:errorBounds} is due to Markov's inequality used with \eqref{eq:achievableEPeb} and \eqref{eq:achievableEQRE}.
For some $0<\epsilon<1$, $\varsigma_n^{(1)}=(1-\epsilon)\zeta_n^{(1)}$ and $\varsigma_n^{(2)}=(1-\epsilon)\zeta_n^{(2)}$, the RHS of \eqref{eq:errorBounds} converges to unity as $n\to\infty$.
Thus, there exists at least one coding scheme that, for sufficiently large $n$ satisfies \eqref{eq:reliabilityCond} and \eqref{eq:covertnessCond} with high probability.
\end{IEEEproof}
\subsection{Converse} \label{sec:converse}
In this section we determine the upper bound on the covert capacity of the memoryless classical-quantum channel, as well as characterize the lower bound on pre-shared secret key requirement. Let's outline  the key concepts in the proof that follows. First, denote by $\mathbf{X}$ the classical random vector describing Alice's input $n$-symbol codeword and by $p_{\mathbf{X}}(\mathbf{x})$ its  distribution. Define the random variable $\tilde{X}$ with distribution:
\begin{align}
p_{\tilde{X}}(x) = \frac{1}{n} \sum_{k=1}^n p_{X_k} (x), \label{eq:tildeDist}
\end{align}
where $p_{X_k}(x)$ is the $k^{\text{th}}$ symbol's marginal density of $p_{\mathbf{X}} (\mathbf{x})$. Letting $\hat{\bar \sigma}_{k} = \sum_{x \in \mathcal{X}} p_{X_k} (x) \hat{\sigma}_x$ and $\hat{\bar \rho}_{k} = \sum_{x \in \mathcal{X}} p_{X_k}(x) \hat{\rho}_x$, the outputs of Bob's and Willie's channels induced by $p_{\tilde{X}}(x)$ are, respectively, $\hat{\tilde{\sigma}} = \frac{1}{n} \sum_{k = 1}^n \hat{\bar \sigma}_{k}$ and $\hat{\tilde{\rho}} = \frac{1}{n} \sum_{k = 1}^n \hat{\bar \rho}_{k}$. Bob's expected state induced by \eqref{eq:tildeDist} is,
\begin{align}
\hat{\tilde{\sigma}} = (1-\mu_n)\hat{\sigma}_0 + \mu_n \sum_{x\in\mathcal{X}\setminus\{0\}}\tilde{\pi}_x\hat{\sigma}_x,
\end{align}
where $\mu_n \equiv 1-p_{\tilde{X}}(0)$ is the average probability of transmitting a non-innocent symbol and $\tilde{\pi}_x=\frac{p_{\tilde{X}}{(x)}}{\mu_n}$. Similarly, Willie's expected state induced by \eqref{eq:tildeDist} is,
\begin{align}
\hat{\tilde{\rho}} = (1-\mu_n)\hat{\rho}_0 + \mu_n \sum_{x\in\mathcal{X}\setminus\{0\}}\tilde{\pi}_x\hat{\rho}_x.
\end{align}
We show that maintaining the covertness constraint described in Section \ref{sec:hypothesis_testing} necessarily constrains $\mu_n\in o\left(\frac{1}{\sqrt{n}}\right)$. This establishes the upper bound on $\log M$ that matches \eqref{eq:achL} from Corollary \ref{cor:scalingConstants}. Furthermore, when we employ the average marginal non-innocent state distribution that maximizes the covert rate, we establish the lower bound on $\log K$ that matches the upper bound \eqref{eq:achJ}. The proof adapts \cite[Sec. VI]{bloch15covert} and \cite{wang16cq-srlconverse}.

\begin{thm}[Converse] \label{thm:converse}
Consider a covert memoryless classical-quantum channel such that, for inputs $x\in\mathcal{X}=\left\{0,1,2,\ldots,N\right\}$, the output state $\hat{\rho}_0$ corresponding to innocent input $x=0$ is not a mixture of non-innocent ones $\left\{\hat{\rho}_x\right\}_{x\in\mathcal{X}\setminus\{0\}}$, and $\forall x\in\mathcal{X}\setminus\{0\}$, $\supp (\hat{\sigma}_x) \subseteq \supp (\hat{\sigma}_0)$ and $\supp(\hat{\rho}_x) \subseteq \supp (\hat{\rho}_0)$. For a sequence of covert communication schemes with increasing blocklength $n$, such that
\begin{align}
\lim_{n \to \infty} P_{\rm e}^{(b)} = 0 \text{ and } \lim_{n \to \infty} D(\rhoBarN \| \rhoNotN) = 0,
\end{align}
we have,
\begin{align}
L_{\rm SRL} \leq \frac{\sum_{x\in \mathcal{X}\setminus\{0\}} \pi^\ast_x D(\hat{\sigma}_x \| \hat{\sigma}_0)} {\sqrt{\frac{1}{2} \eta (\hat{\rho}^\ast_{\neg 0} \| \hat{\rho}_0) }}\label{eq:convL},
\end{align}
where $\{{\pi}^\ast_x\}=\argmax_{\{{\pi}_x\}} \frac{\sum_{x\in\mathcal{X}\setminus\{0\}}\pi_x D(\hat{\sigma}_x \| \hat{\sigma}_0)}{\sqrt{\frac{1}{2}\eta(\hat{\rho}_{\neg 0} \| \hat{\rho}_0)}}$,  $\hat{{\rho}}^\ast_{\neg 0}\triangleq\sum_{x\in\mathcal{X}\setminus\{0\}}{\pi}^\ast_x\hat{\rho}_x$ is  Willie's corresponding average non-innocent state, and $\hat{\bar{\rho}}^n$ is Willie's average state defined in \eqref{eq:W_transmissionstate}.
Furthermore, suppose the average marginal non-innocent state distribution $\{\tilde{\pi}_x\}\to\{{\pi}^\ast_x\}$ as $n\to\infty$. Then,
\begin{align}
J_{\rm SRL}&\geq \frac{\left[ \sum_{x\in\mathcal{X}\setminus\{0\}}\pi^\ast_x  \left( D(\hat{\rho}_x \| \hat{\rho}_0) - D(\hat{\sigma}_x \| \hat{\sigma}_0) \right) \right]^+ }{\sqrt{\frac{1}{2} \eta (\hat{\rho}^\ast_{\neg{0}} \| \hat{\rho}_0)}}.\label{eq:convJ}
\end{align}
\end{thm}

\begin{IEEEproof}[Proof (Theorem \ref{thm:converse})]
 Consider a sequence of covert communication schemes that, for an $n$-symbol input, maintain $P_{\rm e}^{(b)} \leq \epsilon_n$ and $D(\rhoBarN \| \rhoNotN) \leq \delta_n$ such that $\lim_{n \to \infty} \epsilon_n = 0$ and $\lim_{n \to \infty} \delta_n = 0$. Furthermore, as in the proof of \cite[Th.~3]{bloch15covert}, suppose that $\log M$ takes the maximum value such that $\lim_{n \to \infty} \log M = \infty$. This is a reasonable assumption since Theorem \ref{thm:achievability} proves the existence of a sequence of covert communication schemes meeting these conditions. Let $W$ be the random variable describing the message, $S$ be the random variable describing the pre-shared secret key, $\mathbf{X}$ be the classical random vector describing Alice's input $n$-symbol codeword, and $\mathbf{Y}$ be the classical random vector describing Bob's measurement of his output of the channel. 
First, we analyze $\log M$ to upper-bound $L_{\rm SRL}$. 
We have:
\begin{align}	
\log M &= H(W) \label{eq:logMbegin}\\
&= I(W ; \mathbf{Y} S) + H(W | \mathbf{Y} S) \\
&\leq I(W ; \mathbf{Y} S) + 1 + \epsilon_n \log M, \label{eq:fanosIneq}
\end{align}
where \eqref{eq:fanosIneq} follows by Fano's inequality. As the message and pre-shared secret key are independent, $I(W;S)=0$. Thus, $I(W;\mathbf{Y}|S) = I(W;\mathbf{Y}S) - I(W;S)=I(W;\mathbf{Y}S)$, yielding:
\begin{align}
\log M &\leq I(W ; \mathbf{Y}|S) + 1 + \epsilon_n \log M \\
&\leq I(WS;\mathbf{Y}) + 1 + \epsilon_n \log M \label{eq:chainRuleMI} \\
&\leq I(\mathbf{X};\mathbf{Y}) + 1 + \epsilon_n \log M, \label{eq:dataProcessing}
\end{align}
where \eqref{eq:chainRuleMI} follows from the chain rule for mutual information and \eqref{eq:dataProcessing} follows from the data processing inequality. Let $\chi \left( \left\{ p_{\mathbf{X}}(\mathbf{x}), \sigmaNx \right\} \right)$ be the Holevo information of the ensemble $\left\{ p_{\mathbf{X}}(\mathbf{x}), \sigmaNx \right\}$. Application of the Holevo bound \cite[Th. 12.1]{nielsen00quantum} to \eqref{eq:dataProcessing} yields:
\begin{align}
\log M &\leq \chi \left( \left\{ p_{\mathbf{X}}(\mathbf{x}), \sigmaNx \right\}\right) + 1 + \epsilon_n \log M \\ 
&= H\left(\sum_\mathbf{x} p_{\mathbf{X}}(\mathbf{x})\hat{\sigma}^n(\mathbf{x})\right) - \sum_\mathbf{x'}p_{\mathbf{X}}(\mathbf{x'})H(\hat{\sigma}^n(\mathbf{x'})) + 1 + \epsilon_n \log M.\label{eq:HolevoDefnLogMBound}
\end{align}
By subadditivity of von Neumann entropy, 
\begin{align}
    H\left(\sum_\mathbf{x} p_{\mathbf{X}}(\mathbf{x})\hat{\sigma}^n(\mathbf{x})\right) &\leq \sum_{k=1}^n H\left(\sum_\mathbf{x} p_{\mathbf{X}}(\mathbf{x})\Tr_{\neg k}\left[\hat{\sigma}^n(\mathbf{x})\right]\right) \label{eq:subaddvne1}\\
    &=\sum_{k=1}^nH\left(\sum_\mathbf{x} p_{\mathbf{X}}(\mathbf{x})\Tr_{\neg k}\left[\bigotimes_{i=1}^n\hat{\sigma}_{x_i}\right]\right)\\
    &= \sum_{k=1}^nH\left(\sum_{x}p_{X_k}(x)\hat{\sigma}_{x}\right),\label{eq:SubaddVNE}
\end{align}
where $\Tr_{\neg k}\left[\cdot\right]$ denotes the partial trace over $(n-1)$ subsystems that are not the $k^{\text{th}}$ channel use.
Combining \eqref{eq:HolevoDefnLogMBound} and \eqref{eq:SubaddVNE} yields
\begin{align}
    \log M &\leq \sum_{k=1}^nH\left(\sum_{x}p_{X_k}(x)\hat{\sigma}_x\right) - \sum_\mathbf{x}p_{\mathbf{X}}(\mathbf{x})H(\hat{\sigma}^n(\mathbf{x})) + 1 + \epsilon_n \log M \label{eq:rev1comment14-1} \\
    &= \label{eq:HolevoDefnProdState}\sum_{k=1}^n\left(H\left(\sum_{x}p_{X_k}(x)\hat{\sigma}_x\right) - \sum_{x'}p_{X_k}(x')H(\hat{\sigma}_{x'})\right) + 1 + \epsilon_n \log M
    \\
    &= \sum_{k = 1}^n \chi \left( \left\{ p_{X_k} (x), \hat{\sigma}_{x} \right\} \right) + 1 + \epsilon_n \log M, \label{eq:chiProdState}
\end{align}
where \eqref{eq:HolevoDefnProdState} follows from $\sigmaNx$ being a product state.  As the Holevo information is concave in the input distribution, \eqref{eq:chiProdState} is upper bounded using Jensen's inequality as:
\begin{align}
\log M &\leq n \chi \left( \left\{ p_{\tilde{X}} (x), \hat{\sigma}_x \right\} \right) + 1 + \epsilon_n \log M, \label{eq:BobJensen}
\end{align}
where $p_{\tilde{X}}(x)$ is defined in \eqref{eq:tildeDist}. Rearranging the terms yields:
\begin{align}
\log M &\leq \frac{1}{1-\epsilon_n} \left( n \chi \left( \left\{ p_{\tilde{X}} (x), \hat{\sigma}_x \right\} \right) + 1 \right). \label{eq:bobDeltaFrac}
\end{align}
Expanding the Holevo information yields:
\begin{align}
\chi \left( \left\{ p_{\tilde{X}}, \hat{\sigma}_x \right\} \right) &=\mu_n\sum_{x\in\mathcal{X}\setminus\{0\}}\tilde{\pi}_x D(\hat{\sigma}_x \| \hat{\sigma}_0)-D(\hat{\tilde{\sigma}}\|\hat{\sigma}_0) \label{eq:Holevoexpand}\\ 
&\leq \mu_n\sum_{x\in\mathcal{X}\setminus\{0\}}\tilde{\pi}_x D(\hat{\sigma}_x \| \hat{\sigma}_0),\label{eq:noneg}
\end{align}
where \eqref{eq:noneg} follows from the QRE being non-negative. 
Since we assume that, for all $x\in\mathcal{X}$, $\supp (\hat{\sigma}_x) \subseteq \supp(\hat{\sigma}_0)$, $ D(\hat{\sigma}_x \| \hat{\sigma}_0 ) < \infty$. 
Combining \eqref{eq:bobDeltaFrac} and \eqref{eq:noneg} gives us
\begin{align}
\log M \leq \frac{1}{1-\epsilon_n} \left( n \mu_n\sum_{x\in\mathcal{X}\setminus\{0\}}\tilde{\pi}_x D(\hat{\sigma}_x \| \hat{\sigma}_0)  + 1 \right). \label{eq:muNinfty}
\end{align}
By Theorem \ref{thm:achievability}, there exists a sequence of covert communication schemes such that $\lim_{n \to \infty} \log M = \infty$, which together with \eqref{eq:muNinfty} implies $\lim_{n\to\infty} n\mu_n=\infty$. Thus, we have 
\begin{align}
    \frac{1}{n\mu_n}\in o(1)\label{eq:limNMu}.
\end{align}
From the covertness condition, we have,
\begin{align}
\delta_n &\geq D(\rhoBarN \| \rhoNotN ) \\
&=-H\left(\sum_\mathbf{x} p_{\mathbf{X}}(\mathbf{x})\hat{\rho}^n(\mathbf{x})\right) - \Tr \left[\sum_\mathbf{x'} p_{\mathbf{X}}(\mathbf{x'})\hat{\rho}^n(\mathbf{x'})\log\hat{\rho}_0^{\otimes n }\right] \\
&\geq \label{eq:SubaddCovCrit}- \sum_{k=1}^nH\left(\sum_{x}p_{X_k}(x)\hat{\rho}_{x}\right) -\Tr \left[\sum_\mathbf{x} p_{\mathbf{X}}(\mathbf{x})\hat{\rho}^n(\mathbf{x})\log\hat{\rho}_0^{\otimes n }\right]\\
&=\sum_{k=1}^n\left(H\left(\sum_{x}p_{X_k}(x)\hat{\rho}_{x}\right) -\Tr \left[\sum_\mathbf{x} p_{\mathbf{X}}(\mathbf{x})\left(\bigotimes_{l=1}^{k-1}\hat{\rho}_{x_l}\right)\otimes\hat{\rho}_{x_k}\log\hat{\rho}_0\otimes\left(\bigotimes_{j=k+1}^n\hat{\rho}_{x_j}\right)\right]\right)\label{eq:rev1comment14}\\
&= \sum_{k=1}^n\left(-H\left(\sum_{x}p_{X_k}(x)\hat{\rho}_{x}\right)-\Tr\left[\sum_{x'}p_{X_k}(x')\hat{\rho}_{x'}\log\hat{\rho_0}\right]\right)
\\
&= \sum_{k=1}^nD(\hat{\bar\rho}_{x_k}\|\hat{\rho}_0)
\\
&\geq n D ( \hat{\tilde{\rho}} \| \hat{\rho}_0 ) \label{eq:QREconvex},
\end{align}
where \eqref{eq:SubaddCovCrit} follows from \eqref{eq:SubaddVNE} and \eqref{eq:QREconvex} follows from the convexity of the QRE. 
Note that, using the quantum Pinsker's inequality,
\begin{align}
D(\hat{\tilde{\rho}} \| \hat{\rho}_0 ) \geq \frac{\|\hat{\tilde{\rho}} - \hat{\rho}_0 \|_1^2 }{2\log 2} = \mu_n^2 \frac{\|\hat{\tilde{\rho}}_{\neg 0} - \hat{\rho}_0 \|_1^2}{2\log 2} , \label{eq:qPinskerCovertness}
\end{align}
where $\hat{\tilde{\rho}}_{\neg 0}\triangleq\sum_{x\in\mathcal{X}\setminus\{0\}}\tilde{\pi}_x\hat{\rho}_x$.  Combined with \eqref{eq:QREconvex}, \eqref{eq:qPinskerCovertness} implies that the covertness condition is maintained only when:
\begin{align}
\lim_{n \to \infty} \mu_n \sqrt{n} = 0, \label{eq:muNcondition}
\end{align} 
which, in turn, implies that $\mu_n\in o\left(\frac{1}{\sqrt{n}}\right)$. 
Also by \eqref{eq:QREconvex} we have,
\begin{align}
\frac{1}{\mu_n} D(\hat{\tilde{\rho}} \| \hat{\rho}_0 ) \leq \frac{\delta_n}{n\mu_n},\label{eq:nDlessThanDeltan}
\end{align}
implying that $\lim_{n \to \infty} \frac{1}{\mu_n} D(\hat{\tilde{\rho}} \| \hat{\rho}_0 ) = 0$ by \eqref{eq:limNMu}.

Now, dividing both sides of \eqref{eq:bobDeltaFrac} by $\sqrt{n D ( \rhoBarN \| \rhoNotN )}$ and applying \eqref{eq:noneg} yields:
\begin{align}
\frac{\log M}{\sqrt{n D ( \rhoBarN \| \rhoNotN )}} & \leq \frac{n\mu_n\sum_{x\in\mathcal{X}\setminus\{0\}}\tilde{\pi}_x D(\hat{\sigma}_x \| \hat{\sigma}_0) + 1 }{(1-\epsilon_n) \sqrt{n D ( \rhoBarN \| \rhoNotN )}} \label{eq:logMBound} \\
&\leq \frac{n\mu_n\sum_{x\in\mathcal{X}\setminus\{0\}}\tilde{\pi}_x D(\hat{\sigma}_x \| \hat{\sigma}_0) + 1 }{(1-\epsilon_n) \sqrt{n^2 D( \hat{\tilde{\rho}} \| \hat{\rho}_0)}} \label{eq:QREprodstate}, 
\end{align}
where \eqref{eq:QREprodstate} follows from \eqref{eq:QREconvex}. Application of Lemma \ref{lem:etaDefn} to \eqref{eq:QREprodstate} with $\alpha = \mu_n$ yields:
\begin{align}
\frac{\log M}{\sqrt{n D ( \rhoBarN \| \rhoNotN )}}
&\leq \frac{n\mu_n\sum_{x\in\mathcal{X}\setminus\{0\}}\tilde{\pi}_x D(\hat{\sigma}_x \| \hat{\sigma}_0) + 1 }{(1-\epsilon_n)\sqrt{\frac{n^2\mu_n^2}{2}\eta(\hat{\tilde{\rho}}_{\neg 0} \| \hat{\rho}_0) +n^2R(\mu_n)
}}  \label{eq:chiUpperBound} \\
&= \frac{ \sum_{x\in\mathcal{X}\setminus\{0\}}\tilde{\pi}_x D(\hat{\sigma}_x \| \hat{\sigma}_0) + \frac{1}{n \mu_n} }{(1-\epsilon_n)\sqrt{\frac{1}{2}\eta(\hat{\tilde{\rho}}_{\neg 0} \| \hat{\rho}_0) + R(\mu_n)/\mu_n^2}}.\label{eq:preLimitLBound}
\end{align}
Taking the limit of both sides of \eqref{eq:preLimitLBound} yields:
\begin{align}
    L_{\rm SRL} &\leq \lim_{n\to \infty}\frac{ \sum_{x\in\mathcal{X}\setminus\{0\}}\tilde{\pi}_x D(\hat{\sigma}_x \| \hat{\sigma}_0) + \frac{1}{n \mu_n} }{(1-\epsilon_n)\sqrt{\frac{1}{2}\eta(\hat{\tilde{\rho}}_{\neg 0} \| \hat{\rho}_0) + R(\mu_n)/\mu_n^2}} \\
    &\leq \limsup_{n\to \infty}\frac{ \sum_{x\in\mathcal{X}\setminus\{0\}}\tilde{\pi}_x D(\hat{\sigma}_x \| \hat{\sigma}_0) + \frac{1}{n \mu_n} }{(1-\epsilon_n)\sqrt{\frac{1}{2}\eta(\hat{\tilde{\rho}}_{\neg 0} \| \hat{\rho}_0) + R(\mu_n)/\mu_n^2}} \label{eq:Llimsup}.
\end{align}
Now, recall that $\frac{1}{n\mu_n},\epsilon_n,\frac{R(\mu_n)}{\mu_n^2} \in o(1)$. Thus, for every $\epsilon>0$, there is an $N$ such that:
\begin{align}
    \frac{ \sum_{x\in\mathcal{X}\setminus\{0\}}\tilde{\pi}_x D(\hat{\sigma}_x \| \hat{\sigma}_0) + \frac{1}{n \mu_n} }{(1-\epsilon_n)\sqrt{\frac{1}{2}\eta(\hat{\tilde{\rho}}_{\neg 0} \| \hat{\rho}_0) + R(\mu_n)/\mu_n^2}}
    &\leq\frac{\sum_{x\in\mathcal{X}\setminus\{0\}}{\pi}^\ast_x D(\hat{\sigma}_x \| \hat{\sigma}_0)}{\sqrt{\frac{1}{2}\eta(\hat{{\rho}}^\ast_{\neg 0} \| \hat{\rho}_0)}} + \epsilon \label{eq:Llimsupbound}
\end{align}
for every $n>N$. The assumption that the innocent state cannot be reconstructed from non-innocent states ensures $\eta(\hat{{\rho}}^\ast_{\neg 0} \| \hat{\rho}_0)>0$, which implies that our bound is not vacuous. \eqref{eq:Llimsup} together with \eqref{eq:Llimsupbound} yields the desired result in \eqref{eq:convL}.

Now we analyze $\log M + \log K$ to lower-bound $J_{\rm SRL}$. Generalizing \cite[Sec. 5.2.3]{hou2014coding} to classical-quantum channels, we have,
\begin{align}
\log M + \log K &= H(\mathbf{X}) \label{eq:logKlogMbegin}\\
&\geq H(\hat{\bar{\rho}}^n) - \sum_{\mathbf{x} \in \mathcal{X}^n} p_{\mathbf{X}}(\mathbf{x}) H\left(\hat{\rho}^n (\mathbf{x}) \right), \label{eq:entropyNonNeg}
\end{align}
where \eqref{eq:entropyNonNeg} follows from \cite[Ex. 11.9.4]{wilde16quantumit2ed}. Using the covertness condition with \eqref{eq:entropyNonNeg}, we write:
\begin{align}
\log M + \log K &\geq H(\hat{\bar{\rho}}^n) - \sum_{\mathbf{x} \in \mathcal{X}^n} p_{\mathbf{X}}(\mathbf{x}) H\left(\hat{\rho}^n (\mathbf{x}) \right) + D(\rhoBarN \| \rhoNotN ) -\delta_n, \label{eq:covertnessCondMK}
\end{align}
Expanding the QRE in \eqref{eq:covertnessCondMK}, we have,
\begin{align}
\log M + \log K &\geq H(\rhoBarN) - \sum_{\mathbf{x} \in \mathcal{X}^n} p_{\mathbf{X}}(\mathbf{x})H\left(\hat{\rho}^n (\mathbf{x}) \right) - H(\rhoBarN) - \Tr \left\{\rhoBarN \log \rhoNotN \right\} - \delta_n \\
&= - \sum_{\mathbf{x} \in \mathcal{X}^n} p_{\mathbf{X}}(\mathbf{x})H\left(\hat{\rho}^n (\mathbf{x}) \right) - \sum_{\mathbf{x^\prime }\in \mathcal{X}^n} p_{\mathbf{X}}(\mathbf{x^\prime}) \Tr \left\{ \hat{\rho}^n (\mathbf{x^\prime}) \log \rhoNotN \right\} - \delta_n \\
&= \sum_{\mathbf{x} \in \mathcal{X}^n} p_{\mathbf{X}}(\mathbf{x}) \bigg(- H \left(\hat{\rho}^n(\mathbf{x}) \right) - \Tr \{ \hat{\rho}^n(\mathbf{x}) \log \rhoNotN \} \bigg) - \delta_n \\
&= \sum_{x \in \mathcal{X}} \sum_{k=1}^n p_{X_k}(x) \bigg(- H \left(\hat{\rho}_{x} \right) - \Tr \{ \hat{\rho}_{x} \log \hat{\rho}_0 \} \bigg) - \delta_n, \label{eq:rhoProdState}
\end{align}
where \eqref{eq:rhoProdState} follows from $\hat{\rho}^n (\mathbf{x})$ and $\rhoNotN$ being product states. Substituting \eqref{eq:tildeDist}, we have,
\begin{align}
\log M + \log K &\geq n \sum_{x \in \mathcal{X}} p_{\tilde{X}}(x) \bigg(- H \left(\hat{\rho}_x \right) - \Tr \{ \hat{\rho}_x \log \hat{\rho}_0 \} \bigg) - \delta_n \\
&= -n \sum_{x \in \mathcal{X}} p_{\tilde{X}}(x) H(\hat{\rho}_x) - n \Tr \{ \hat{\tilde{\rho}} \log \hat{\rho}_0 \}  - \delta_n \\
&\geq -n \sum_{x \in \mathcal{X}} p_{\tilde{X}}(x) H(\hat{\rho}_x) - n \Tr \{ \hat{\tilde{\rho}} \log \hat{\rho}_0 \} - n D(\hat{\tilde{\rho}} \| \hat{\rho}_0) - \delta_n \label{eq:QREnotNeg} \\
&= -n \sum_{x \in \mathcal{X}} p_{\tilde{X}}(x) H(\hat{\rho}_x) +n H(\hat{\tilde{\rho}}) - \delta_n \\
&= n \chi \left( \left\{ p_{\tilde{X}}(x), \hat{\rho}_x \right\} \right) - \delta_n, \label{eq:logMlogK_chi}
\end{align}
where \eqref{eq:QREnotNeg} follows from the non-negativity of QRE. 
Now, as in \eqref{eq:Holevoexpand},  
\begin{align}
\chi \left( \left\{ p_{\tilde{X}}, \hat{\rho}_x \right\} \right) =\mu_n\sum_{x\in\mathcal{X}\setminus\{0\}}\Tilde{\pi}_x D(\hat{\rho}_x \| \hat{\rho}_0) - D(\hat{\tilde{\rho}} \| \hat{\rho}_0). \label{eq:willieHolevoInf}
\end{align}
Dividing both sides of \eqref{eq:logMlogK_chi} by $\sqrt{nD(\rhoBarN \| \rhoNotN)}$ and combining with \eqref{eq:willieHolevoInf} yields:
\begin{align}
\frac{\log M + \log K}{\sqrt{nD(\rhoBarN \| \rhoNotN)}} &\geq  \frac{n \mu_n  \sum_{x\in\mathcal{X}\setminus\{0\}}\tilde{\pi}_x D(\hat{\rho}_x \| \hat{\rho}_0) - n D(\hat{\tilde{\rho}} \| \hat{\rho}_0) - \delta_n}{\sqrt{nD(\rhoBarN \|\rhoNotN)}} \label{eq:willieHolevoSubstitute}\\
&\geq \frac{n\mu_n}{\sqrt{nD(\rhoBarN \|\rhoNotN)}} \left( \sum_{x\in\mathcal{X}\setminus\{0\}}\tilde{\pi}_x D(\hat{\rho}_x \| \hat{\rho}_0) - \frac{2\delta_n}{n\mu_n} \right)\label{eq:beforeSubBetaTerm},
\end{align}
where \eqref{eq:beforeSubBetaTerm} is because $\frac{D(\hat{\tilde{\rho}} \| \hat{\rho}_0)}{\mu_n} \leq \frac{\delta_n}{n\mu_n}$ by \eqref{eq:nDlessThanDeltan}.
 We lower-bound the left-hand side of \eqref{eq:willieHolevoSubstitute} further by employing the approach from the proof of \cite[Th.~3]{bloch15covert}: by Corollary \ref{cor:scalingConstants}, there exists a sequence of covert communication schemes such that, for large enough $n$ and $\vartheta_n\in o(1)$, 
\begin{align}
\frac{\log M}{\sqrt{nD(\rhoBarN \| \rhoNotN)}}&\geq  \frac{(1-\vartheta_n)\sum_{x\in\mathcal{X}\setminus\{0\}}{\pi}^\ast_x D(\hat{\sigma}_x \| \hat{\sigma}_0) }{\sqrt{\frac{1}{2}\eta(\hat{\rho}^\ast_{\neg 0} \| \hat{\rho}_0)}}.\label{eq:logMcodingbound}
\end{align}
Combining \eqref{eq:logMcodingbound} with \eqref{eq:logMBound} and rearranging yields:
\begin{align}
\frac{n\mu_n}{\sqrt{nD(\rhoBarN \|\rhoNotN)}} & \geq \frac{(1-\vartheta_n)(1-\epsilon_n)\sum_{x\in\mathcal{X}\setminus\{0\}}{\pi}^\ast_x D(\hat{\sigma}_x \| \hat{\sigma}_0) }{\sqrt{\frac{1}{2}\eta(\hat{\rho}^\ast_{\neg 0} \| \hat{\rho}_0)}\left(\sum_{x\in\mathcal{X}\setminus\{0\}}\tilde{\pi}_x D(\hat{\sigma}_x \| \hat{\sigma}_0) + \frac{1}{n\mu_n}\right)}.\label{eq:normalizednmunlowerbound}
\end{align}
Combining \eqref{eq:beforeSubBetaTerm} and \eqref{eq:normalizednmunlowerbound}, we obtain: 
\begin{align}
\frac{\log M + \log K}{\sqrt{nD(\rhoBarN \| \rhoNotN)}} &\geq \frac{(1-\vartheta_n)(1-\epsilon_n)\sum_{x\in\mathcal{X}\setminus\{0\}}{\pi}^\ast_x D(\hat{\sigma}_x \| \hat{\sigma}_0) }{\sqrt{\frac{1}{2}\eta(\hat{\rho}^\ast_{\neg 0} \| \hat{\rho}_0)}\left(\sum_{x\in\mathcal{X}\setminus\{0\}}\tilde{\pi}_x D(\hat{\sigma}_x \| \hat{\sigma}_0) + \frac{1}{n\mu_n}\right)}\nonumber\\
&\phantom{\geq}\times\left( \sum_{x\in\mathcal{X}\setminus\{0\}}\tilde{\pi}_x D(\hat{\rho}_x \| \hat{\rho}_0) - \frac{2\delta_n}{n\mu_n} \right).\label{eq:finallogMlogKbound}
\end{align}
Subtracting \eqref{eq:preLimitLBound} from \eqref{eq:finallogMlogKbound} yields:
\begin{align}
\frac{\log K}{\sqrt{nD(\rhoBarN \| \rhoNotN)}} &\geq \frac{(1-\vartheta_n)(1-\epsilon_n)\sum_{x\in\mathcal{X}\setminus\{0\}}{\pi}^\ast_x D(\hat{\sigma}_x \| \hat{\sigma}_0) }{\sqrt{\frac{1}{2}\eta(\hat{\rho}^\ast_{\neg 0} \| \hat{\rho}_0)}\left(\sum_{x\in\mathcal{X}\setminus\{0\}}\tilde{\pi}_x D(\hat{\sigma}_x \| \hat{\sigma}_0) + \frac{1}{n\mu_n}\right)}\nonumber\\
&\phantom{\geq-}\times\left( \sum_{x\in\mathcal{X}\setminus\{0\}}\tilde{\pi}_x D(\hat{\rho}_x \| \hat{\rho}_0) - \frac{2\delta_n}{n\mu_n} \right)\nonumber \\
&\phantom{\geq}-\frac{ \sum_{x\in\mathcal{X}\setminus\{0\}}\tilde{\pi}_x D(\hat{\sigma}_x \| \hat{\sigma}_0) + \frac{1}{n \mu_n} }{(1-\epsilon_n)\sqrt{\frac{1}{2}\eta(\hat{\tilde{\rho}}_{\neg 0} \| \hat{\rho}_0) +R(\mu_n)/\mu_n^2}}\label{eq:finallogKbound}
\end{align}
We obtain the lower bound on $J_{\rm SRL}$ stated in \eqref{eq:convJ} by taking the limit of both sides of \eqref{eq:finallogKbound} as $n\to\infty$, with \eqref{eq:limNMu}, and again noting that $\mu_n\in o\left(\frac{1}{\sqrt{n}}\right)$ by \eqref{eq:muNcondition}, $R(\mu_n)\in\mathcal{O}(\mu_n^3)$ by Lemma \ref{lem:etaDefn}, and $\{\tilde{\pi}_x\}\to\{\pi_x^\ast\}$ by assumption.
\end{IEEEproof}
\begin{rk}
    Although the upper bound on the covert capacity did not require any assumptions on the innocent distribution, we require the assumption $\{\tilde{\pi}_x\}\to\{\pi_x^\ast\}$ for the lower bound on the pre-shared secret key requirement. This assumption is unnecessary in the case where there is only one non-innocent state and the non-innocent distribution is trivial. We believe that we can dispense with this assumption in general by adapting \cite{shamai97goodcodes}. We defer this to future work. 
\end{rk}

\section{Special Cases} \label{sec:cornercases}

\subsection{Constant rate covert communication} \label{subsec:constantrate}
Suppose Willie's state induced by the innocent input, $\hat{\rho}_0$, is a mixture of $\{ \hat{\rho}_x \}_{x \in \mathcal{X} \setminus \{0\}}$, i.e., there exists a distribution $\pi (x)$ on non-innocent symbols where $\sum_{x \in \mathcal{X} \setminus \{0\}} \pi(x) = 1$ such that $\hat{\rho}_0 = \sum_{x \in \mathcal{X} \setminus \{0\}} \pi (x) \hat{\rho}_x$. Define the probability distribution:
\begin{align}
p_X(x) &=
\begin{cases}
	1- \alpha, & x = 0 \\
	\alpha \pi(x), & x \neq 0,\\
\end{cases}
\end{align}
where $0 < \alpha \leq 1$ is the probability of using a non-innocent symbol. Using $\{ p_X (x) \}$ on input symbols results in an ensemble $\{ p_X (x), \hat{\sigma}_x \}$, at Bob's output that has positive Holevo information by the Holevo-Schumacher-Westmoreland (HSW) theorem \cite[Ch. 19]{wilde16quantumit2ed}. Thus, Alice can simply draw her codewords from the set of states using $\{p_X (x)\}$ and transmit at the positive rate undetected by Willie. Therefore, in the previous sections, we assume that $\hat{\rho}_0$ is not a mixture of the non-innocent symbols.

\subsection{$\mathcal{O}(\sqrt{n}\log n)$ covert communication} \label{subsec:sqrtnlogn}

Now we consider the case where part of the output-state support for at least one non-innocent input lies outside Bob's innocent-state support while lying inside Willie's innocent-state support.  While the resulting $\sqrt{n} \log n$ scaling is identical to that for entanglement-assisted covert communications over the bosonic \cite[Sec. III.C]{gagatsos20codingcovcomm} and qubit depolarizing \cite{zlotnick23eacovcommdepol} channels, the scaling constants have different form. We leave investigation of connection between these results to future work.

In the following theorems, we assume binary inputs.
Our approach can be generalized to more than two inputs, but care must be taken in delineating the supports of the resulting quantum states.
This does not affect the fundamental scaling law, but increases the complexity of the proof.
Therefore, to simplify the exposition, we analyze only the binary-input scenario.

\begin{thm} Consider a covert memoryless classical-quantum channel such that binary inputs $\mathcal{X}=\{0,1\}$ induce outputs $\{\hat{\sigma}_0, \hat{\sigma}_1 \}$ and $\{\hat{\rho}_0, \hat{\rho}_1\}$ at Bob and Willie, respectively.
Suppose the output state $\hat{\rho}_0$ corresponds to innocent output $x=0$, and $\supp (\hat{\rho}_{1}) \subseteq \supp(\hat{\rho}_0)$ while $\supp (\hat{\sigma}_{1}) \not\subseteq \supp(\hat{\sigma}_0)$. For $\alpha_n \triangleq \frac{\gamma_n}{\sqrt{n}}$ with $\gamma_n$ defined in \eqref{eq:gamman}, and any $\varsigma_n\in o(1)\cap\omega\left(\sqrt{\gamma_n}n^{\frac{1}{4}}\right)$, there exist sequences $\varsigma_n^{(1)} \in \omega\left(\gamma_n\sqrt{n}\right), \varsigma_2 \in \omega\left(\gamma_n^{\frac{3}{2}} n^{\frac{1}{4}}\right)$ such that, \label{thm:Onsqrtn_achievability}
\begin{align}
\log M &\geq (1 - \varsigma_n)\kappa\gamma_n \sqrt{n}\log n \left( \frac{1}{2} + \frac{\log \gamma_n^{-1}}{\log n} \right),\label{eq:secondLogMDef} \\
\log K &= 0,
\end{align}
where $\kappa=1-\Tr\left[\hat{\Pi}_0\hat{\sigma}_1\right]$, with $\hat{\Pi}_0$ being the projection onto the support of $\hat{\sigma}_0$, and
\begin{align}
P_{\rm{e}}^{(b)} &\leq e^{-\varsigma_n^{(1)} \gamma_n \sqrt{n}}, \\
\abs{D(\hat{\bar{\rho}}^n\| \rhoNotN) - D(\rhoAlphaN \| \rhoNotN)} &\leq e^{-\varsigma_n^{(2)} \gamma_n^{\frac{3}{2}} n^{\frac{1}{4}}} 
\end{align}
where  Willie's average state $\hat{\bar{\rho}}^n$ is defined in \eqref{eq:W_transmissionstate}.
\end{thm}

\begin{IEEEproof} [Proof (Theorem \ref{thm:Onsqrtn_achievability})]
{\bf Construction:}
For each $m$, where $m\in\left\{1,\ldots, M\right\}$ is a message, Alice generates an i.i.d. random sequence $\mathbf{x} (m)\in \mathcal{X}^n$ from the binary distribution 
\begin{align}
p_X(x)&=\left\{\begin{array}{ll}1-\alpha_n,& x=0\\ \alpha_n,&x=1\end{array}\right.,
\end{align}
where $\alpha_n$ satisfies the requirements for a covert quantum-secure state given in Section \ref{sec:QSCS}. 
Alice chooses a codeword $\mathbf{x}(m)$ based on the message $m$ she wants to send. The codebook is used only once for a single shot transmission. Willie's and Bob's output states corresponding to a single use of the classical-quantum channel by Alice are given in \eqref{eq:rhoalphan} and \eqref{eq:bobState}.

\noindent \textbf{Reliability analysis:}
Let Bob use a POVM $\{(\hat{I}-\hat{\Pi}_0), \hat{\Pi}_0\}$ on each of his $n$ received states, where $\hat{\Pi}_0$ is a projection onto the support of $\hat{\sigma}_0$. This induces a classical DMC with unity probability of correctly identifying $\hat{\sigma}_0$ and probability $\kappa=1-\Tr\left[\hat{\Pi}_0 \hat{\sigma}_1 \right]$ of correctly identifying $\hat{\sigma}_1$. This construction admits the reliability analysis in the proof of \cite[Th. 7]{bloch15covert}, and results in \eqref{eq:secondLogMDef}.

\noindent \textbf{Covertness analysis:} The same analysis as performed in Theorem \ref{thm:achievability} can still be applied here, with choice of $\log M$ in \eqref{eq:secondLogMDef} being adequate to ensure that,
\begin{align}
E_{\mathcal{C}} [ D(\rhoBarN \| \rhoAlphaN ) ] \leq e^{-\varsigma_n^{(2)} \gamma_n^{\frac{3}{2}} n^{\frac{1}{4}}} ,
\end{align}
for appropriate choice of constant $\varsigma_2 > 0$.
\end{IEEEproof}

The scaling constant of $\log M$ is the square root-log law covert capacity:

\begin{defn}[Square root-log law covert rate]
The rate $R_{n, \log}$ of a covert communication scheme $\mathcal{K}_n$ over the memoryless square root-log law classical-quantum channel is 
\begin{align}
    R_{n, \log} = \frac{\log M}{\sqrt{n \log n D(\rhoBarN \| \rhoNotN)}},
\end{align}
where $M$ is the size of the message set and $n$ is blocklength. 
\end{defn}

\begin{defn}[Square root-log law achievable covert rate]
    A covert rate $R_{\log}$ is achievable if there exists a sequence of covert communication schemes $\left(\mathcal{K}_n\right)$ such that 
    \begin{align}
        \label{eq:Rlog}
        R_{\log}&=\lim_{n\to\infty} R_{n, \log} \text{, with }\\
\label{eq:Rlogconds}\lim_{n \to \infty} D(\rhoBarN \| \rhoNotN) &= 0, \lim_{n \to \infty} P_{\rm e}^{(b)} = 0.
    \end{align}
\end{defn}

\begin{defn}[Square root-log law covert capacity]
    \label{def:Llog}The capacity $L_{\log}$ of covert communication over the memoryless square root-log law classical-quantum channel is the supremum of achievable covert rates defined in \eqref{eq:Rlog}.
\end{defn}

The following characterizes the square root-log law covert capacity:

\begin{cor} \label{cor:scalingConstM}
Consider a covert memoryless classical-quantum channel defined in the statement of Theorem \ref{thm:Onsqrtn_achievability}. Then, there exists a covert communication code such that
\begin{align}
\lim_{n \to \infty} D(\rhoBarN \| \rhoNotN) = 0, \lim_{n \to \infty} P_{\rm e}^{(b)} = 0,
\end{align}
and 
\begin{align}
L_{\log}& \geq \kappa \sqrt{\frac{2}{\eta ( \hat{\rho}_1 \| \hat{\rho}_0)}} \left( \frac{1}{2} + \xi_1 \right),\label{eq:Onsqrtn_achievability}
\end{align}
where $\xi_1 = \lim_{n \to \infty} \frac{\log \gamma_n ^{-1} }{\log n}$, $\kappa$ is defined in the statement of Theorem \ref{thm:Onsqrtn_achievability}, and  Willie's average state $\hat{\bar{\rho}}^n$ is defined in \eqref{eq:W_transmissionstate}.
\end{cor}
\begin{IEEEproof} [Proof (Corollary \ref{cor:scalingConstM})]
Combining \eqref{eq:QREUpperBound}, \eqref{eq:QRErhoBarNrhoNotN}, and \eqref{eq:secondLogMDef}, we have, for $\varsigma_n \in o(1)$:
\begin{align}
\lim_{n \to \infty} \frac{\log M}{\sqrt{n D( \rhoBarN \| \rhoNotN )} \log n} &\geq \lim_{n \to \infty} \frac{(1 - \varsigma_n)\kappa\gamma_n \sqrt{n}\log n \left( \frac{1}{2} + \frac{\log \gamma_n^{-1}}{\log n} \right)}{\sqrt{ n\frac{\gamma_n^2}{2} \eta(\hat{\rho}_{\neg 0} \| \hat{\rho}_0)  + n^2R(\alpha_n)+ne^{-\varsigma_2 \gamma_n^{\frac{3}{2}}n^{\frac{1}{4}}}
}\log n}. 
\end{align}
Using the definition of $\gamma_n$ in \eqref{eq:gamman} and that $R(\alpha_n) \in \mathcal{O}(\alpha_n^3)$ yields \eqref{eq:Onsqrtn_achievability} and the corollary.
\end{IEEEproof}

Now we prove the following converse result:
\begin{thm} Consider a covert memoryless classical-quantum channel such that binary inputs $\mathcal{X}=\{0,1\}$ induce outputs $\{\hat{\sigma}_0, \hat{\sigma}_1 \}$ and $\{\hat{\rho}_0, \hat{\rho}_1\}$ at Bob and Willie, respectively.
Suppose the output state $\hat{\rho}_0$ corresponds to innocent input $x=0$, and $\supp (\hat{\rho}_{1}) \subseteq \supp(\hat{\rho}_0)$ while $\supp (\hat{\sigma}_{1}) \not\subseteq \supp(\hat{\sigma}_0)$.  For a sequence of covert communication codes with increasing blocklength $n$ such that $\lim_{n \to \infty} P_{\rm e}^{(b)} = 0$ and  $\lim_{n \to \infty} D(\rhoBarN \| \rhoNotN) = 0$, we have:
\label{thm:Onsqrtn_converse}
\begin{align}
L_{\log}&\leq \kappa \sqrt{\frac{2}{\eta (\hat{\rho}_1 \| \hat{\rho}_0 )}} \left(\frac{1}{2} +  \xi_2 \right),\label{eq:Onsqrtn_converse}
\end{align}
where $\kappa=1-\Tr\left[\hat{\Pi}_0\hat{\sigma}_1\right]$,  Willie's average state $\hat{\bar{\rho}}^n$ is defined in \eqref{eq:W_transmissionstate}, and $\xi_2 = \lim_{n \to \infty} \frac{\log \varpi_n^{-1}}{\log n}$ with $\hat{\Pi}_0$ being the projection onto the support of $\hat{\sigma}_0$ and $\varpi_n = o(1) \cap \omega \left( \frac{1}{\sqrt{n}\log n} \right)$.
\end{thm}
\begin{IEEEproof}[Proof (Theorem \ref{thm:Onsqrtn_converse})] 
As in the proof of Theorem \ref{thm:converse}, consider a sequence of $n$ input symbols with $P_{\rm e}^{(b)} \leq \epsilon_n$ and $D(\rhoBarN \| \rhoNotN) \leq \delta_n$ such that $\lim_{n \to \infty} \epsilon_n = 0$ and $\lim_{n \to \infty} \delta_n = 0$, and suppose that $\log M$ takes the maximum value (such that $\lim_{n \to \infty} \log M = \infty$ per Theorem \ref{thm:Onsqrtn_achievability}). We can apply the results and notation in \eqref{eq:tildeDist}-\eqref{eq:chiProdState} and \eqref{eq:logKlogMbegin}-\eqref{eq:logMlogK_chi} here, as they do not rely on the supports of Bob's received states. However, since $\supp(\hat{\sigma}_x) \nsubseteq \supp (\hat{\sigma}_0)$, the bound on the Holevo information of Bob's average state in \eqref{eq:noneg} cannot be used. Instead, we employ the projection $\hat{\Pi}_0$ onto the support of $\hat{\sigma}_0$. We expand the Holevo information by rearranging terms as follows:
\begin{align}
    \chi \left( \left\{ p_{\tilde{X}}, \hat{\sigma}_x \right\} \right) &= (1 - \mu_n) D(\hat{\sigma}_0 \| \hat{\tilde{\sigma}} ) + \mu_n D(\hat{\sigma}_1 \| \hat{\tilde{\sigma}} ) \label{eq:sqrtnStart}.
\end{align}
Starting with the second term of \eqref{eq:sqrtnStart}, 
\begin{align}
    \mu_n D(\hat{\sigma}_1 \| \hat{\tilde{\sigma}} ) &= \mu_n\Tr \left[\hat{\sigma}_1\log\hat{\sigma}_1\right] - \mu_n\Tr \left[\hat{\sigma}_1\log\hat{\tilde{\sigma}}\right] \\
    &=\mu_n\Tr \left[\hat{\sigma}_1\log\hat{\sigma}_1\right]- \mu_n\Tr \left[\sqrt{\hat{\sigma}_1}\log\hat{\tilde{\sigma}}\sqrt{\hat{\sigma}_1}\right] \\
    &=\mu_n\Tr \left[\hat{\sigma}_1\log\hat{\sigma}_1\right] - \mu_n\Tr \left[\hat{\Pi}_0\sqrt{\hat{\sigma}_1}\log\hat{\tilde{\sigma}}\sqrt{\hat{\sigma}_1}\right] \nonumber
    \\
    &\phantom{=}- \mu_n\Tr \left[(\hat{I}-\hat{\Pi}_0)\sqrt{\hat{\sigma}_1}\log\hat{\tilde{\sigma}}\sqrt{\hat{\sigma}_1}\right] 
    \\
    &=\mu_n\Tr \left[\hat{\sigma}_1\log\hat{\sigma}_1\right] - \mu_n\Tr \left[\hat{\Pi}_0\sqrt{\hat{\sigma}_1}\log\hat{\tilde{\sigma}}\sqrt{\hat{\sigma}_1}\hat{\Pi}_0\right] 
    \nonumber \\
    &\phantom{=}- \mu_n\Tr \left[(\hat{I}-\hat{\Pi}_0)\sqrt{\hat{\sigma}_1}\log\hat{\tilde{\sigma}}\sqrt{\hat{\sigma}_1}(\hat{I}-\hat{\Pi}_0)\right].
\end{align}
Since $\hat{\tilde{\sigma}} = (1-\mu_n)\hat{\sigma}_0+\mu_n\hat{\sigma}_1 \succeq \mu_n\hat{\sigma}_1$ and logarithm is operator monotone, we have 
\begin{align}
    \label{eq:converse_nsqrtn_revised}\mu_n D(\hat{\sigma}_1 \| \hat{\tilde{\sigma}} ) &\leq  \mu_n\Tr \left[\hat{\sigma}_1\log\hat{\sigma}_1\right] - \mu_n\Tr \left[\hat{\Pi}_0\sqrt{\hat{\sigma}_1}\log\hat{\tilde{\sigma}}\sqrt{\hat{\sigma}_1}\hat{\Pi}_0\right] 
    \nonumber \\
    &\phantom{\leq}- \mu_n\Tr \left[(\hat{I}-\hat{\Pi}_0)\sqrt{\hat{\sigma}_1}\log(\mu_n\hat{{\sigma}}_1)\sqrt{\hat{\sigma}_1}(\hat{I}-\hat{\Pi}_0)\right] \\
    &=  \mu_n\Tr \left[\hat{\sigma}_1\log\hat{\sigma}_1\right] - \mu_n\Tr \left[\hat{\Pi}_0\sqrt{\hat{\sigma}_1}\log\hat{\tilde{\sigma}}\sqrt{\hat{\sigma}_1}\hat{\Pi}_0\right] 
    \nonumber \\
    &\phantom{\leq}- \mu_n\Tr \left[(\hat{I}-\hat{\Pi}_0){\hat{\sigma}_1}\log(\mu_n\hat{{\sigma}}_1)\right]\label{eq:sigma1commutes}\\
    &=  \mu_n\Tr \left[\hat{\Pi}_0\hat{\sigma}_1\log\hat{\sigma}_1\right] - \mu_n\Tr \left[\hat{\Pi}_0\sqrt{\hat{\sigma}_1}\log\hat{\tilde{\sigma}}\sqrt{\hat{\sigma}_1}\hat{\Pi}_0\right] 
    \nonumber \\
    &\phantom{\leq}+ \mu_n\Tr \left[(\hat{I}-\hat{\Pi}_0){\hat{\sigma}_1}\left(\log\hat{\sigma}_1-\log(\mu_n\hat{{\sigma}}_1)\right)\right]\\
    &=\mu_n\Tr \left[\hat{\Pi}_0\hat{\sigma}_1\log\hat{\sigma}_1\right] - \mu_n\Tr \left[\hat{\Pi}_0\sqrt{\hat{\sigma}_1}\log\hat{\tilde{\sigma}}\sqrt{\hat{\sigma}_1}\hat{\Pi}_0\right] 
    \nonumber \\
    &\phantom{\leq}-\mu_n\Tr \left[(\hat{I}-\hat{\Pi}_0){\hat{\sigma}_1}\right]\log\mu_n \\
    &= \mu_n\Tr \left[\hat{\Pi}_0\hat{\sigma}_1\log\hat{\sigma}_1\right] - \mu_n\Tr \left[\hat{\Pi}_0\sqrt{\hat{\sigma}_1}\log\hat{\tilde{\sigma}}\sqrt{\hat{\sigma}_1}\hat{\Pi}_0\right]+\mu_n\kappa\log\frac{1}{\mu_n},\\
    &\leq \mu_n\Tr \left[\hat{\Pi}_0\hat{\sigma}_1\log\hat{\sigma}_1\right] - \mu_n\Tr \left[\hat{\Pi}_0\sqrt{\hat{\sigma}_1}\log((1-\mu_n)\hat{\sigma}_0)\sqrt{\hat{\sigma}_1}\hat{\Pi}_0\right]+\mu_n\kappa\log\frac{1}{\mu_n}\label{eq:logoperatormonotone}\\
    &=\mu_n\Tr \left[\hat{\Pi}_0\left(\hat{\sigma}_1\log\hat{\sigma}_1-\sqrt{\hat{\sigma}_1}\log\hat{\sigma}_0\sqrt{\hat{\sigma}_1}\right)\right]+\mu_n\kappa\log\frac{1}{\mu_n}\label{eq:logterm2expand}
\end{align}
where \eqref{eq:sigma1commutes} is because $\hat{\sigma}_1$ commutes with $\log\hat{\sigma}_1$ and \eqref{eq:logoperatormonotone} is because $\hat{\tilde{\sigma}} = (1-\mu_n)\hat{\sigma}_0+\mu_n\hat{\sigma}_1 \succeq (1-\mu_n)\hat{\sigma}_0$ and logarithm is operator monotone.
Applying a similar strategy to the first term of \eqref{eq:sqrtnStart} yields: 
\begin{align}
    (1 - \mu_n) D(\hat{\sigma}_0 \| \hat{\tilde{\sigma}} ) &= (1-\mu_n)\Tr\left[\hat{\sigma}_0\left(\log\hat{\sigma}_0-\log\hat{\tilde{\sigma}}\right)\right] 
    \\
    &\leq (1-\mu_n) \Tr\left[\hat{\sigma}_0\left(\log\hat{\sigma}_0-\log\left((1-\mu_n)\hat{\sigma}_0\right)\right)\right] \label{eq:logoperatermonotone2}\\
    &= -(1-\mu_n)\log(1-\mu_n)
\label{eq:logterm1expand}
\end{align}
Substituting \eqref{eq:logterm2expand} and \eqref{eq:logterm1expand} into \eqref{eq:sqrtnStart} yields:
\begin{align}
    \chi \left( \left\{ p_{\tilde{X}}, \hat{\sigma}_x \right\} \right) &\leq \log\frac{1}{1-\mu_n} + \mu_n\Tr \left[\hat{\Pi}_0\left(\hat{\sigma}_1\log\hat{\sigma}_1-\sqrt{\hat{\sigma}_1}\log\hat{\sigma}_0\sqrt{\hat{\sigma}_1}\right)\right]+\mu_n\kappa\log\frac{1}{\mu_n}\label{eq:sqrtnBobUpperBound}.
\end{align}

Dividing both sides of \eqref{eq:bobDeltaFrac} by $\sqrt{n D ( \rhoBarN \| \rhoNotN )}\log n$ and applying Lemma \ref{lem:etaDefn} with $\alpha = \mu_n$ yields:
\begin{align}
\frac{\log M}{\sqrt{n D ( \rhoBarN \| \rhoNotN )} \log n} &\leq \frac{n \chi (p_{\tilde{X}}(x), \hat{\sigma}_x ) + 1 }{(1-\epsilon_n) \sqrt{\frac{n^2\mu_n^2}{2}\eta(\hat{\rho}_1 \| \hat{\rho}_0)+n^2 R(\mu) } \log n} \\
\label{eq:applysqrtnBobUpperBound}&\leq \frac{\log\frac{1}{1-\mu_n} + \mu_n\Tr \left[\hat{\Pi}_0\left(\hat{\sigma}_1\log\hat{\sigma}_1-\sqrt{\hat{\sigma}_1}\log\hat{\sigma}_0\sqrt{\hat{\sigma}_1}\right)\right]+\mu_n\kappa\log\frac{1}{\mu_n} + \frac{1}{n} }{(1-\epsilon_n)\sqrt{\frac{\mu_n^2}{2}\eta(\hat{\rho}_1 \| \hat{\rho}_0) + R(\mu)} \log n} \\
&= \frac{\frac{-\log(1-\mu_n)}{\mu_n \log n} + \frac{\Tr \left[\hat{\Pi}_0\left(\hat{\sigma}_1\log\hat{\sigma}_1-\sqrt{\hat{\sigma}_1}\log\hat{\sigma}_0\sqrt{\hat{\sigma}_1}\right)\right]}{\log n}  + \frac{\kappa  \log \left(\frac{1}{\mu_n}\right)}{\log n} + \frac{1}{n \mu_n \log n} }{(1-\epsilon_n)\sqrt{\frac{1}{2}\eta(\hat{\rho}_1 \| \hat{\rho}_0) + R(\mu)} }, \label{eq:BobSqrtNBound}
\end{align}
where \eqref{eq:applysqrtnBobUpperBound} is because of \eqref{eq:sqrtnBobUpperBound}.
By \eqref{eq:muNcondition}, $\mu_n = \frac{\varpi_n}{\sqrt{n}}$, where $\varpi_n = o(1)$. 
Combining \eqref{eq:bobDeltaFrac} and \eqref{eq:sqrtnBobUpperBound} yields:
\begin{align}
n &\left( \log\frac{1}{1-\mu_n} + \mu_n\Tr \left[\hat{\Pi}_0\left(\hat{\sigma}_1\log\hat{\sigma}_1-\sqrt{\hat{\sigma}_1}\log\hat{\sigma}_0\sqrt{\hat{\sigma}_1}\right)\right]  + \mu_n \kappa \log\left(\frac{1}{\mu_n}\right) \right) \nonumber\\&\phantom{\log\frac{1}{1-\mu_n} + \mu_n\Tr \left[\hat{\Pi}_0\left(\hat{\sigma}_1\log\hat{\sigma}_1-\sqrt{\hat{\sigma}_1}\log\hat{\sigma}_0\sqrt{\hat{\sigma}_1}\right)\right]+===}\geq n \chi \left( \left\{ p_{\tilde{X}}, \hat{\sigma}_x \right\} \right) \label{eq:lhsasymptoticdom}\\
&\phantom{\log\frac{1}{1-\mu_n} + \mu_n\Tr \left[\hat{\Pi}_0\left(\hat{\sigma}_1\log\hat{\sigma}_1-\sqrt{\hat{\sigma}_1}\log\hat{\sigma}_0\sqrt{\hat{\sigma}_1}\right)\right]+===}\geq (1-\epsilon_n)\log M -1. \label{eq:doubleBoundSqrtN}
\end{align}
The term $\mu_n \kappa \log\left(\frac{1}{\mu_n}\right)$ is the asymptotically dominant term on the left-hand side of \eqref{eq:lhsasymptoticdom}. Thus, in order to have $\lim_{n \to \infty}\log M = \infty$,
\begin{align}
\lim_{n\to \infty} n \mu_n \log\left(\frac{1}{ \mu_n}\right) = \lim_{n\to\infty} \sqrt{n} \varpi_n \left( \frac{1}{2} \log n + \log \varpi_n^{-1} \right) = \infty,\label{eq:varpirequirement}
\end{align}
which requires that $\varpi_n \in \omega \left(\frac{1}{\sqrt{n} \log n} \right)$. Hence, we have $\varpi_n \in o(1) \cap \omega \left(\frac{1}{\sqrt{n} \log n} \right)$.
This, along with reliability condition $\epsilon_n\to 0$, and $R(\mu)\in\mathcal{O}(\mu^3)$ allows evaluation of the limit of both sides of \eqref{eq:BobSqrtNBound} as $n\to\infty$ to yield \eqref{eq:Onsqrtn_converse} and the theorem.
\end{IEEEproof}

Note that just as the corresponding results in \cite[Cor.~4 and Th.~8]{bloch15covert}, the scaling constants $\xi_1$ and $\xi_2$ in Corollary \ref{cor:scalingConstM} and Theorem \ref{thm:Onsqrtn_converse} depend on the choice of $\gamma_n$ and $\varpi_n$. This is unlike Corollary \ref{cor:scalingConstants} and Theorem \ref{thm:converse} where the scaling constant remains the same for \emph{all} choices of $\gamma_n$ and scaling of $\mu_n$ meeting the condition in \eqref{eq:muNcondition}.
However, unlike in \cite[Cor.~4 and Th.~8]{bloch15covert}, our $\xi_1$ and $\xi_2$ do not match.
For $\gamma_n$ defined in \eqref{eq:gamman}, $\gamma_n^{-1}\in o\left(\frac{n^{\frac{1}{6}}}{\left(\log n\right)^{\frac{2}{3}}}\right)$.
Thus, for $n$ large enough, 
\begin{align}
\frac{\log \gamma_n^{-1}}{\log n}&\leq \frac{\log\left(\frac{n^{\frac{1}{6}}}{\left(\log n\right)^{\frac{2}{3}}}\right)}{\log n}\leq\frac{1}{6},\label{eq:rev2comment6}
\end{align}
which implies that $0<\xi_1\leq\frac{1}{6}$. 
On the other hand, for $\varpi_n \in \omega \left( \frac{1}{\sqrt{n}\log n} \right)$ and $n$ large enough,
\begin{align}
\frac{\log \varpi_n^{-1}}{\log n} \leq \frac{\log(\sqrt{n} \log n)}{\log n} = \frac{1}{2} + \frac{\log\log n}{\log n},
\end{align}
so that $0<\xi_2 \leq \frac{1}{2}$. 
We note that the gap between $\xi_1$ and $\xi_2$ enters solely through the constraint on $\gamma_n$ in the proof of Lemma \ref{lem:trace-expected}. 
This gap also affects the results on entanglement-assisted covert communication over qubit depolarizing channel in \cite{zlotnick23eacovcommdepol}.
This can be addressed by evolving the channel resolvability analysis to reduce the restrictions on $\gamma_n$.
Although we defer this to future work, closing the gap between $\xi_1$ and $\xi_2$ would imply that a product measurement is optimal in the setting of Theorems \ref{thm:Onsqrtn_achievability} and \ref{thm:Onsqrtn_converse}.
This is unlike the setting of Theorems \ref{thm:achievability} and \ref{thm:converse}, where achieving the covert capacity likely requires Bob to employ joint-detection receiver on his $n$ received states.

\subsection{No covert communication} 
\label{subsec:nocovcomms}
We specialize the channel model from Section \ref{sec:channel_model} to show that covert communication is impossible when \emph{all} of Alice's non-innocent inputs result in Willie's output states having support outside of innocent state's support. 
Consider a general quantum channel from Alice to Willie $\mathcal{N}_{A^n \to W^n}$ that may not be memoryless across $n$ channel uses.
Denote by $\hat{\phi}^n\in\mathcal{D}\left(\mathcal{H}_A^{\otimes n}\right)$ Alice's $n$-channel-use input state, and by $\hat{\rho}^n=\mathcal{N}_{A^n \to W^n}\left(\hat{\phi}^n\right)\in\mathcal{D}\left(\mathcal{H}_W^{\otimes n}\right)$ the corresponding Willie's output.
$\mathcal{H}_A$ and $\mathcal{H}_W$ can be infinite-dimensional.
We designate a pure state $\hat{\phi}_0=\ket{0}\bra{0}$ as Alice's innocent input (noting that any mixed state can be purified by adding dimensions to $\mathcal{H}_A$).
The $n$-channel-use innocent input state is $\hat{\phi}_0^{\otimes n}=\ket{\mathbf{0}}\bra{\mathbf{0}}$, where $\ket{\mathbf{0}} = \ket{0} \otimes \ket{0} \otimes \cdots \otimes \ket{0}$, with the corresponding output state $\hat{\rho}_0^n=\mathcal{N}_{A^n \to W^n}\left(\hat{\phi}_0^{\otimes n}\right)$.
We now generalize \cite[Th.~1]{bash15covertbosoniccomm} as follows.

\begin{thm}
\label{thm:nocovcomms} Suppose that, for any of Alice's non-innocent input state $\hat{\phi}^n\neq\hat{\phi}^{\otimes n}_0$, the corresponding output state at Willie $\hat{\rho}^n=\mathcal{N}_{A^n \to W^n}\left(\hat{\phi}^n\right)$ has support $\supp\left(\hat{\rho}^n\right)\not\subseteq \supp\left(\hat{\rho}^n_0\right)$.  Then, there exists a constant $\delta_0>0$, such that maintaining Willie's detection error probability $P_{\rm e}^{(w)}\geq\frac{1}{2}-\delta$ for $\delta\in (0,\delta_0)$ results in Bob's decoding error probability $P_{\rm e}^{(b)}\geq\frac{1}{4}\left(1-\sqrt{\frac{\delta}{\delta_0}}\right)$ for any number of channel uses $n$.
\end{thm}

\begin{IEEEproof}
Alice sends one of $M$ (equally likely) $\log M$-bit messages by choosing an element from an arbitrary codebook $\{ \hat{\phi}^n_m, m=1, \ldots, M\}$, where a general pure state $\hat{\phi}^n_m = \ket{\psi^n_m} \bra{\psi^n_m}$ encodes a $\log M$-bit message $m$ in $n$ channel uses and $\ket{\psi^n_m} \in \mathcal{H}_A^{\otimes n}$.  
We limit our analysis to pure input states since, by convexity, using mixed states as inputs can only degrade the performance (since that is equivalent to transmitting a randomly chosen pure state from an ensemble and discarding the knowledge of that choice).
Denoting the set of non-negative integers by $\mathbb{N}_0$, we can construct a complete orthonormal basis (CON) $\mathcal{B}(\mathcal{H}_A) = \{\ket{b}, b\in \mathbb{N}_0 \}$ of $\mathcal{H}_A$ such that, for the innocent state $\hat{\phi}_0=\ket{0}\bra{0}$, $\ket{0}\in \mathcal{B}(\mathcal{H}_A)$.
We can then express $\ket{\psi^n_m} = \sum_{\mathbf{b} \in \mathbb{N}^n_0} a_{\mathbf{b}}(m) \ket{\mathbf{b}}$, where $\ket{\mathbf{b}} \equiv \ket{b_1} \otimes \ket{b_2} \otimes \cdots \otimes \ket{b_n}$ with $\ket{b_k}\in\mathcal{B}(\mathcal{H}_A)$. 

When $m$ is transmitted, Willie's hypothesis test reduces to discriminating between the states $\hat{\rho}_0^{n}$ and $\hat{\rho}^{n}_m$,
where $\hat{\rho}^{n}_m=\mathcal{N}_{A^n\rightarrow W^n}\left(\hat{\phi}^{n}_m\right)$.
Let Willie use a detector that is given by the POVM $\left\{\hat{\Pi}^n_0,\Hat{I}-\hat{\Pi}^n_0\right\}$, where $\hat{\Pi}^n_0$ is the projection onto the support of the innocent state $\hat{\rho}_0^{n}$.
Thus, Willie's average error probability is:
\begin{align}
\label{eq:pewnogo}P_{\rm e}^{(w)} &=\frac{1}{2M}\sum_{m=1}^{M}\Tr\left[\hat{\Pi}^n_0\hat{\rho}^{n}_m\right],
\end{align}
since messages are sent equiprobably. Note that the error is entirely because of missed codeword detections, as Willie's receiver never raises a false alarm because the completely innocent state is an eigenstate of $\hat{\Pi}^n_0$. By assumption that $\supp\left(\hat{\rho}_m^n\right)\not\subseteq \supp\left(\hat{\rho}^n_0\right)$, we have that for some $b_m>0$
\begin{align}
    P_{\rm e}^{(w)} &=\frac{1}{2M}\sum_{m=1}^{M}\Tr\left[\hat{\Pi}^n_0\hat{\rho}^{n}_m\right]\\
    &\leq \frac{1}{2}-\frac{1}{2M}\sum_{m=1}^{M}b_m\\
    &\leq\frac{1}{2}-\frac{c_{\min}}{2}\left(1-\frac{1}{M}\sum_{m=1}^{M}\left|a_{\mathbf{0}}(m)\right|^2\right),
\end{align}
where $c_{\min}=\min_{m,m>0} \frac{b_m}{1-\left|a_{\mathbf{0}}(m)\right|^2}$, and note that $c_{\min}>0$.
Thus, to ensure $P_{\rm e}^{(w)} \geq\frac{1}{2}-\delta$, Alice must use a codebook such that:
\begin{align}
\label{eq:restrict_a0}\frac{1}{M}\sum_{m=1}^{M}\left|a_{\mathbf{0}}(m)\right|^2&\geq1-\frac{2\delta}{c_{\min}}.
\end{align}
We can restate \eqref{eq:restrict_a0} as follows:
\begin{align}
\label{eq:restrict_not_a0}\frac{1}{M}\sum_{m=1}^{M}\left(1-\left|a_{\mathbf{0}}(m)\right|^2\right)&\leq\frac{2\delta}{c_{\min}}.
\end{align}Bob's decoding error analysis follows from the corresponding analysis in the proof of \cite[Th. 1]{bash15covertbosoniccomm}, starting from \cite[Supp.~Note~3,~Eq.~(27)]{bash15covertbosoniccomm}, with minor substitutions: $m\to u$, $M\to 2^M$, $c_{\min}\to\eta_{\rm w}$, $\delta\to\epsilon$, $\hat{\phi}^n_m\to\hat{\rho}^{A^n}_u$ and by replacing the unity-transmissivity pure-loss channel from Alice to Bob discussed above \cite[Supp.~Note~3,~Eq.~(32)]{bash15covertbosoniccomm} with a trivial identity channel. The main proof idea is to show that  any code construction satisfying \eqref{eq:restrict_not_a0} have non-zero probability of decoding error even when Alice and Bob can use the trivial identity channel. This choice of channel allows us to lower bound Bob's probability of error using \cite[Eq.~(9.172)]{wilde16quantumit2ed} and the pure state definition $\ket{\psi^n_m} = \sum_{\mathbf{b} \in \mathbb{N}^n_0} a_{\mathbf{b}}(m) \ket{\mathbf{b}}$, giving us a lower bound in terms of $|a_{\mathbf{0}}(m)|^2$.  This, together with \eqref{eq:restrict_not_a0}, yields the desired result.
\end{IEEEproof}

\section{Discussion}\label{sec:discussion}

\subsection{Non-vanishing $D\left(\hat{\bar{\rho}}^n\middle\|\hat{\rho}_0^{\otimes n}\right)$}

Let's relax our covertness constraint and only require that
\begin{align}
\lim_{n\to\infty}D\left(\hat{\bar{\rho}}^n\middle\|\hat{\rho}_0^{\otimes n}\right)=\delta,\label{eq:non-vanishing}
\end{align}for some constant $\delta>0$. Adapting the results of Theorems \ref{thm:achievability} and \ref{thm:converse} amounts to selecting a sequence of $\gamma_n$ such that $\lim_{n\to\infty}\gamma_n =\gamma_0$ for some $\gamma_0>0$ satisfying \eqref{eq:non-vanishing}. We arrive at the same square root law scaling as previously, specifically, $\log M \in \Theta(\sqrt{n})$ and $\log K \in \Theta(\sqrt{n})$. The special cases follow from a similar alteration to the sequence $\gamma_n$. When Willie's innocent output state is a mixture of non-innocent output states, as in Section \ref{subsec:constantrate}, $\log M \in \Theta(n)$. When at least one non-innocent input lies outside Bob's innocent-state support while remaining inside Willie's innocent-state support, as in Section \ref{subsec:sqrtnlogn}, $\log M \in \Theta(\sqrt{n}\log n)$.

\subsection{Bob restricted to product measurement}

Now consider the practical scenario where Bob is restricted to a specific symbol-by-symbol measurement described by a POVM $\left\{\hat{\Pi}_y\right\}$ while Willie remains unconstrained in his measurement choice. This induces a classical channel from Alice to Bob with transition probability
\begin{align}
    p_{Y|X}(y|x)=\Tr \left[\hat{\sigma}_x\hat{\Pi}_y\right], \label{eq:classicalpmf}
\end{align}
where $Y$ is the random variable corresponding to Bob's measurement outcome defined over output space $\mathcal{Y}$, and $X$ is the random variable corresponding to Alice's input to the channel. Such restriction on Bob can only reduce the information throughput between Alice and Bob. It is equivalent to giving Bob classical output states of the form 
\begin{align}
    \hat{\sigma}_x^{(\mathrm{c})} \triangleq \sum_{y\in\mathcal{Y}}p_{Y|X}(y|x)\ket{y}\bra{y}, \label{eq:commutingoutputstates}
\end{align}
that commute for each $x \in \mathcal{X}$. 
Let $P_x$ be the probability distribution of Bob's measurement outcome conditioned on Alice's input $x\in\mathcal{X}$. The following corollary demonstrates that the SRL holds for this scenario.
\begin{cor}\label{cor:prodmeas}
Consider a covert communication scenario in which the Alice-to-Bob channel is a classical DMC with $P_x$ absolutely continuous with respect to $P_0$ $\forall x\in \mathcal{X}$, and the Alice-to-Willie channel is memoryless and classical-quantum, where the output state $\hat{\rho}_0$ corresponding to innocent input $x=0$ is not a mixture of non-innocent ones $\{\hat{\rho}_x\}_{x\in\mathcal{X}\setminus\{0\}}$ and $\forall x\in\mathcal{X}$, $\supp (\hat{\rho}_x) \subseteq \supp(\hat{\rho}_0)$. Then, there exists a sequence of covert communication codes such that 
\begin{align}
\lim_{n\to\infty}D(\hat{\bar{\rho}}^n\|\hat{\rho}_0^{\otimes n})=0, \lim_{n\to\infty} P_\mathrm{e}^{(b)}=0,
\end{align}
with covert capacity and pre-shared secret key requirement:
\begin{align}
    \lim_{n\to \infty} \frac{\log M}{\sqrt{n D(\hat{\bar{\rho}}^n\|\hat{\rho}_0^{\otimes n}})} &= \frac{\sum_{x\in\mathcal{X}\setminus\{0\}}\pi_x D(P_x\|P_0)}{\sqrt{\frac{1}{2}\eta(\hat{\rho}_{\neg 0}\|\hat{\rho}_0})}\\
    \lim_{n\to \infty} \frac{\log K}{\sqrt{n D(\hat{\bar{\rho}}^n\|\hat{\rho}_0^{\otimes n}})}&= \frac{\left[\sum_{x\in\mathcal{X}\setminus\{0\}}\pi_x \left(D(\hat{\rho}_x\|\hat{\rho}_0)-D(P_x\|P_0)\right)\right]^+}{\sqrt{\frac{1}{2}\eta(\hat{\rho}_{\neg 0}\|\hat{\rho}_0})},
\end{align}
where  Willie's average state $\hat{\bar{\rho}}^n$ is defined in \eqref{eq:W_transmissionstate}.
\end{cor}
\begin{IEEEproof}[Proof (Corollary \ref{cor:prodmeas})]
This is a special case of Theorems \ref{thm:achievability} and \ref{thm:converse}, where Bob's output states are defined in \eqref{eq:commutingoutputstates}. Thus, the corollary immediately follows after noting that 
\begin{align}
    D\left(\hat{\sigma}_x^{(\mathrm{c})}\|\hat{\sigma}_0^{(\mathrm{c})}\right) = D(P_x\|P_0),
\end{align}
for each $x\in\mathcal{X}$.
\end{IEEEproof}

\subsection{Future Work}
In this paper, we proved the SRL for classical-quantum channels and found the exact expressions for the covert capacity and pre-shared secret key requirement. However, the second-order asymptotic results for covert communication over classical-quantum channels are yet to be derived. These have been characterized for bosonic channels \cite{gagatsos20codingcovcomm} by using position-based coding \cite{oskouei18unionbound}. Adapting our code construction in Section \ref{sec:achievability} to use position-based coding allows for message rate bounds to be derived via perturbation theory \cite{daleckii1965,grace22perturbationtheory}, though this does not admit a tractable method for bounding secret key size while maintaining relative entropy as the covertness metric. To solve this problem, one can use trace distance as covertness metric instead. This allows a more straight-forward use of convex splitting \cite{anshu17convexsplit}, as was done recently in \cite{wang22tracedistance} for the bosonic channel. Furthermore, this has an important benefit of bounding Willie's probability of error exactly, per \eqref{eq:error}. However, adapting the converse argument in Section \ref{sec:converse} to a trace-distance metric presents a slew of technical challenges that are the subject of our future work. The authors of \cite{wang22tracedistance} report similar challenges for adapting the converse in the covert bosonic channel setting. 

\appendices

\section{} \label{ap:pinchingMapConvex}
\begin{IEEEproof}[Proof (Lemma \ref{lem:pinchingMapConvex})]
Beginning with the left-hand side of \eqref{eq:pinchingMapConvex}, we have 
\begin{align}
    \Tr\left[\left(\mathcal{E}_{\hat{A}}(\hat{B})\right)^n\hat C\right] &= \Tr\left[\left(\prod_{i=1}^n\sum_{k_i}\label{eq:pMCDefs}\hat     P_{k_i}    \hat{B}\hat P_{k_i}\right)\sum_l \gamma_l \hat P_l\right]\\
    &=\Tr\left[\left(\sum_{k_1,\ldots,k_n}\prod_{i=1}^n\hat  P_{k_i} \hat{B}\hat P_{k_i}\right)\sum_l \gamma_l \hat P_l\right]\\ 
    \label{eq:pMCProjs1}
    &=\Tr\left[\sum_k \left(\hat P_k \hat{B}\right)^n\hat P_k \sum_l \gamma_l \hat P_l\right]\\ 
    \label{eq:pMCProjs2}
    &= \Tr\left[\sum_k \left(\hat P_k \hat{B}\right)^n\gamma_k\hat P_k\right]\\
    &=\Tr\left[\sum_k \left(\hat P_k \gamma_k^{1/n}\hat{B}\right)^n\hat P_k\right]\label{eq:pMCpause},
\end{align}
where \eqref{eq:pMCDefs} is from the definitions of the pinching map $\mathcal{E}_{\hat A}(\hat B)=\sum_i \hat P_i \hat B \hat P_i$ and the operator $\hat C = \sum_i \gamma_i \hat P_i$,  \eqref{eq:pMCProjs1} and \eqref{eq:pMCProjs2} follow from the projectors $\{\hat{P}_i\}$ being mutually orthogonal, and \eqref{eq:pMCpause} follows from the fact that $\gamma_k$ are non-negative. Continuing from \eqref{eq:pMCpause},
\begin{align}
   \Tr\left[\left(\mathcal{E}_{\hat{A}}(\hat{B})\right)^n\hat C\right] &= \Tr\left[\sum_k \left(\hat P_k \gamma_k^{1/n}\hat{B}\right)^n\hat P_k\right] \\
   &=\Tr\left[\sum_{k_1,\ldots,k_n}\prod_{i=1}^n\hat  P_{k_i} \gamma_{k_i}^{1/n}\hat{B}\hat P_{k_i}\right]\\
   &= \Tr\left[\prod_{i=1}^n\sum_{k_i}\hat     P_{k_i} \gamma_{k_i}^{1/n} \hat{B}\hat P_{k_i}\right]\\
   &= \Tr\left[\left(\sum_k \hat P_k \gamma_k^{1/n} \hat B \hat P_k\right)^n\right]\\
   &\leq \Tr\left[\sum_k \hat P_k \left(\gamma_k^{1/n} \hat B \right)^n\hat P_k\right]\label{eq:pMCTraceIneq}\\
   &=\Tr\left[\sum_k\hat P_k  \hat B^n\hat P_k\sum_l \gamma_l \hat P_l\right] \\
   &= \Tr\left[\mathcal{E}_{\hat A}(\hat B^n) \hat C\right]
\end{align}
where \eqref{eq:pMCTraceIneq} is Jensen's trace inequality \cite[Th. 2.4]{hansen2003jensen} since scalar function $(\cdot)^n$ for natural numbers $n$ is convex on interval $[0,\infty)$ containing the spectra of $\gamma_k^{1/n}\hat B$ for all $k$. 
\end{IEEEproof}

\section{} \label{ap:phi_bound}

\begin{IEEEproof}[Proof (Lemma~\ref{lem:phi_bound})]
Let $ \phi(s,  p) \triangleq \log \big( \sum_{x\in\mathcal{X}}p_X(x)\left(\Tr\left[\hat{\rho}_x^{1-s}\hat{\rho}_{ p}^s\right]\right)\big)$, where $s \leq 0 $, \begin{align}
p_X(x)=\begin{cases} 1-p,& x=0 \\ pq_x, &x=1,\ldots,N\end{cases}
\end{align} for $p,q_x\in[0,1]$, $\sum_{x\in\mathcal{X}\setminus\{0\}}q_x=1$, and $\hat{\rho}_p =\sum_{x\in\mathcal{X}}p_{X}(x)\hat{\rho}_x= (1-p)\hat{\rho}_0+p\sum_{x\in\mathcal{X}\setminus\{0\}}q_x\hat{\rho}_x$.
Recall that $s_0\leq s\leq 0$ for an arbitrary constant $s_0<0$.
We first state the following claim, which we use to prove the lemma.
\begin{claim} \label{claim:smoothness} Function $\phi: ]-\infty, 0] \times [0, 1] \to \mathbb{R}$ is smooth, i.e.,  its partial derivatives of any order exist.
\end{claim}
We defer the proof of Claim \ref{claim:smoothness} to the end of this appendix. We use the notation 
\begin{align}
    D^{(i, j)} \phi \triangleq \frac{\partial^{i+j}\phi}{\partial^i s \partial^j p}
\end{align}
for non-negative integers $i$ and $j$.

Applying Taylor's theorem to $\phi(s, p)$ as a function of $s$ at $s=0$ for a fixed $p$, we obtain
\begin{align} \phi(s, p) = \phi(0, p) + D^{(1, 0)}\phi(0, p) s + \frac{1}{2} D^{(2, 0)}\phi(0, p)s^2 + D^{(3, 0)}\phi(\eta, p) s^3, 
\end{align}
for $\eta \in [s,0]$. Applying Taylor's theorem again to $D^{(2, 0)}\phi(0, p)$ as a function of $p$ at $ p = 0$ yields
\begin{align}
 D^{(2, 0)}\phi(0, p) &=D^{(2, 0)}\phi(0, 0)+D^{(2, 1)}\phi(0, \tau) p,
\end{align}
for $\tau \in [0, p]$. Note that 
\begin{align}
    \phi(0, p) &= D^{(2, 0)}\phi(0, 0) = 0\\
    D^{(1, 0)}\phi(0, p) &= -\chi\left(\left\{p_X(x),\hat{\rho}_x\right\}\right)
\end{align}
where $\chi\left(\left\{p_X(x),\hat{\rho}_x\right\}\right)$ is the Holevo information of the ensemble $\left\{p_X(x),\hat{\rho}_x\right\}$ defined in Section \ref{sec:lemmas}.
The Taylor expansion of $\phi(s, p)$ at $s=0$ can then be expressed as
\begin{align}
\phi(s, p) = -\chi\left(\left\{p_X(x),\hat{\rho}_x\right\}\right) s + \frac{1}{2}D^{(2, 1)}\phi(0, \tau) ps^2 + \frac{1}{6} D^{(3, 0)}\phi(\eta, p)s^3. 
\end{align}
We set 
\begin{align}
    \vartheta_1 &\triangleq \frac{1}{2} \max_{\tau' \in [0, 1]}\abs{D^{(2, 1)}\phi(0, \tau')}\\
    \vartheta_2 &\triangleq \frac{1}{6} \max_{\eta' \in [s_0, 0], p'\in [0, 1]}\abs{D^{(3, 0)}\phi(\eta', p')}\\
\end{align}
which are finite by the continuity of all derivatives of $\phi(s,p)$.
This implies that
\begin{align}
\label{eq:phiinequal}\phi(s, p) \leq -\chi\left(\left\{p_X(x),\hat{\rho}_x\right\}\right)s+\vartheta_1 ps^2- \vartheta_2 s^3. 
\end{align}By expanding the Holevo information, we also have
\begin{align} 
\label{eq:holevoexpanded}\chi\left(\left\{p_X(x),\hat{\rho}_x\right\}\right)& =  p\sum_{x\in\mathcal{X}\setminus\{0\}}q_x D(\hat{\rho}_x \| \hat{\rho}_0) -D(\hat{\rho}_p \| \hat{\rho}_0)\\
\label{eq:holevoupperbound}&\leq p\sum_{x\in\mathcal{X}\setminus\{0\}}q_x D(\hat{\rho}_x \| \hat{\rho}_0),
\end{align}
where \eqref{eq:holevoupperbound} is because $D(\hat{\rho}_p \| \hat{\rho}_0) \geq 0$.
Since $s<0$, using \eqref{eq:holevoupperbound} yields an upper bound for \eqref{eq:phiinequal}.
Substituting $\alpha_n$ for $p$ and $\pi_x$ for $q_x$ yields \eqref{eq:phi_bound} and the lemma.
\end{IEEEproof}

\begin{IEEEproof}[Proof of Claim \ref{claim:smoothness}]
    Define the functions
\begin{align}
A(s,  p) &\triangleq \left((1- p)(\hat{\rho}_0)^{1-s}+ p\sum_{x\in\mathcal{X}\setminus\{0\}}q_x(\hat{\rho}_x)^{1-s}\right)\left((1-p)\hat{\rho}_0+ p\sum_{x\in\mathcal{X}\setminus\{0\}}q_x\hat{\rho}_x\right)^s, \\
g(\hat{M}) &\triangleq  \Tr\left[\hat{M}\right], \\
\psi(x) &\triangleq \log(x).
\end{align}
$\phi(s, p)$ is a composition of these functions s.t. $\phi(s, p) = \psi \circ g \circ A (s, p)$. We use Taylor's theorem in order to find an upper bound on $\phi(s, p)$. To apply Taylor's theorem, we must first show $\phi(s, p)$ is smooth. We show the above functions are infinitely many times differentiable. The operators that compose $A(s, p)$ have the following $i$-th partial derivatives with respect to $s$:
\begin{align}
\frac{\partial^i}{\partial s^i} \left[(1- p)(\hat{\rho}_0)^{1-s}\right] &= (1- p)(\hat{\rho}_0)^{1-s}(-\log(\hat{\rho}_0))^i , \label{eq:E1} \\
\frac{\partial^i}{\partial s^i} \left[ p\sum_{x\in\mathcal{X}\setminus\{0\}}q_x(\hat{\rho}_x)^{1-s}\right] &=  p\sum_{x\in\mathcal{X}\setminus\{0\}}q_x(\hat{\rho}_x)^{1-s}(-\log(\hat{\rho}_x))^i , \label{eq:E2} \\ 
\frac{\partial^i}{\partial s^i} \left[ \left((1- p)\hat{\rho}_0+ p\sum_{x\in\mathcal{X}\setminus\{0\}}q_x\hat{\rho}_x\right)^s \right] = \frac{\partial^i}{\partial s^i} \left[ (\hat{\rho}_{ p})^s \right] &= (\hat{\rho}_{ p})^s(\log(\hat{\rho}_{ p}))^i . \label{eq:E3}
\end{align}
As the composition of $i$-th differentiable functions is also $i$-th differentiable, $A$ is $i$ times differentiable with respect to $s$. Taking the $j$-th partial derivative with respect to $ p$ of ($\ref{eq:E1}$) yields
\begin{align}
\frac{\partial^j}{\partial  p^j} \left[(1- p)(\hat{\rho}_0)^{1-s}(-\log(\hat{\rho}_0))^i \right] = \begin{cases}
 -(\hat{\rho}_0)^{1-s}(-\log(\hat{\rho}_0))^i, & j = 1\\
0, & j> 1.
\end{cases}
\end{align}
Similarly, taking the $j$-th partial derivative with respect to $p$ of ($\ref{eq:E2}$) yields
\begin{align}
\frac{\partial^j}{\partial  p^j} \left[p\sum_{x\in\mathcal{X}\setminus\{0\}}q_x(\hat{\rho}_x)^{1-s}(-\log(\hat{\rho}_x))^i \right] = \begin{cases}
\sum_{x\in\mathcal{X}\setminus\{0\}}q_x(\hat{\rho}_x)^{1-s}(-\log(\hat{\rho}_x))^i, & j = 1 \\
0, & j>1.
\end{cases}
\end{align}
Thus there exists a $j$-th partial derivative of ($\ref{eq:E1}$) and ($\ref{eq:E2}$) with respect to $ p$. To show ($\ref{eq:E3}$) is $j$-th differentiable with respect to $ p$, consider its components individually. First, the $j$-th derivative with respect to $p$ of its first component $(\hat{\rho}_p)^s$ can be calculated directly as
\begin{align}
\frac{\partial^j}{\partial  p^j} \left[(\hat{\rho}_{ p})^s \right] = s^j(\hat{\rho}_{ p})^{s-j}(\hat{\rho}_{\neg 0}-\hat{\rho}_0)^j. 
\end{align}
The derivative of the second component $\frac{\partial^j}{\partial  p^j} \Big[(\log(\hat{\rho}_{ p}))^i \Big]$ is itself a composition of functions that are $j$-th differentiable with respect to $ p$, and so must also be $j$-th differentiable with respect to $ p$. The above implies that $A(s,p)$ is also $j$ times differentiable with respect to $ p$.
$g(\hat{M})$ is a trace, which is $k$ times differentiable due to linearity. $\psi(x)$ is a smooth function. Therefore, $\phi(s, p)$ is differentiable by the chain rule.  
\end{IEEEproof}

\section{} \label{ap:lambda-min-alpha}

\begin{IEEEproof}[Proof (Lemma \ref{lem:lambda-min-alpha})]
    We have
    \begin{align}
        \lambda_{\min}(\rhoAlpha) 
        &= \min_{\ket{\phi}\in \supp(\rhoAlpha): \norm{\ket{\phi}} = 1} \bra{\phi} \rhoAlpha \ket{\phi}\\
        &\label{eq:supp-rho}\geq\min_{\ket{\phi}\in \supp(\rhoNot): \norm{\ket{\phi}} = 1} \bra{\phi} \rhoAlpha \ket{\phi}\\
        &\geq\min_{\ket{\phi}\in \supp(\rhoNot): \norm{\ket{\phi}} = 1} \bra{\phi} (1-\alpha_n)\rhoNot \ket{\phi}\\
        &= (1-\alpha_n) \lambda_{\min}(\rhoNot)
    \end{align}
    where \eqref{eq:supp-rho} follows since $\supp(\hat{\rho}_x) \subseteq \supp\pra{\rhoNot}$ for all $x\in \mathcal{X}\setminus \set{0}$.
\end{IEEEproof}
\section{}\label{ap:ECPebBoundDerivation}
Here, we adapt the methods of \cite{hayashi2003general} to obtain \eqref{eq:expectationIID}. For a given instance of the codebook, Bob's decoding error probability is:
\begin{align}
P_{\rm e}^{(b)} &= \frac{1}{KM}  \sum_{k=1}^K \sum_{m=1}^M \left(1 - \Tr \left[\hat{\sigma}^n\left(m,k\right) \hat{\Lambda}_{m,k}^n \right] \right)  \label{eq:bobErrorProb}	\\
&= \frac{1}{KM}  \sum_{k=1}^K \sum_{m=1}^M \Tr\left[\hat{\sigma}^n\left(m,k\right) \left(\hat{I} - \hat{\Lambda}_{m,k}^n \right)  \right], \label{eq:bobErrProb2}
\end{align}
where \eqref{eq:bobErrProb2} is because the trace of the density operator is unity and the trace is linear. Now,
\begin{align}
\hat{I} - \hat{\Lambda}_{m,k}^n 
&= \hat{I} - \left( \sum_{m'=1}^M \hat{\Pi}_{m',k}^n \right)^{-1/2} \hat{\Pi}_{m,k}^n \left( \sum_{m'=1}^M \hat{\Pi}_{m',k}^n \right)^{-1/2}\label{eq:defOfPOVM} \\
&\preceq  2(\hat{I} - \hat{\Pi}_{m,k}^n) + 4 \sum_{m'\neq m}^M \hat{\Pi}_{m',k}^n \label{eq:appOfLemHayashi},
\end{align}
where \eqref{eq:appOfLemHayashi} follows from application of Lemma \ref{lem:hayashinagaoka} with $\hat{A}=\hat{\Pi}_{m,k}^n$, $\hat{B} = \sum_{m' \neq m}^M \hat{\Pi}_{m',k}^n$, and $c = 1$. Since density operators have a Hermitian square root, and $\Tr\left[\hat{A}\hat{B}\hat{A}\right]\leq \Tr\left[\hat{A}\hat{C}\hat{A}\right]$ for Hermitian $\hat{A}$ and $\hat B \preceq \hat C$, \eqref{eq:bobErrProb2} can be upper bounded with \eqref{eq:appOfLemHayashi} as
\begin{align}
P_{\rm e}^{(b)} &\leq 
\frac{1}{KM} \sum_{k=1}^K \sum_{m=1}^M \left[2\Tr \left[ \hat{\sigma}^n\left(m,k\right) (\hat{I} - \hat{\Pi}_{m,k}^n)\right] + 4\sum_{m' \neq m}^M \Tr \left[\hat{\sigma}^n\left(m,k\right) \hat{\Pi}_{m',k}^n\right]\right], \label{eq:upperBoundLinearity}	
\end{align}
which follows from the linearity of the trace. We now upper bound the expectation of \eqref{eq:upperBoundLinearity} taken with respect to the codebook in \eqref{eq:codebook}. We have
\begin{align}
E_{\mathcal{C}} \left[ P_{\rm e}^{(b)} \right] &\leq 
\frac{1}{KM} \sum_{k,m}\left(2E_{\mathcal{C}}\left[\Tr \left[ \hat{\sigma}^n\left(m,k\right) (\hat{I} - \hat{\Pi}_{m,k}^n)\right]\right] +4E_\mathcal{C}\left[ \sum_{m' \neq m}^M \Tr \left[\hat{\sigma}^n\left(m,k\right) \hat{\Pi}_{m',k}^n\right]\right]\right) \label{eq:twoCBexpectations}
\end{align}
which is due to linearity of expectation.
Beginning with the first expectation in \eqref{eq:twoCBexpectations}, we have 
\begin{align}
    E_{\mathcal{C}}\left[\Tr \left[ \hat{\sigma}^n\left(m,k\right) (\hat{I} - \hat{\Pi}_{m,k}^n)\right]\right] &= \sum_{\mathbf{x}(m,k)\in\mathcal{X}^n}p_{\mathbf{X}(m,k)}\left(\mathbf{x}(m,k)\right)\left[\Tr \left[ \hat{\sigma}^n\left(\mathbf{x}(m,k)\right) (\hat{I} - \hat{\Pi}_{\mathbf{x}(m,k)}^n)\right]\right] \\
    &= \sum_{\mathbf{x}\in\mathcal{X}^n}p_{\mathbf{X}}\left(\mathbf{x}\right)\left[\Tr \left[ \hat{\sigma}^n\left(\mathbf{x}\right) (\hat{I} - \hat{\Pi}_{\mathbf{x}}^n)\right]\right] \label{eq:independentmk}.
\end{align}
where \eqref{eq:independentmk} is because $\mathcal{C}$ consists of i.i.d. random vectors independent of message-secret key pair $(m,k)$. Similarly, with the second expectation over the codebook in \eqref{eq:twoCBexpectations}, we have 
\begin{align}
    E_\mathcal{C}\left[ \sum_{m' \neq m}^M \Tr \left[\hat{\sigma}^n\left(m,k\right) \hat{\Pi}_{m',k}^n\right]\right] &= \sum_{\mathbf{x},\mathbf{x'}\in\mathcal{X}^n}\sum_{m' \neq m}^M p_{\mathbf{X}}\left(\mathbf{x}\right)p_{\mathbf{X'}}\left(\mathbf{x'}\right)\Tr \left[\hat{\sigma}^n\left(\mathbf{x}\right) \hat{\Pi}_{\mathbf{x'}}^n\right]\\
&=(M-1)\sum_{\mathbf{x},\mathbf{x'}\in\mathcal{X}^n}p_{\mathbf{X}}\left(\mathbf{x}\right)p_{\mathbf{X'}}\left(\mathbf{x'}\right)\Tr \left[\hat{\sigma}^n\left(\mathbf{x}\right) \hat{\Pi}_{\mathbf{x'}}^n\right].\label{eq:secondindependentmk}
\end{align}
Substituting \eqref{eq:independentmk} and \eqref{eq:secondindependentmk} into \eqref{eq:twoCBexpectations} yields \eqref{eq:expectationIID}.

\section{} \label{ap:lemc1}
\begin{IEEEproof}[Proof (Lemma \ref{lem:c1})]
    Let $C_1\triangleq\sum_{\mathbf{x}} p_{\mathbf{X}} (\mathbf{x}) \Tr \left[ \sigmaNx (\hat{I} - \hat{\Pi}_{\mathbf{x}}^n)\right]$ be the left hand side of \eqref{eq:C1simple}. 
Now consider a random variable $L$ indicating the number of non-innocent symbols in $\mathbf{X}$. We define a set $\mathcal{C}_{\nu_n^{(1)}}^n$ as in \cite[Eq.~(52)]{bloch15covert}:
\begin{align}
\mathcal{C}_{\nu_n^{(1)}}^n \triangleq \left\{l \in \mathbb{N} : l > \left(1-\sqrt{\nu_n^{(1)}}\right)\gamma_n \sqrt{n} \right\}, \label{eq:setNonInnocent}
\end{align}
where $\sqrt{\nu_n^{(1)}} \in o(1)\cap\omega\left(\frac{1}{\log(n)^{\frac{2}{3}}n^{1/6}}\right)$. Applying the law of iterated expectations using $L$ to \eqref{eq:sigmaAlphaNDef} yields:
\begin{align}
 C_1 & = \sum_{\mathbf{x}, l \in \mathcal{C}_{\nu_n^{(1)}}^n} p_{\mathbf{X} | L} (\mathbf{x} | l) \Tr \left[ \sigmaNx (\hat{I} - \hat{\Pi}_{\mathbf{x}}^n)\right] p_{L}(l) \nonumber \\
&\phantom{\leq} + \sum_{\mathbf{x}, l \notin \mathcal{C}_{\nu_n^{(1)}}^n} p_{\mathbf{X} | L} (\mathbf{x} | l) \Tr \left[ \sigmaNx (\hat{I} - \hat{\Pi}_{\mathbf{x}}^n)\right] p_{L}(l) \label{eq:lawOfExpec}
\end{align}
Since $\Tr \left[ \sigmaNx (\hat{I} - \hat{\Pi}_{\mathbf{x}}^n)\right] \leq 1$, \eqref{eq:lawOfExpec} is upper bounded as:
\begin{align}
C_1&\leq  \sum_{\mathbf{x}, l \in \mathcal{C}_{\nu_n^{(1)}}^n} p_{\mathbf{X} | L} (\mathbf{x} | l) \Tr \left[ \sigmaNx (\hat{I} - \hat{\Pi}_{\mathbf{x}}^n)\right] p_{L}(l) + \sum_{l \notin \mathcal{C}_{\nu_n^{(1)}}^n}p_{L}(l)\\
&\leq  \sum_{\mathbf{x}, l \in \mathcal{C}_{\nu_n^{(1)}}^n} p_{\mathbf{X} | L} (\mathbf{x} | l) \Tr \left[ \sigmaNx (\hat{I} - \hat{\Pi}_{\mathbf{x}}^n)\right] p_{L}(l) + e^{-\nu_n^{(1)}\gamma_n \sqrt{n}/2},\label{eq:chernoffBound}
\end{align}
where \eqref{eq:chernoffBound} follows from applying a Chernoff bound on $\sum_{l \notin \mathcal{C}_{\nu_n^{(1)}}^n} p_L (l)$ as in \cite[Eq.~(54)]{bloch15covert}.
Now, by Lemma \ref{lem:ogawaHayashi} for $\hat{\phi}^n = \sigmaNx$, $\hat{\tau}^n = \sigmaNotN$, $t=e^a$, and the definition of $\hat{\Pi}_{\mathbf{x}}^n$ in \eqref{eq:projX}, we have,
\begin{align}
\Tr \left[\sigmaNx  (\hat{I} - \hat{\Pi}_\mathbf{x}^n ) \right] &= \Tr \left[\sigmaNx \left\{ \mathcal{E}_{\sigmaNotN} \left( \sigmaNx \right) - e^a \sigmaNotN \preceq 0 \right\} \right] \label{eq:defnIMinPi} \\
&\leq (n+1)^{d_b} e^{ar} \Tr \left[ \sigmaNx (\sigmaNotN)^{r/2}(\sigmaNx)^{-r}(\sigmaNotN)^{r/2} \right], \label{eq:stmtOfLemHayashi}
\end{align}
where $d_b=\dim\left(\mathcal{H}_B\right)$ 
and $0\leq r\leq 1$. 
Substitution of \eqref{eq:stmtOfLemHayashi} in \eqref{eq:chernoffBound} yields: 
\begin{align}
 C_1 &\leq \sum_{\mathbf{x}, l \in \mathcal{C}_{\nu_n^{(1)}}^n} p_{\mathbf{X} | L} (\mathbf{x} | l) (n+1)^{d_b} e^{ar}  \Tr \left[ \sigmaNx (\sigmaNotN)^{r/2} (\sigmaNx)^{-r} (\sigmaNotN)^{r/2} \right]   p_L (l)   + e^{-\nu_n^{(1)}\gamma_n \sqrt{n}/2} \\
&= \sum_{\mathbf{x}, l \in \mathcal{C}_{\nu_n^{(1)}}^n} p_{\mathbf{X} | L} (\mathbf{x} | l) (n+1)^{d_b} \exp\left(ar + \sum_{k=1}^n \log \left[ \Tr \left[ \hat{\sigma}(x_k) \sigma_0^{r/2} (\hat{\sigma}(x_k))^{-r} \sigma_0^{r/2}   \right] \right] \right) p_L (l)  \nonumber \\
&\phantom{=} + e^{-\nu_n^{(1)}\gamma_n \sqrt{n}/2} ,\label{eq:memorylessChan}
\end{align}
where \eqref{eq:memorylessChan} follows from the channel being memoryless. Let us define the following function:
\begin{align}
\varphi (\hat{\sigma}(x_k),r) = -\log \left[ \Tr \left[ \hat{\sigma}(x_k) \sigma_0^{r/2} (\hat{\sigma}(x_k))^{-r} \sigma_0^{r/2} \right] \right]. \label{eq:reliabilityPhiFunc}
\end{align}
We also state the following claim, which is used later in the proof. 
\begin{claim}\label{claim:varphicont}
    $\frac{\varphi(\hat{\sigma}_x,r)}{r}$ is continuous with respect to $r\in[0,1]$ and $\lim_{r\to0}\frac{\varphi(\hat{\sigma}_x,r)}{r}= D(\hat{\sigma_x} \| \hat{\sigma}_0)$. 
\end{claim} 
We defer the proof of Claim \ref{claim:varphicont} to the end of this appendix.
Substitution of $\varphi (\hat{\sigma}(x_k),r)$ in \eqref{eq:memorylessChan} yields:
\begin{align}
  C_1  &\leq
  \sum_{\mathbf{x}, l \in \mathcal{C}_{\nu_n^{(1)}}^n} p_{\mathbf{X} | L} (\mathbf{x} | l) (n+1)^{d_b}\exp\left(ar - \sum_{k=1}^n \varphi (\hat{\sigma}(x_k),r) \right) p_L (l)  + e^{-\nu_n^{(1)}\gamma_n \sqrt{n}/2}\label{eq:insertVarphi} \\
&=  \sum_{\mathbf{x}, l \in \mathcal{C}_{\nu_n^{(1)}}^n} p_{\mathbf{X} | L} (\mathbf{x} | l) (n+1)^{d_b} \exp\left(ar - \sum_{k=1,x_k\neq0}^n\varphi\left(\hat{\sigma}(x_k),r\right) \right) p_L (l)  + e^{-\nu_n^{(1)}\gamma_n \sqrt{n}/2}
  \label{eq:nonZeroVarphi} \\
&\leq  (n+1)^{d_b} \exp\left(ar - \left(1-\sqrt{\nu_n^{(1)}}\right) \gamma_n \sqrt{n}\left(\sum_{x\in\mathcal{X}\setminus\{0\}}\pi_x\varphi\left(\hat{\sigma}_x,r\right)-\delta_n\right) \right)   + e^{-\nu_n^{(1)}\gamma_n \sqrt{n}/2}.
\label{eq:insertDefOfL}
\end{align}
where \eqref{eq:nonZeroVarphi} follows by noting that, since $\varphi(\hat{\sigma}_0,r) = 0$, only terms with $x_k \neq 0$ contribute to the inner summation in \eqref{eq:insertVarphi}. Now, \eqref{eq:insertDefOfL} follows from the number of non-innocent symbols in the inner sum $l>\left(1-\sqrt{\nu_n^{(1)}}\right)\gamma_n\sqrt{n}$ per definition of the set $\mathcal{C}_{\nu_n^{(1)}}^n$ in \eqref{eq:setNonInnocent}, the fact that $p_L(l)<1$, and noting that
\begin{multline}
\label{eq:sumnoninnocentHoeffding} P\left(\left|\frac{1}{\left(1-\sqrt{\nu_n^{(1)}}\right)\gamma_n\sqrt{n}}\sum_{k=1,x_k\neq0}^n\varphi\left(\hat{\sigma}(x_k),r\right) -E_{X|X\in\mathcal{X}\setminus\{0\}}\left[\varphi\left(\hat{\sigma}_X,r\right)\middle|X\in\mathcal{X}\setminus\{0\}\right]\right|\geq\delta_n\right) \\ \leq 2\exp\left(-c_{\delta}\delta_n^2\left(1-\sqrt{\nu_n^{(1)}}\right)\gamma_n\sqrt{n}\right)
\end{multline} 
by Hoeffding's inequality for constant $c_{\delta}=\frac{2}{\max_{x\in\mathcal{X}\setminus\{0\}}\varphi\left(\hat{\sigma}_x,r\right)}$ and $\delta_n\in o\left(\frac{1}{(\log n)^{2/3}n^{1/8}}\right) \cap \omega\left(\frac{1}{(\log n)^{2/3}n^{1/6}}\right)$. We have \eqref{eq:insertDefOfL} since the expected value conditioned on $x$ being non-innocent symbol 
\begin{align}
E_{X|X\in\mathcal{X}\setminus\{0\}}\left[\varphi\left(\hat{\sigma}_X,r\right)\middle|X\in\mathcal{X}\setminus\{0\}\right]&=\sum_{x\in\mathcal{X}\setminus\{0\}}\pi_x\varphi\left(\hat{\sigma}_x,r\right).
\end{align}
Note that $\delta_n$ vanishes in \eqref{eq:insertDefOfL} and the bound in \eqref{eq:sumnoninnocentHoeffding} approaches zero as $n\to\infty$.
By Claim \ref{claim:varphicont}, for $r_n \in o(1)\cap \omega\left(\frac{1}{n^{1/8}}\right)$ there exists a sequence  $\nu_n^{(2)} \in o(1)$ such that, for all $x\in\mathcal{X}\setminus\{0\}$,
\begin{align}
\abs{ \frac{\varphi (\hat{\sigma}_x,r_n) }{r_n} - D(\hat{\sigma}_x \| \hat{\sigma}_0) } &< \nu_n^{(2)} D(\hat{\sigma}_x \| \hat{\sigma}_0)\\
\Rightarrow\varphi(\hat{\sigma}_x,r_n) &> (1-\nu_n^{(2)})r_nD(\hat{\sigma}_x \| \hat{\sigma}_0). \label{eq:varphiBound}
\end{align}
Substituting \eqref{eq:varphiBound} in \eqref{eq:insertDefOfL} completes the proof.
\end{IEEEproof}

\begin{IEEEproof}[Proof of Claim \ref{claim:varphicont}]
    Here, we show that $\left.\frac{\partial}{\partial r}\varphi(\hat{\sigma}_x,r)\right|_{r=0}=D(\hat{\sigma}_x\|\hat{\sigma}_0)$, where $\varphi(\hat{\sigma}_x,r)$ is defined in \eqref{eq:reliabilityPhiFunc}.
First, note for operator $\hat{A}$ and scalars $r$ and $c$,
\begin{align}
\frac{\partial}{\partial r}\hat{A}^{cr}=\frac{\partial}{\partial r}e^{cr\log \hat{A}}=c(\log \hat{A}) \hat{A}^{cr}.
\end{align}
Now, consider the derivative of $\varphi(\hat{\sigma}_x,r)$:
\begin{align}
\frac{\partial}{\partial r}\varphi(\hat{\sigma}_x,r)
&= -\frac{\partial}{\partial r}\log \Tr\left[\hat{\sigma}_x\hat{\sigma}_0^{r/2}\hat{\sigma}_x^{-r}\hat{\sigma}_0^{r/2}\right]=-\frac{\frac{\partial}{\partial r} \Tr\left[\hat{\sigma}_x\hat{\sigma}_0^{r/2}\hat{\sigma}_x^{-r}\hat{\sigma}_0^{r/2}\right]}{\Tr\left[\hat{\sigma}_x\hat{\sigma}_0^{r/2}\hat{\sigma}_x^{-r}\hat{\sigma}_0^{r/2}\right]}.\label{eq:a101}
\end{align}
Trace and derivative in \eqref{eq:a101} can be interchanged since $\hat{\sigma}_x\hat{\sigma}_0^{r/2}\hat{\sigma}_x^{-r}\hat{\sigma}_0^{r/2}$ is finite-dimensional. We have,
\begin{align}
\frac{\partial}{\partial r}\hat{B}^{\frac{r}{2}}\hat{A}^{-r}\hat{B}^{\frac{r}{2}}
&\nonumber=\left( \frac{\partial}{\partial r}\hat{B}^{\frac{r}{2}}\right) \hat{A}^{-r}\hat{B}^{\frac{r}{2}}+\hat{B}^{\frac{r}{2}}\left( \frac{\partial}{\partial r}\hat{A}^{-r}\right) \hat{B}^{\frac{r}{2}}+\hat{B}^{\frac{r}{2}}\hat{A}^{-r}\left( \frac{\partial}{\partial r}\hat{B}^{\frac{r}{2}}\right) \\
&=\frac{1}{2}(\log \hat{B})\hat{B}^{\frac{r}{2}}\hat{A}^{-r}\hat{B}^{\frac{r}{2}}-\hat{B}^{\frac{r}{2}}(\log \hat{A})\hat{A}^{-r}B^{\frac{r}{2}}+\frac{1}{2}
\hat{B}^{\frac{r}{2}}\hat{A}^{-r}(\log \hat{B})\hat{B}^{\frac{r}{2}}\label{eq:varphiderivative}.
\end{align}
Applying \eqref{eq:varphiderivative} to \eqref{eq:a101} with $A=\hat{\sigma}_x$ and $B=\hat{\sigma}_0$ yields:
\begin{align}
&\frac{\partial}{\partial r}\varphi(\hat{\sigma}_x,r)=\frac{\Tr\left[\hat{\sigma}_x^{-r}\hat{\sigma}_0^{\frac{r}{2}}\hat{\sigma}_x\hat{\sigma}_0^{\frac{r}{2}}\log \hat{\sigma}_x
	-\frac{1}{2}\left( \hat{\sigma}_0^{\frac{r}{2}}\hat{\sigma}_x^{-r}\hat{\sigma}_0^{\frac{r}{2}}\hat{\sigma}_x
	+\hat{\sigma}_0^{\frac{r}{2}}\hat{\sigma}_x\hat{\sigma}_0^{\frac{r}{2}}\hat{\sigma}_x^{-r}\right)
	 \log \hat{\sigma}_0\right]}
{\Tr\left[\hat{\sigma}_x\hat{\sigma}_0^{r/2}\hat{\sigma}_x^{-r}\hat{\sigma}_0^{r/2}\right]},\label{eq:a107}
\end{align}
\normalsize
which is  continuous with respect to $r\in[0,1]$.
Setting $r=0$ in \eqref{eq:a107} yields the desired result.
\end{IEEEproof}

\section{} \label{ap:lemc2}

\begin{IEEEproof}[Proof (Lemma \ref{lem:c2})]
Let $C_2\triangleq\sum_{\mathbf{x}} p_{\mathbf{X}} (\mathbf{x})\Tr \left[\hat{\sigma}_{\alpha_n}^{\otimes n} \hat{\Pi}_{\mathbf{x}}^n \right] $ be the term on the left hand side of \eqref{eq:exponentialBound}. As $\sigmaNotN$ and $\hat{\Pi}_{\mathbf{x}}^n$ commute,
\begin{align}
 C_2&\leq  \Tr \left[ \sum_{\mathbf{x'}}  p_{\mathbf{X'}} (\mathbf{x'}) \mathcal{E}_{\sigmaNotN}(\sigmaAlphaN) \hat{\Pi}_{\mathbf{x'}}^n \right] \label{eq:applyPinchCommute} \\
&= \sum_{\mathbf{x'}}  p_{\mathbf{X'}} (\mathbf{x'})  \Tr \left[  \mathcal{E}_{\sigmaNotN}(\sigmaAlphaN) \hat{\Pi}_{\mathbf{x'}}^n \right] \label{eq:pinchingLinearity} \\
&= \sum_{\mathbf{x'}}  p_{\mathbf{X'}} (\mathbf{x'}) \Tr \left[  \left( \sigmaNotN \right)^{-1}\sigmaNotN \mathcal{E}_{\sigmaNotN}(\sigmaAlphaN) \hat{\Pi}_{\mathbf{x'}}^n \right]\label{eq:sigmaNotInverse}\\ 
&=\sum_{\mathbf{x'}}  p_{\mathbf{X'}} (\mathbf{x'}) \Tr \left[ \left( \sigmaNotN \right)^{-1} \mathcal{E}_{\sigmaNotN}(\sigmaAlphaN)\sigmaNotN  \hat{\Pi}_{\mathbf{x'}}^n \right], \label{eq:sigmaNotInversecommutes}
\end{align}
where \eqref{eq:applyPinchCommute} follows from application of Lemma \ref{lem:pinchingMapTr} for $\hat{A} = \sigmaNotN$, $\hat{B} = \sigmaAlphaN$, and $\hat{C} = \hat{\Pi}_{\mathbf{x'}}^n$, \eqref{eq:pinchingLinearity} follows from the linearity of the trace, and \eqref{eq:sigmaNotInversecommutes} follows from the fact that $\sigmaNotN$ and $\mathcal{E}_{\sigmaNotN}(\sigmaAlphaN)$ commute. Now, $\left(\hat{\sigma}_{0}^{\otimes n}\right)^{-1}\mathcal{E}_{\sigmaNotN}\succeq0$ and $\left(\mathcal{E}_{\sigmaNotN}(\hat{\sigma}_{\mathbf{x'}}^n)-e^a\sigmaNotN\right)\hat{\Pi}_{\mathbf{x'}}^n\succeq0$
imply that 
$\Tr \left[\left(\hat{\sigma}_{0}^{\otimes n}\right)^{-1}\mathcal{E}_{\sigmaNotN}(\sigmaAlphaN)\left(\mathcal{E}_{\sigmaNotN}(\hat{\sigma}_{\mathbf{x'}}^n)-e^a\sigmaNotN\right)\hat{\Pi}_{\mathbf{x'}}^n\right] \geq 0$. Linearity of the trace implies 
\begin{align}
\Tr\left[\left(\hat{\sigma}_{0}^{\otimes n}\right)^{-1}\mathcal{E}_{\sigmaNotN}(\sigmaAlphaN)\sigmaNotN\hat{\Pi}_{\mathbf{x'}}^n\right]   \leq e^{-a}\Tr\left[\left(\hat{\sigma}_{0}^{\otimes n}\right)^{-1}\mathcal{E}_{\sigmaNotN}(\sigmaAlphaN)\mathcal{E}_{\sigmaNotN}(\hat{\sigma}_{\mathbf{x'}}^n)\hat{\Pi}_{\mathbf{x'}}^n\right].  \label{eq:lem7linearity}
\end{align}
We can further bound the right-hand side of \eqref{eq:lem7linearity} as 
\begin{align}
    e^{-a}\Tr\left[\left(\hat{\sigma}_{0}^{\otimes n}\right)^{-1}\right.&\left.\mathcal{E}_{\sigmaNotN}(\sigmaAlphaN)\mathcal{E}_{\sigmaNotN}(\hat{\sigma}_{\mathbf{x'}}^n)\hat{\Pi}_{\mathbf{x'}}^n\right]\nonumber\\ &\leq e^{-a}\sqrt{\Tr\left[\mathcal{E}_{\sigmaNotN}(\hat{\sigma}_{\mathbf{x'}}^n)\left(\mathcal{E}_{\sigmaNotN}(\sigmaAlphaN)\right)^2\left(\hat{\sigma}_{0}^{\otimes n}\right)^{-2}\right]\Tr\left[\mathcal{E}_{\sigmaNotN}(\hat{\sigma}_{\mathbf{x'}}^n)\left(\hat{\Pi}_{\mathbf{x'}}^n\right)^2\right]}\label{eq:CauchySchwarz}
    \\
    &\leq e^{-a}\sqrt{\Tr\left[\mathcal{E}_{\sigmaNotN}(\hat{\sigma}_{\mathbf{x'}}^n)\left(\mathcal{E}_{\sigmaNotN}(\sigmaAlphaN)\right)^2\left(\hat{\sigma}_{0}^{\otimes n}\right)^{-2}\right]},\label{eq:CSProjInequality}
\end{align}
where \eqref{eq:CauchySchwarz} is the Cauchy-Schwarz inequality for trace inner products and follows from the fact that $\mathcal{E}_{\sigmaNotN}(\hat{\sigma}_{\mathbf{x'}}^n)$ is Hermitian, and \eqref{eq:CSProjInequality} is because $\Tr\left[\mathcal{E}_{\sigmaNotN}(\hat{\sigma}_{\mathbf{x'}}^n)\left(\hat{\Pi}_{\mathbf{x'}}^n\right)^2\right]=\Tr\left[\mathcal{E}_{\sigmaNotN}(\hat{\sigma}_{\mathbf{x'}}^n)\hat{\Pi}_{\mathbf{x'}}^n\right]\leq\Tr\left[\mathcal{E}_{\sigmaNotN}(\hat{\sigma}_{\mathbf{x'}}^n)\right]\leq 1$. 

Combining \eqref{eq:sigmaNotInversecommutes},  \eqref{eq:lem7linearity} and \eqref{eq:CSProjInequality} yields:

\begin{align}
 C_2&\leq e^{-a}\sum_{\mathbf{x'}}  p_{\mathbf{X'}} (\mathbf{x'}) \sqrt{\Tr\left[\mathcal{E}_{\sigmaNotN}(\hat{\sigma}_{\mathbf{x'}}^n)\left(\mathcal{E}_{\sigmaNotN}(\sigmaAlphaN)\right)^2\left(\hat{\sigma}_{0}^{\otimes n}\right)^{-2}\right]}\\
&\leq e^{-a} \sqrt{\sum_{\mathbf{x'}}  p_{\mathbf{X'}}(\mathbf{x'}) \Tr\left[\mathcal{E}_{\sigmaNotN}(\hat{\sigma}_{\mathbf{x'}}^n)\left(\mathcal{E}_{\sigmaNotN}(\sigmaAlphaN)\right)^2\left(\hat{\sigma}_{0}^{\otimes n}\right)^{-2}\right]}\label{eq:JensenSqrt}\\
&= e^{-a} \sqrt{\Tr\left[\mathcal{E}_{\sigmaNotN}\left(\sum_{\mathbf{x'}}  p_{\mathbf{X'}}(\mathbf{x'}) \hat{\sigma}_{\mathbf{x'}}^n\right)\left(\mathcal{E}_{\sigmaNotN}(\sigmaAlphaN)\right)^2\left(\hat{\sigma}_{0}^{\otimes n}\right)^{-2}\right]}\label{eq:LinearityTrPinch}\\
&= e^{-a} \sqrt{\Tr\left[\left(\mathcal{E}_{\sigmaNotN}(\sigmaAlphaN)\right)^3\left(\hat{\sigma}_{0}^{\otimes n}\right)^{-2}\right]}\label{eq:DefnSigmaAlphaN2},
\end{align}
where \eqref{eq:JensenSqrt} follows from Jensen's inequality, \eqref{eq:LinearityTrPinch} is due to the linearity of the trace and pinching, and \eqref{eq:DefnSigmaAlphaN2} follows from the definition of $\sigmaAlphaN$.
Application of Lemma \ref{lem:pinchingMapConvex} to the term inside the square root of \eqref{eq:DefnSigmaAlphaN2} yields:
\begin{align}
 C_2&\leq e^{-a}\sqrt{\Tr\left[\mathcal{E}_{\sigmaNotN}\left(\left(\sigmaAlphaN\right)^3\right)\left(\hat{\sigma}_{0}^{\otimes n}\right)^{-2}\right]}\label{eq:pinchingProp3} \\
&=  e^{-a}\sqrt{\Tr \left[ \left(\sigmaAlphaN\right)^3 \left(\sigmaNotN\right)^{-2} \right]}\label{eq:pinchingProp2},
\end{align}
where \eqref{eq:pinchingProp2} follows from Lemma \ref{lem:pinchingMapTr}. The properties of the tensor product and its trace yield:
\begin{align}
 C_2 &\leq  e^{-a} \sqrt{\Tr \left[ \left(\hat{\sigma}_{\alpha_n}^3\right)^{\otimes n} \left(\hat{\sigma}_{0}^{-2}\right)^{\otimes n} \right]} \\
&= e^{-a} \left(\Tr \left[  \hat{\sigma}_{\alpha_n}^3  \hat{\sigma}_0^{-2} \right] \right)^{n/2} \\
&=  e^{-a} \left(1 -3\alpha_n^2+ (1-\alpha_n)\alpha_n^2\left( \Tr \left[\hat{\sigma}_{\neg 0}\hat{\sigma}_0\hat{\sigma}_{\neg 0}\hat{\sigma}_0^{-2}\right] +2\Tr\left[\hat{\sigma}_{\neg 0}^2\hat{\sigma}_0^{-1}\right]\right)\right.\nonumber\\&\phantom{=+e^{-a}((}\left.+ \alpha_n^3\left(\Tr\left[\hat{\sigma}_{\neg 0}^3\hat{\sigma}_0^{-2}\right]+2\right)\right)^{n/2}\label{eq:algebra}\\
&\leq e^{-a} \bigg(1 + \alpha_n^2\underbrace{\left(\Tr \left[\hat{\sigma}_{\neg 0}\hat{\sigma}_0\hat{\sigma}_{\neg 0}\hat{\sigma}_0^{-2}\right] + 2\Tr\left[\hat{\sigma}_{\neg 0}^2\hat{\sigma}_0^{-1}\right]\right)}_{\beta_1}+ \alpha_n^3\underbrace{\left(\Tr\left[\hat{\sigma}_{\neg 0}^3\hat{\sigma}_0^{-2}\right] +2\right)}_{\beta_2}\bigg)^{n/2} \label{eq:betadef}
\end{align}
where we define $\hat{\sigma}_{\neg 0}\triangleq \sum_{x\in\mathcal{X}\setminus\{0\}}\pi_x \hat{\sigma}_x$, \eqref{eq:algebra} follows by expanding $\hat{\sigma}_{\alpha_n}$, \eqref{eq:betadef} follows from dropping the negative terms, and constants $\beta_1$, $\beta_2$ are positive since they are traces of the product of two positive operators. 
Now, using the fact that $\log(1+x)\leq x$ for $x\geq -1$, we have:
\begin{align}
 C_2 &\leq  e^{-a} \left(1 + \alpha_n^2 \beta_1 + \alpha_n^3 \beta_2\right)^{n/2} \\
&\leq  e^{-a+\frac{n}{2}\left(\alpha_n^2\beta_1 + \alpha_n^3 \beta_2 \right)}. \label{eq:exponentialBound}
\end{align}
 Substituting the definition of $\alpha_n$ from Section \ref{sec:QSCS} in \eqref{eq:exponentialBound} completes the proof.
\end{IEEEproof}

\section*{Acknowledgement}
The authors are grateful to Dennis Goeckel, Don Towsley, Uzi Pereg, and Elyakim Zlotnick for helpful discussions, and to Matthieu Bloch for answering the many questions about \cite{bloch15covert}.  The authors also thank Mark Wilde for encouraging us to continue working on this paper.\bibliographystyle{IEEEtran}
\bibliography{papers}

\end{document}